\documentclass[preprint,onecolumn,a4paper]{elsarticle}

\usepackage[T1]{fontenc}
\usepackage{lmodern}
\usepackage[english]{babel}

\usepackage{mathrsfs}
\usepackage{amsfonts}
\usepackage{amssymb}
\usepackage{graphicx}% Include figure files
\usepackage{dcolumn} % Align table columns on decimal point
\usepackage{amsmath}
\usepackage{bm}      % bold math
\usepackage{graphicx}
\usepackage{color}
\usepackage{epsfig,psfrag,subfigure,subeqnarray}
\usepackage{relsize}

\def\lan{\langle}
\def\ran{\rangle}
\def\va{\varepsilon}
\def\dag{\dagger}

\def\vk{{\bf k}}
\def\vR{{\bf R}}

\def\vp{{\bf p}}

\def\vr{{\bf r}}

\def\v0{{\bf 0}}

\def\vK{{\bf K}}

\newcommand{\bd}{\begin{equation}}
\newcommand{\ed}{\end{equation}}
\newcommand{\be}{\begin{equation}}
\newcommand{\ee}{\end{equation}}
\newcommand{\bt}{\begin{split}}
\newcommand{\et}{\end{split}}

\newcommand{\bn}{\begin{align}}
\newcommand{\en}{\end{align}}
\newcommand{\bea}{\begin{eqnarray}}
\newcommand{\eea}{\end{eqnarray}}
\newcommand{\ba}{\begin{array}}
\newcommand{\ea}{\end{array}}
\newcommand{\nn}{\nonumber}

\DeclareMathAlphabet\mathbfcal{OMS}{cmsy}{b}{n}

\begin{document}
\begin{frontmatter}
\title{A fresh view on Frenkel excitons  \\
Electron-hole \textit{pair} exchange and many-body formalism}

\author[1]{Shiue-Yuan Shiau}
\ead{shiau.sean@gmail.com}
\author[2]{Monique Combescot}
\ead{monique.combescot@insp.upmc.fr}
\address[1]{Physics Division, National Center for Theoretical Sciences, 10617 Taipei, Taiwan}
\address[2]{Institut des NanoSciences de Paris, Sorbonne Universit\'e, CNRS, 75005 Paris, France}

\begin{abstract}

Excitons are coherent excitations that travel over the semiconductor sample. Two types are commonly distinguished: Wannier excitons that are formed in inorganic materials, and Frenkel excitons that are formed in organic materials and rare-gas crystals.

\noindent (\textit{i}) Wannier excitons are made from \textit{delocalized} conduction electrons and \textit{delocalized} valence holes. The Coulomb interaction acts in two different ways. The \textit{intraband} processes bind a free electron and a free hole into a Wannier exciton wave, while the \textit{interband} processes separate the optically bright excitons made of spin-singlet electron-hole pairs from dark excitons made of spin-triplet pairs, with an additional transverse-longitudinal splitting that results from a nonanalytical Coulomb scattering which depends on the exciton wave vector direction with respect to the crystal axes.

\noindent (\textit{ii}) Frenkel excitons are made from \textit{highly localized} excitations. The Coulomb interaction then acts in \textit{one way only}, the intralevel Coulomb processes between sites, analogous to intraband processes, being negligible for on-site excitations in the tight-binding limit. The \textit{interlevel} Coulomb processes, analogous to interband processes, produce both, the Frenkel exciton wave and its singular splitting, through electron-hole \textit{pair} exchange between sites: the excitonic wave is produced by delocalizing the on-site excitations through the recombination of an electron-hole pair on one site and its creation on another site. These interlevel processes also split the exciton level in exactly the same singular way as for Wannier excitons. Importantly, since the interlevel Coulomb interaction only acts on spin-singlet pairs, just like the electron-photon interaction, optically dark pairs do not form excitonic waves; as a result, Frenkel excitons are optically bright in the tight-binding limit.

We here present a fresh approach to Frenkel excitons in cubic semiconductor crystals, with a special focus on the spin and spatial degeneracies of the electronic states. This approach uses a second quantization formulation of the problem in terms of creation operators for electronic states on \textit{all} lattice sites --- their creation operators being true fermion operators in the tight-binding limit valid for semiconductors hosting Frenkel excitons. This operator formalism avoids using cumbersome ($6N_s\times 6N_s$) Slater determinants --- 2 for spin, 3 for spatial degeneracy and $N_s$ for the number of lattice sites --- to represent state wave functions out of which the Frenkel exciton eigenstates are derived. A deep understanding of the tricky Coulomb physics that takes place in the Frenkel exciton problem, is a prerequisite for possibly diagonalizing this very large matrix analytically. This is done in three steps:

\noindent (\textit{i}) the first diagonalization, with respect to lattice sites, follows from transforming excitations on the $N_s$ lattice sites $\textbf{R}_\ell$ into $N_s$ exciton waves $\textbf{K}_n$, by using appropriate phase prefactors; 

\noindent (\textit{ii}) the second diagonalization, with respect to spin, follows from the introduction of spin-singlet and spin-triplet electron-hole pair states, through the commonly missed sign change when transforming electron-absence operators into hole operators; 

\noindent (\textit{iii}) the third diagonalization, with respect to threefold spatial degeneracy, leads to the splitting of the exciton level into one longitudinal and two transverse modes, that result from the singular interlevel Coulomb scattering in the small $\textbf{ K}_n$ limit.

To highlight the advantage of the second quantization approach to Frenkel exciton we here propose, over the standard first quantization procedure used for example in the seminal book by R. Knox, we also present detailed calculations of some key results on Frenkel excitons when formulated in terms of Slater determinants.

Finally, as a way forward, we show how many-body effects between Frenkel excitons can be handled through a composite boson formalism appropriate to excitons made from electron-hole pairs with zero spatial extension. Interestingly, this Frenkel exciton study led us to reformulate the dimensionless parameter that controls exciton many-body effects --- first understood in terms of the Wannier exciton Bohr radius driven by Coulomb interaction --- as a parameter entirely driven by the Pauli exclusion principle between the exciton fermionic components. This formulation is not only valid for Wannier excitons but also for composite bosons like Cooper pairs.

\end{abstract}
\date{\today}

\end{frontmatter}
\tableofcontents

\section{Introduction\label{sec1}}

 Excitons have been very early identified as ``quantum of electronic excitations traveling in a periodic structure, whose motion is characterized by a wave vector''\cite{Dexter1981}. These electronic excitations, commonly produced by photon absorption in a semiconductor, exist in two configurations, known as Wannier excitons\cite{Wannier1937} and Frenkel excitons\cite{Frenkel1931a,Frenkel1931b}. Although they both result from electron-hole pairs correlated by the Coulomb interaction, this interaction acts in opposite ways because the pairs on which  Wannier or Frenkel excitons are made have a totally different spatial extension\cite{Knox1963,Monicbook}. The  electronic picture for semiconductors hosting Wannier excitons starts with itinerant conduction and valence electrons, with energies separated by a gap\cite{Bastardbook,Cardona}. Through \textit{intraband} Coulomb processes, a delocalized conduction electron and a delocalized valence-electron absence, \textit{i.e}, a valence hole, end  by forming a bound state having a delocalized center of mass. By contrast, the electronic picture for semiconductors hosting Frenkel excitons starts with excitations on the lattice sites of a periodic crystal\cite{Agra1961,Agrabook,Davydov1971}. These highly localized excitations are delocalized through \textit{interlevel} Coulomb processes, to also end as electron-hole pairs with a center of mass delocalized over the whole sample.

In short, the Coulomb interaction localizes a delocalized electron and a delocalized hole to  form a Wannier exciton bound state, through \textit{intraband} processes, while it delocalizes an electron and a hole localized on the same lattice site, to form a Frenkel exciton wave, through \textit{interlevel} processes.

 \textit{Intralevel} Coulomb processes, analogous to intraband processes, \textit{a priori} exist for Frenkel excitons. However, as they require finite overlaps between the electronic wave functions of different lattice sites, these processes are negligible in the tight-binding limit, \textit{i.e.}, no wave function overlap between sites, an approximation valid for semiconductors hosting Frenkel excitons. In addition, keeping these overlaps prevents using a second quantization procedure based on a clean set of fermionic operators, as we here propose, whereas the introduction of finite wave function overlaps does not bring any significant effect.

 Similarly, \textit{interband} Coulomb processes, analogous to interlevel processes, also exist for Wannier excitons. They physically correspond to one conduction electron returning to the valence band, while a valence electron is excited to the conduction band. These processes are small compared to intraband processes; but they cannot be neglected because they bring two significant effects:
 
 \noindent \textit{(i)} the interband Coulomb interaction only acts on electron-hole pairs that are in a spin-singlet  state, just like the electron-photon interaction. As a direct consequence, the interband Coulomb interaction participates in the splitting between bright and dark excitons\cite{monic_epl2022}, that ultimately drives the exciton Bose-Einstein condensation to occur in a dark state\cite{alloing}.
 
 \noindent \textit{(ii)} the scattering associated with interband processes is singular in the limit of small exciton center-of-mass wave vector. For cubic crystals, this singularity induces a splitting of the degenerate Wannier exciton level, which has a direct link to the transverse-longitudinal splitting in the exciton-polariton problem\cite{Onodera,Andreani1,Andreani2}.

In short, the interband and interlevel Coulomb processes are physically similar: they both correspond to the recombination of an electron-hole pair along with the excitation of another pair. In the case of Frenkel excitons, these processes are essential: they are the ones responsible for the exciton formation, that is, the delocalization of on-site excitations into a wave quantum over the whole sample. By contrast, the interband processes for Wannier excitons are secondary: they just split the otherwise degenerate exciton level.

\textbf{The purpose of this manuscript is to provide a microscopic understanding}, starting from scratch, of the Coulomb processes that are responsible not only for the Frenkel exciton formation, but also for the splitting of its degenerate levels when the electronic levels are not only spin but also spatially degenerate. This understanding brings a fresh view to its Wannier exciton analog, known as ``electron-hole exchange'' --- an improper name because different fermions do not quantum exchange: this effect just comes from interband Coulomb processes. Our goal is to catch the interplay between the spatial part of the problem that enters the Coulomb scatterings through the electronic wave functions, and the spin part conserved in a Coulomb process. Discussing these two parts separately enlightens the ``electron-hole exchange'' splitting discussed in papers dealing with Wannier excitons, and its so-called ``short-range'' and ``nonanalytical long-range'' contributions. 

Frenkel excitons follow from the diagonalization of the system Hamiltonian in a ($2\times 3 \times N_s$)-fold excitation subspace, $2$ for the spin degeneracy of the excited electron, $3$ for the spatial degeneracy of the unexcited electron level, and $N_s$ for the number of lattice sites on which the excitation can take place. The analytical diagonalization of the resulting ($6N_s\times 6N_s$) matrix from which the Frenkel exciton eigenstates are obtained, is a formidable mathematical task that requires a deep physical understanding of the problem, to possibly solve it analytically. The best way to reach this understanding is to separately study its three parts --- lattice site degeneracy, spin degeneracy and spatial degeneracy --- before putting them together.

Numerous Frenkel exciton-based applications have been proposed over the past several decades. In this manuscript, we have chosen not  to enter into these applications. Still, for interested readers, we rather  list some references devoted to quantum computing\cite{Castellanos,Yurke}, photosynthesis\cite{Fassioli,Blankenship}, light harvesting devices\cite{Jiang,Minami}, light emitting devices\cite{Spano,Zhao,emitting}, and solar cells\cite{Gunes,GangLi,chargecarrier}, in all of which the Frenkel exciton plays a key role.

\textbf{The paper is organized as follows}:

In Sec.~\ref{sec2}, we analyze the whole Frenkel exciton problem step by step.

In Sec.~\ref{sec3}, we forget spin and spatial degeneracies. This allows us to identify the phase prefactor that transforms electronic excitations on any lattice sites into coherent wave excitations. This also highlights the necessity for the electronic ground and excited levels to have a different parity in order to possibly form a Frenkel exciton. As a direct consequence, taking into account the state spatial degeneracy is mandatory.

In Sec.~\ref{sec4}, we introduce the spin but not yet the spatial degeneracy. This part highlights  the importance of formulating the problem in terms of electrons and holes: indeed, this formulation leads us, in a natural way, to distinguish spin-triplet from spin-singlet pairs and to readily catch that spin-triplet electron-hole pairs do not suffer interlevel Coulomb processes; so, these pairs do not participate in the Frenkel exciton formation.

In Sec.~\ref{sec5}, we consider the spatial degeneracy of the electronic level but we forget the spin degeneracy. This allows us to pin down the effect of this degeneracy on the interlevel Coulomb processes and the splitting of the exciton level that comes from the singular behavior  of the Coulomb scattering in the limit of small exciton center-of-mass wave vector.

In Sec.~\ref{sec6}, we consider Frenkel excitons made of electronic states having both a threefold spatial degeneracy and a twofold spin degeneracy. The previous sections provide the necessary help to comprehend the formation of Frenkel excitons in its full complexity.

In Sec.~\ref{Sota}, we discuss the presentation of Frenkel excitons given in the representative exciton ``Bible''  written by R. Knox\cite{Knox1963}. It relies on a first quantization formulation of the problem that makes use of Slater determinants for many-body state wave functions. It is well known that calculations involving Slater determinants are very cumbersome. Many results in this book are qualified as ``easy to find'', somewhat dismissive to our opinion. This is why we find it useful here to provide some detailed derivations. These derivations once more demonstrate the great superiority of the second quantization formalism when dealing with a many-body problem, which fundamentally is what the Frenkel exciton problem is, in spite of the fact that we ultimately end with one electron-hole pair only. 

In Sec.~\ref{sec8}, we briefly show how to handle many-body effects between Frenkel excitons through a composite boson formalism for excitons having a ``size'' equal to zero. Wannier excitons, characterized by two quantum indices, namely their center-of-mass wave vector and their relative-motion index, have a finite size, their Bohr radius. By contrast, Frenkel excitons have one quantum index only, their wave vector, the size of these excitons being vanishingly small because they are made of on-site excitations. The ``sizelessness''  in the case of Frenkel excitons forced us to reconsider the physics of the dimensionless parameter that controls exciton many-body effects, first understood in terms of the Coulomb-driven exciton overlap through the exciton Bohr radius. We ultimately understood that this parameter is entirely controlled by the Pauli exclusion principle between the fermionic components of the excitons --- an understanding that also extends to Cooper pairs which are composite bosons made of opposite-spin electrons.

We then conclude.

\section{The Frenkel exciton problem, step by step \label{sec2}}

The Frenkel exciton problem at its root is a math problem: the diagonalization of a $6N_s\times 6N_s$ matrix for electrons on the $N_s$ lattice sites of a semiconductor crystal, the electronic states having a twofold spin degeneracy and a ground level with a threefold spatial degeneracy. No doubt, a good guess of the form of the eigenstates, based on wise physical considerations, is necessary to possibly solve this formidable math problem analytically. The purpose of this section is to guide the reader to the result, through a convoluted journey that we hope shall ultimately appear as an ``easy ride'', once the physics of each step is revealed. This journey experiences four different physical landscapes:
 
 \noindent (\textit{i}) the electronic levels for atoms or molecules located on a periodic lattice, in the absence of spin and spatial degeneracies;
 
  \noindent  (\textit{ii}) these electronic levels with spin, but no spatial degeneracy; 
  
   \noindent(\textit{iii}) these electronic levels with no spin but a threefold spatial degeneracy either for the excited level or for the ground level; 
   
    \noindent(\textit{iv}) these electronic levels with both, spin and spatial degeneracies.

\subsection{In the absence of spin and  spatial degeneracies}

Let us begin with the simplest problem, to establish the procedure\cite{Monicprb2008}: $N_s$ electrons with charge $-|e|$ and $N_s$ ions with charge $|e|$ located at the nodes of a periodic lattice.

The first step is to find the physically relevant basis for one-electron states that will be used to define the one-electron operators for a quantum formulation of the problem. The whole spectrum made of the $\nu$ eigenstates for one electron in the presence of one ion located at the lattice site $\textbf{R}_\ell$, comes across as a possible basis. Indeed, being Hamiltonian eigenstates, the ($\nu;\textbf{R}_\ell$) states for a particular $\textbf{R}_\ell$ but different $\nu$'s form a complete basis that can in principle be used to describe electrons located on any other lattice site, provided that enough $\nu$ states are included into the description. It is however clear that a better idea is to use a basis in which enter all $\textbf{R}_\ell$ sites. This can be done by restricting the $\nu$ states to the  ground and lowest-excited levels, $(g,e)$, provided that the $(g,e)$ states from different lattice sites have a very small wave function overlap, as for materials hosting Frenkel excitons: in the tight-binding limit, that is, no wave function overlap between different sites, the $(\nu;\textbf{R}_\ell)$ electronic states for $\nu=(g,e)$ and $\ell=(1,\cdots,N_s)$, can indeed be used to cleanly define the one-body fermionic operators necessary for a second quantization formulation of the problem.

The second step is to write the system in second quantization using  the creation operators $\hat{a}^\dag_{\nu,\ell}$ for electrons in these $(\nu;\textbf{R}_\ell)$ states. The system ground state essentially corresponds to each $(g;\textbf{R}_\ell)$ ground state of all $N_s$ lattice sites occupied by one electron, while for the lowest set of excited states, one $(g;\textbf{R}_\ell)$ ground state is replaced by the $(e;\textbf{R}_\ell)$ excited state of the \textit{same} lattice site: indeed, a jump to the excited level of another site would lead to a higher-energy excitation due to the electrostatic cost resulting from charge separation; this cost, large in the tight-biding limit, forces the electronic cloud to stay close to the $\textbf{R}_\ell$ site.
With respect to the system ground state, this excited state corresponds to one excited-level electron and one ground-level electron absence, on the same lattice site.

The third step is to turn from excited-level electron and ground-level electron absence to electron and hole. Although this procedure is not mandatory in the absence of spin, it actually corresponds to the proper physical description of the Frenkel exciton problem in terms of on-site electron-hole pair excitations. Turning to electron-hole pairs becomes crucial when the spin is introduced because this formulation naturally goes along with a splitting between spin-singlet and spin-triplet subspaces, neatly defined when speaking in terms of electrons and holes.

We are then left with the diagonalization of a $N_s\times N_s$ matrix for one electron-hole pair excitation on each of the $\ell=(1,2,\cdots, N_s)$ lattice sites, the coupling between lattice sites being mediated by interlevel Coulomb processes in which the excited electron of one lattice site returns to the ground level of the same site, while another site is excited. This diagonalization is easy to perform by turning from $N_s$ pairs localized on the $\vR_\ell$ sites, to $N_s$ correlated pairs, characterized by a wave vector $\vK_n$ with $n=(1,\cdots, N_s)$, that are linear combinations of the $\vR_\ell$ pairs, with a prefactor which is just a phase 
\be\label{0}
e^{i\vK_n \cdot\vR_\ell}
\ee  
This linear combination, known as Frenkel exciton, fundamentally corresponds to delocalizing the on-site excitations over the whole sample.

Importantly, the interlevel Coulomb scatterings responsible for the pair delocalization, differ from zero provided that the ground and excited electronic levels have a different parity. For this reason, it is mandatory to bring  the spatial degeneracy of the electronic states into the problem, in order to possibly explain the Frenkel exciton formation.

\subsection{With spin but no spatial degeneracy}

Before introducing this spatial degeneracy, let us tackle a generic difficulty associated with the electronic state degeneracy that comes from the electron spin. Each ground level then is occupied by two electrons having opposite spins. This goes along with the fact that the ion on each lattice site must carry a $2|e|$ charge, as required by the system neutrality. After some thoughts that will be detailed later on, we have reached the conclusion that the relevant electronic states for second quantization are not the ones of an electron in the presence of a $2|e|$ ion, but still the ones of an electron in the presence of a $|e|$ charge, this electron possibly having an up or down spin. The lowest set of excited states then corresponds to one of the two ground-level electrons jumping with its ($\pm1/2$) spin, to the excited level of the same lattice site: indeed, the Coulomb interaction or the electron-photon interaction that can produce such electronic excitation, conserves the spin. The system then has two possible excited states on each of the $N_s$ lattice sites of the crystal.  The Frenkel excitons are constructed by diagonalizing the resulting ($2N_s \times 2N_s$) matrix that represents the system Hamiltonian in this lowest excited subspace.

The appropriate way to catch the physics of the Coulomb coupling between these excited states is to turn to electron-hole pairs and to write these pairs in their spin-triplet and spin-singlet configurations. The phase factor that appears between electron destruction operator and hole creation operator\cite{SS2021,JPCM2021}, differentiates spin-singlet from spin-triplet pairs. We then easily find that the pairs in the spin-singlet configuration are the only ones that suffer the interlevel Coulomb processes responsible for the delocalization of on-site excitations over the whole sample. So, by simply writing the problem in terms of spin-triplet and spin-singlet  states, the ($2N_s\times 2N_s$) matrix splits into a diagonal ($N_s\times N_s$) matrix for the spin-triplet subspace, and a nondiagonal ($N_s\times N_s$) matrix for the spin-singlet subspace. The diagonalization of the spin-singlet matrix is then performed, as in the absence of spin, by turning from pairs localized on the $\vR_\ell$ lattice sites to delocalized $\vK_n$ pairs, through the same phase factor as the one given in Eq.~(\ref{0}).

\subsection{With spatial degeneracy but no spin}

\noindent $\bullet$ Let us first consider that the ground $(g)$ level is nondegenerate and  the excited $(e)$ level is threefold degenerate because this degeneracy configuration is simpler. When excited, one electron in the ground-level state $(g; \vR_\ell)$ of the $\vR_\ell$ lattice site, jumps into one of the three excited states $(\mu,e;\vR_\ell)$ of the same site, with $\mu=(x,y,z)$ along the cubic crystal axes.
The system then has $3N_s$ possible excited states, each of which corresponds to an electron in one of the three excited states $\mu$ on one of the $N_s$ lattice sites and an empty ground level on the same site.

A first diagonalization of the resulting ($3N_s\times 3N_s$) matrix is performed with respect to lattice sites, by using the $e^{i\vK_n \cdot\vR_\ell}$ phase factor of Eq.~(\ref{0}). This delocalizes the excitations on the $\vR_\ell$ lattice sites as $\vK_n$ wave over the whole sample.

We are left with $N_s$ submatrices ($3\times3$) that are associated with the $N_s$ different $\vK_n$ wave vectors.  It turns out that the interlevel Coulomb scatterings that enters their nondiagonal matrix elements, are singular in the small $\vK_n$ limit. This explains why their diagonalization leads to a splitting of the $\vK_n$ Frenkel exciton wave, into one ``longitudinal'' and two ``transverse'' levels with respect to the  $\vK_n$ direction, the energy splitting depending on the direction of the $\vK_n/K_n$ vector with respect to the cubic axes.

\noindent $\bullet$ The situation seems at first more complicated when the ground level is threefold and the excited level is nondegenerate because the ion on the $\vR_\ell$ lattice site now hosts three ground-level electrons; so, its charge is $3|e|$, due to crystal neutrality. When excited, one of the three ground electrons $(\mu,g;\vR_\ell)$ jumps to the unique excited level $(e;\vR_\ell)$.

The smart way to understand these excited states is not to see them as a $3|e|$ ion with two ground electrons and one excited electron, but in terms of electronic excitations: the $\vR_\ell$ site then hosts three types of excited states that can be labeled by the $\mu$ index of the ground electron that has jumped to the unique excited level. Within this formulation, the calculation  follows straightforwardly the one for a nondegenerate ground level and a threefold excited level: we first perform the diagonalization of the resulting ($3N_s\times 3N_s$) matrix with respect to the $\vR_\ell$ lattice sites, to get $N_s$ excitons $\vK_n$. Next, we perform the diagonalization of the resulting ($3\times 3$) submatrix associated with a particular $\vK_n$, from which we obtain the transverse-longitudinal splitting of the exciton level that arises from the singularity of the interlevel Coulomb scattering in the small $\vK_n$ limit.

\subsection{With spin and  spatial degeneracies}

Having understood the consequences of spin and spatial degeneracies separately, the strategy to handle them all together becomes easier to construct.

We first introduce the one-electron eigenstates in the presence of a $|e|$ ion located on the $\vR_\ell$ lattice site. As a basis, we use the ground and lowest excited levels for all lattice sites $\vR_\ell$, provided that these states are highly localized, \textit{i.e.}, no overlap between wave functions of different lattice sites. We moreover consider that the ground level is spatially threefold and the excited level is nondegenerate.

These ($\nu;\vR_\ell$) states with $\nu=(\mu, g)$ or $\nu=e$, are used to define the one-electron operators, in terms of which we formulate the system in second quantization.    

Next, we turn from  excited electron and ground electron absence to electron and hole. The phase factor that appears in this change differentiates spin-singlet from spin-triplet electron-hole pair subspaces in a straightforward way.

The Frenkel exciton problem begins with electron-hole pairs having a ($2 \times3$)  degeneracy, due to spin and spatial degrees of freedom, that are each localized on one of the $N_s$ lattice sites $\vR_\ell$. So, the matrix we have to diagonalize is ($6N_s\times 6N_s$). We first turn to electrons and holes and then to the spin-singlet and spin-triplet combinations. The part in the spin-triplet subspace readily appears diagonal because interlevel Coulomb processes do not exist for spin-triplet pairs. So, we are left with diagonalizing a ($3N_s\times 3N_s$) matrix in the spin-singlet subspace. A first diagonalization is performed by switching from  $N_s$ lattice sites $\vR_\ell$ to  $N_s$ wave vectors $\vK_n$ through the phase factor given in Eq.~(\ref{0}). We remain with a set of ($3\times 3$) submatrices associated with different $\vK_n$ wave vectors. Due to the singular behavior of the interlevel Coulomb processes in the small $\vK_n$ limit, their diagonalization leads to the same transverse-longitudinal splitting along the $\vK_n$ direction, as the one found in the absence of spin. 

Let us now study these four steps in details, to confirm the above understanding.

\section{Frenkel exciton without spin and spatial degeneracies\label{sec3}}

\subsection{Appropriate basis for quantum formulation}

\subsubsection{System Hamiltonian} 

We consider a neutral system made of $N_s$ free electrons with mass $m_0$, charge $-|e|$, spatial coordinate $\textbf{r}_j$ for $j=(1,\cdots,N_s$), and ${N_s}$ ions with infinite mass, charge $|e|$, located at the $\vR_\ell$ nodes of a periodic lattice for $\ell=(1,\cdots,N_s$). The system Hamiltonian reads in first quantization as
\bea
\label{2'}
H_{N_s}&=&\sum_{j=1}^{N_s} \frac{\vp_j^2}{2m_0}+\sum_{j=1}^{N_s} \sum_{\ell=1}^{N_s} \frac{-e^2}{|\vr_j-\vR_\ell|}+\frac{1}{2}\sum_{j=1}^{N_s} \sum_{j'\not=j}^{N_s} \frac{e^2}{|\vr_j-\vr_{j'}|}\nn\\
%\nn\\
&&+\frac{1}{2}\sum_{\ell=1}^{N_s} \sum_{\ell'\not=\ell}^{N_s} \frac{e^2}{|\vR_\ell-\vR_{\ell'}|}\label{1}
\eea
The first term corresponds to the electron kinetic energy, the second term to the electron-ion attraction, the third term to the electron-electron repulsion.
The last term, which is a constant with respect to the electron motion, ensures the elimination of volume divergent terms coming from the long-range character of the Coulomb interaction when the sample volume $L^3$ goes to infinity. Note that, by considering ${N_s}$ electron-ion pairs, we \textit{de facto} consider that $N_s/L^3$ has a finite value in the large $L$ thermodynamic limit.

We wish to note that the above Hamiltonian corresponds to electrons in the presence of point-charge ions. In reality, the crystal cells are occupied by a finite-size atom or molecule that has to be visualized as a ``core'' plus one electron either in the ground or  excited level. The core includes the nucleus plus the remaining electron cloud, the core total charge being equal to $|e|$ since the atom or molecule is neutral. Taking into account the cloud spatial extension would mean to replace the point-charge potential $-e^2/|\vr_{s,j}-\vR_\ell|$ appearing in Eq.~(\ref{2'}), by a $v_{|e|}(\vr_{s,j}-\vR_\ell)$ potential corresponding to the same $|e|$ charge but somewhat broadened over the $\vR_\ell$ cell. The shape of this potential does not affect the Coulomb physics we here study because, as shown below, the formation of Frenkel excitons is entirely driven by the electron-electron Coulomb interaction. The precise shape of this potential only enters the wave functions of the one-electron states that are used in the quantum formulation of the problem, that is, the numerical values of the electron-electron Coulomb scatterings that appear in the formalism; the  precise $v_{|e|}(\vr_{s,j}-\vR_\ell)$ shape has no effect on the physics of the formalism we present.

For systems represented by the above Hamiltonian, the physically relevant electrons either have wave functions that are highly localized on the ion site at the lattice cell scale, or wave functions that are delocalized over the sample. In the former case, the excitons that are formed are called Frenkel excitons, while in the latter case, they are called Wannier excitons. 
In this work, we concentrate on Frenkel excitons.

The appropriate way to handle many-body states like the ones we here study, is through the second quantization formalism. The very first step is to choose a one-electron basis. Although any basis can be used, choosing a ``good'' basis  facilitates the calculations and enlighten their physics. The ``good basis'' is made of one-body states that contain as much physics as possible. This prompts us to first analyze the problem, with this goal in mind.

\subsubsection{On choosing the good one-electron basis}

When the relevant electron wave functions are highly localized on ions, the $N_s$-electron states fundamentally correspond to one electron on each ion site. So, the good one-electron basis has to be related to the eigenstates for one electron in the presence of one ion. 
The Hamiltonian for one electron and a $|e|$ ion located at $\vR_\ell$ reads
\be\label{2}
h_{\scriptscriptstyle{\vR_\ell}}=\frac{\vp^2}{2m_0}-\frac{e^2}{|\vr-\vR_\ell|}
\ee
Its $\nu$ eigenstates, with energy $\va_\nu$ and wave function 
\be\label{3}
\lan\vr|\nu,\vR_\ell\ran=\lan \vr-\vR_\ell|\nu\ran
\ee
reduce to the hydrogen atom states when the core is replaced by a point charge. As with any Hamiltonian eigenstates, these states, which are orthogonal
\be\label{4}
\lan \nu',\vR_\ell|\nu,\vR_\ell\ran=\delta_{\nu',\nu}
\ee
 form a complete basis. So, the $|\nu,\vR_\ell\ran$ states can \textit{a priori} be used to describe an electron located on any other ion site: indeed, the $\lan \nu',\vR_{\ell'}|\nu,\vR_\ell\ran$ overlaps for $\vR_{\ell'} \neq \vR_\ell$ differ from zero for high-energy extended states. However, this requires using a very large number of $\nu$'s in the state description. Due to this, the $|\nu,\vR_\ell\ran$ states with $\vR_\ell$ fixed, do not constitute the good basis we are looking for, to describe $N_s$-body states with one electron on each  $\vR_\ell$ ion. 

To construct this basis, we note that for highly localized states at the lattice cell scale, the wave function overlaps between different ions, although not exactly zero, are very small, even for nearby ions. So, a physically reasonable idea is to take the tight-binding approximation as strict, that is, no wave function overlap between different ions
\be\label{5}
\lan\nu'|\vr-\vR_{\ell'}\ran\lan\vr-\vR_{\ell}|\nu\ran= 0  \quad \textrm{for} \quad (\ell'\not=\ell) \quad \textrm{and} \quad (\nu,\nu')=(e,g)
\ee
Indeed, the above mathematical limit requires highly localized electronic states, which is physically acceptable for the two lowest-energy states $\nu=(e,g)$ that drive the Frenkel exciton physics. 

\begin{figure}[t]
\centering
\includegraphics[trim=0cm 12cm 7cm 0.5cm,clip,width=4in]{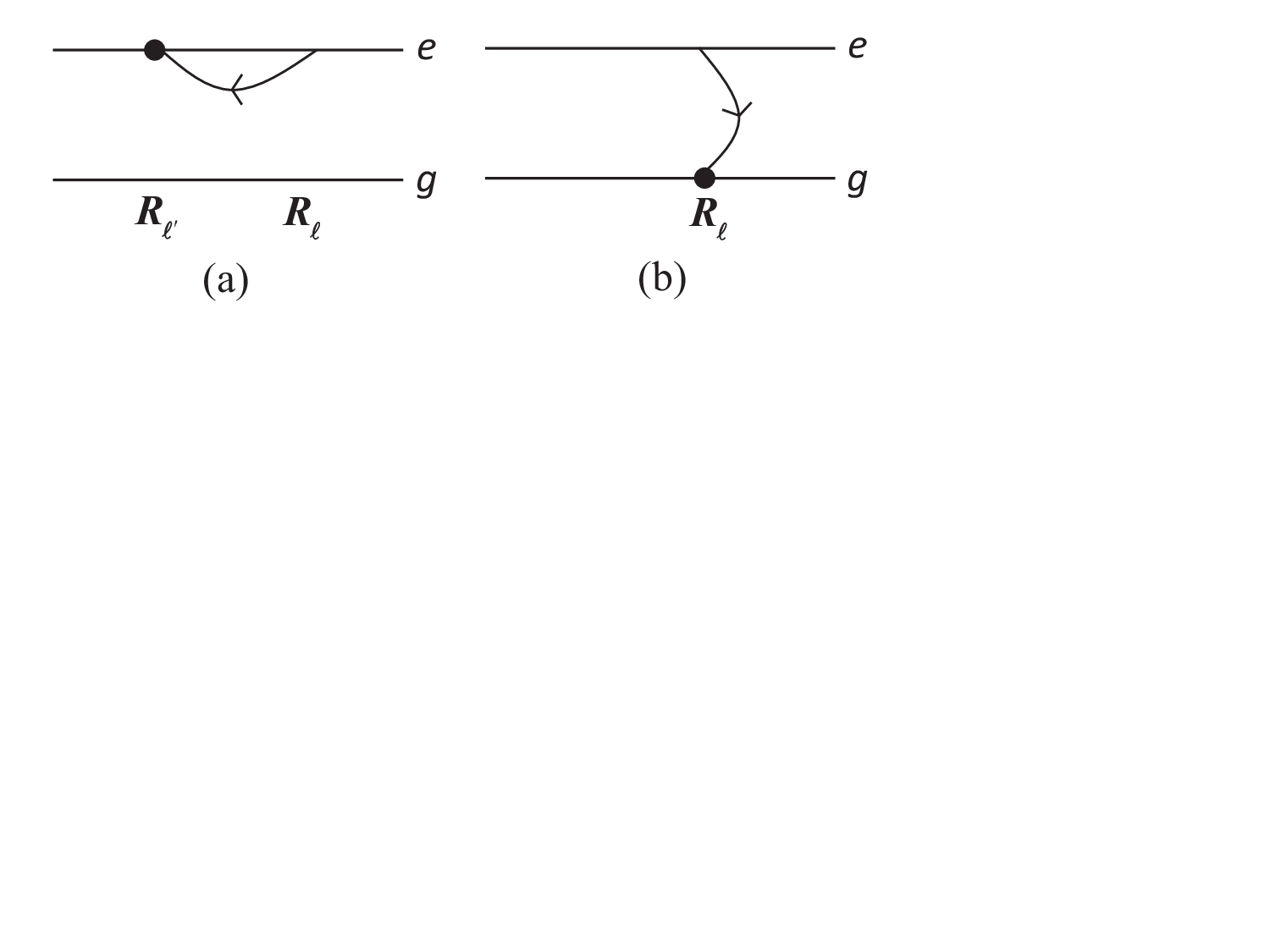}
\vspace{-0.7cm}
\caption{(a) In an \textit{intralevel} Coulomb process, the electron keeps its $\nu$ level while changing lattice site. The associated scattering then contains the wave function overlap between different sites --- which reduces to zero in the tight-binding limit. (b) In an \textit{interlevel} Coulomb process, the electron changes level while staying in the same lattice site.  }
\label{fig0}
\end{figure}

We wish to stress that this tight-binding limit underlies the entire Frenkel exciton story. In particular, it energetically supports taking the semiconductor lowest-excited states as made of one electron in the ground level of the $\vR_\ell$ lattice site, jumping to the lowest excited level of the \textit{same} site, due to the electrostatic energy cost for separating the electron from its ion at a distance large compared to electronic state extension. It also is of importance to note that the tight-binding limit eliminates all intralevel Coulomb processes between different sites (see Fig.~\ref{fig0}(a)), because the associated scatterings contain wave function overlaps $\lan e,\vR_{\ell'}|\vr\ran\lan \vr|e,\vR_\ell\ran$ that in this limit, reduce to zero for $\ell'\not=\ell$. These intralevel processes are unimportant because the Frenkel exciton physics is driven by the interlevel Coulomb interaction (see Fig.~\ref{fig0}(b)), with scatterings that read in terms of on-site overlaps $\lan g,\vR_\ell|\vr\ran\lan \vr|e,\vR_\ell\ran$. This is in stark contrast to Wannier excitons, for which the intraband Coulomb processes, \textit{i.e.}, the counterparts of the intralevel processes, are entirely responsible for the formation of the excitonic wave. Still, by accepting the above relation ($\ref{5}$), we by construction drop all deviations from the tight-binding approximation.  Although these deviations produce electron hopping from ion to ion, they do not affect the formation of the Frenkel exciton wave we here study.

When added to Eq.~(\ref{4}), the equation (\ref{5}) leads to
\be\label{6}
\lan \nu',\vR_{\ell'}|\nu,\vR_{\ell}\ran=\delta_{\nu',\nu}\delta_{\ell',\ell}
\quad \textrm{for} \quad (\nu,\nu')=(e,g)
\ee
This orthogonality allows us to use the $|\nu,\vR_\ell\ran$ states as a one-electron basis to construct the set of creation operators necessary for a quantum description of the Frenkel exciton problem, through
\be\label{7}
|\nu,\vR_\ell\ran=\hat{a}^\dag_{\nu,\ell}|v\ran
\ee
with $|v\ran$ denoting the vacuum state. Indeed, due to Eq.~(\ref{6}), these  operators fulfill the anticommutation relations  for fermion operators  (see \ref{app:M1}), 
\bea
\left[\hat{a}^\dag_{\nu',\ell'},\hat{a}^\dag_{\nu,\ell}\right]_+&=&0\label{8}\\
\left[\hat{a}_{\nu',\ell'},\hat{a}^\dag_{\nu,\ell}\right]_+&=&\delta_{\nu',\nu} \,\, \delta_{\ell',\ell}\label{9}
\eea

\subsection{Hamiltonian in terms of ($\nu,\ell$) electron operators}

Once the good basis of the problem is determined, we can proceed along the second quantization formalism to rewrite the $H_{N_s}$ Hamiltonian for $N_s$ electrons as a $\hat{H}$ operator that reads in terms of the electron creation operators $\hat{a}^\dag_{\nu,\ell}$ associated with the chosen basis. The $\hat{H}$ operator is \textit{a priori} valid whatever $N_s$, which is one of the advantages of the second quantization formalism. However, we must keep in mind that the $\hat{H}$ operator we are going to write is only valid for a physics driven by highly localized states, like $(g,e)$, due to the tight-binding limit $(\ref{5})$ that we have accepted to possibly construct the $\hat{a}^\dag_{\nu,\ell}$ fermionic operators: this is necessary to avoid handling Slater determinants for many-body states. 

The second quantization procedure to transform $H_{N_s}$ into $\hat{H}$ depends on the  nature of the operator at hand, one-body or two-body. Let us consider the various terms of $H_{N_s}$ successively.

 \subsubsection{One-body part}
 
The one-body part of the $H_{N_s}$ Hamiltonian given in Eq.~(\ref{2'}), which includes the electron kinetic energy and the electron-ion attraction, can be written as a sum of one-body Hamiltonians
\be\label{11}
H_{0,{N_s}}=\sum_{j=1}^{N_s}  \left (\frac{\vp^2_j}{2m_0}+\sum_{\ell=1}^{N_s}\frac{-e^2}{|\vr_j-\vR_{\ell}|} \right )  
\equiv \sum_{j=1}^{N_s} h_j
\ee 
Note that $h_j$ differs from the  Hamiltonian $h_{\scriptscriptstyle{\vR_\ell}}$ given in Eq.~(\ref{2}) because the $j$ electron interacts with \textit{all} the ions. As a direct consequence, $H_{0,{N_s}}$ will not appear  diagonal in the $(\nu, \ell)$ basis of Eq.~(\ref{7}). 

According to the second quantization procedure, the $\hat{H}_0$ operator associated with $H_{0,{N_s}}$ reads as
\be\label{13}
\hat{H}_0=\sum_{\nu', \ell'}\sum_{\nu, \ell}h_{\nu', \ell';\nu, \ell}\,\,\hat{a}^\dag_{\nu',\ell'}\hat{a}_{\nu,\ell}
\ee
 The prefactor, given by
 \bea
 \label{14}
h_{\nu', \ell';\nu, \ell}=
\int_{L^3}d^3r\, \lan\nu',\vR_{\ell'}|\vr \ran
 \left(\frac{\hat{\vp}^2}{2m_0}{+}\sum_{\ell''=1}^{N_s}\frac{-e^2}{|\vr{-}\vR_{\ell''}|}\right)\lan\vr|\nu,\vR_{\ell}\ran
\eea
is calculated by first noting that the wave-function product is equal to zero for $\ell'\not=\ell$, due to the  tight-binding limit (\ref{5}). Next, we note that $|\nu,\vR_\ell\ran$ is  $h_{\scriptscriptstyle{\vR_\ell}}$ eigenstate. So, by isolating the $\ell''=\ell$ term from the $\ell''$ sum, we can split this prefactor as
\be\label{15}
h_{\nu', \ell';\nu, \ell}=\delta_{\ell',\ell}\left(\va_\nu\delta_{\nu',\nu}+t_{\nu',\nu}\right)
\ee 
The $t_{\nu',\nu}$ part comes from the electron interaction with all the other ions, $\ell''\not=\ell$. Using Eq.~(\ref{3}) for $\vr-\vR_{\ell}=\vr_\ell$, it precisely reads 
\bea
t_{\nu',\nu}=t^\ast_{\nu,\nu'}
&=& \int_{L^3}d^3r_\ell\, \lan\nu'|\vr_\ell\ran\lan\vr_\ell|\nu\ran \sum_{\ell''\not=\ell}^N\frac{-e^2}{|\vr_\ell+\vR_\ell-\vR_{\ell''}|}\nn\\
&=&\int_{L^3}d^3r\, \lan\nu'|\vr\ran\lan\vr|\nu\ran \sum_{\vR\not=\bf0}\frac{-e^2}{|\vr-\vR|}\label{16}
\eea
the $\ell''\not=\ell$ sum becoming $\ell$-independent when the Born-von Karman boundary condition $f(\vr)=f(\vr+\vR_\ell)$ is used to extend the lattice periodicity to a finite crystal. \

We could think to drop the $t_{\nu',\nu}$ term that comes from the Coulomb interaction of the electron distribution $ \lan\nu'|\vr\ran\lan\vr|\nu\ran$, highly localized on the $\vR=\bf0$ site, with ions located on different lattice sites, where this distribution is very small. Yet, the long-range character of the Coulomb potential renders this dropping questionable in the large sample limit: we will see that the $t_{\nu',\nu}$ term is necessary to properly eliminate spurious large-\textbf{r} singularities that originate from the electron-electron repulsion.

The above equations give the one-body Hamiltonian $\hat{H}_0$ as
\be\label{17}
\hat{H}_0=\sum_{\nu,\ell} \va_\nu  \, \hat{a}^\dag_{\nu,\ell}\hat{a}_{\nu,\ell}+
\sum_{\nu',\nu,\ell} t_{\nu',\nu} \, \hat{a}^\dag_{\nu',\ell}\hat{a}_{\nu,\ell}
\ee 
its second term allowing transitions between electronic levels on the \textit{same} ion site.

 \subsubsection{Two-body electron-electron interaction}

 The $H_{N_s}$ Hamiltonian given in Eq.~(\ref{2'}) also contains a two-body part that corresponds to the electron-electron repulsion. In second quantization, this interaction appears in terms of the electron operators $\hat{a}^\dag_{\nu,\ell}$ as
 \be\label{18}
 \hat{V}_{e-e}=\frac{1}{2}\sum_{\{\nu,\ell\}}V\left(\begin{smallmatrix}
\nu'_2,\ell'_2& \hspace{0.1cm}\nu_2,\ell_2\\ \nu'_1,\ell'_1&\hspace{0.1cm} \nu_1,\ell_1\end{smallmatrix}\right)\hat{a}^\dag_{\nu'_1,\ell'_1}\hat{a}^\dag_{\nu'_2,\ell'_2}\hat{a}_{\nu_2,\ell_2}\hat{a}_{\nu_1,\ell_1}
 \ee
 The prefactor given by
 \bea\label{20}
\lefteqn{V\left(\begin{smallmatrix}
\nu'_2,\ell'_2& \hspace{0.1cm}\nu_2,\ell_2\\ \nu'_1,\ell'_1&\hspace{0.1cm} \nu_1,\ell_1\end{smallmatrix}\right)=}\\
&&\iint_{L^3}d^3r_1 d^3r_2\, \lan\nu'_1,\vR_{\ell'_1}|\vr_1\ran\lan\nu'_2,\vR_{\ell'_2}|\vr_2\ran\frac{e^2}{|\vr_1-\vr_2|}
\lan\vr_2|\nu_2,\vR_{\ell_2}\ran\lan\vr_1|\nu_1,\vR_{\ell_1}\ran\nn
 \eea
 differs from zero for $\ell'_1=\ell_1$ and $\ell'_2=\ell_2$ due to the tight-binding limit (\ref{5}); so, it also reads as
 \be\label{21}
 V\left(\begin{smallmatrix}
\nu'_2,\ell'_2& \nu_2,\ell_2\\\hspace{0.1cm} \nu'_1,\ell'_1& \hspace{0.1cm}\nu_1,\ell_1\end{smallmatrix}\right)=\delta_{\ell'_1,\ell_1}\delta_{\ell'_2,\ell_2} \mathcal{V}_{\vR_{\ell_1}-\vR_{\ell_2}}\left(\begin{smallmatrix}
\nu'_2& \nu_2\\ \hspace{0.1cm}\nu'_1&\hspace{0.1cm} \nu_1\end{smallmatrix}\right)\hspace{3.5cm}
\ee
with the Coulomb part given by
\be
\mathcal{V}_{\vR}\left(\begin{smallmatrix}
\nu'_2& \hspace{0.1cm}\nu_2\\ \nu'_1&\hspace{0.1cm} \nu_1\end{smallmatrix}\right)=\iint_{L^3}d^3r_{\ell_1} d^3r_{\ell_2}\,\lan\nu'_1|\vr_{\ell_1}\ran\lan\nu'_2|\vr_{\ell_2}\ran
%\nn\\
\frac{e^2}{|\vR+\vr_{\ell_1}-\vr_{\ell_2}|}\lan\vr_{\ell_2}|\nu_2\ran\lan\vr_{\ell_1}|\nu_1\ran\label{21_0}
 \ee
As a result, the electron-electron interaction (\ref{18})  appears as
 \bea\label{22}
 \hat{V}_{e-e}=\frac{1}{2}\sum_{\{\nu\}}\sum_{\ell_1,\ell_2}\mathcal{V}_{\vR_{\ell_1}-\vR_{\ell_2}}
 \left(\begin{smallmatrix}
\nu'_2&\hspace{0.1cm} \nu_2\\ \hspace{0.1cm}\nu'_1& \nu_1\end{smallmatrix}\right)
\hat{a}^\dag_{\nu'_1,\ell_1}\hat{a}^\dag_{\nu'_2,\ell_2}\hat{a}_{\nu_2,\ell_2}\hat{a}_{\nu_1,\ell_1}
 \eea
 This interaction destroys two electrons located on the $\ell_1$ and $\ell_2$  sites and recreates them on the \textit{same} sites, either in the same level $\nu'=\nu$, or in a  different level $\nu'\not=\nu$. The fact that the Coulomb interaction is restricted to on-site processes, follows from enforcing the tight-binding limit to the relevant electronic states.

 \subsubsection{Ion-ion interaction}

 The ion-ion interaction, \textit{i.e.}, the last term of the $H_{N_s}$ Hamiltonian (\ref{1}), provides a constant contribution equal to
 \be\label{32}
V_{i-i}=\frac{1}{2}\sum_{\ell=1}^{N_s}\sum_{\ell'\not=\ell}^{N_s}\frac{e^2}{|\vR_{\ell}-\vR_{\ell'}|}=\frac{{N_s}}{2}\sum_{\vR\not=\bf0}\frac{e^2}{|\vR|}
\ee
 due to the lattice periodicity extended through the Born-von Karman boundary condition.

  \subsection{Semiconductor states for ${N_s}$ electrons} 
 
 \subsubsection{Ground state $| \Phi_g\ran$}

\noindent $\bullet$ Let $|g;\vR_\ell\ran=\hat{a}^\dag_{g,\ell}|v\ran$ be the electronic ground state of the $\vR_\ell$ ion, with energy  $\va_g$. The ${N_s}$-electron ground state has one electron in the ground level of each $\vR_\ell$ ion. In second quantization, this state reads
 \be\label{23}
| \Phi_g\ran=\hat{a}^\dag_{g,1}\hat{a}^\dag_{g,2}\cdots \hat{a}^\dag_{g,{N_s}}|v\ran
 \ee 
 In first quantization, this state would be written through its wave function represented by the Slater determinant
  \be\label{24}
  \frac{1}{\sqrt{{N_s}!}}\begin{vmatrix}
  \lan\vr_1-\vR_1|g\ran & \cdots &  \lan\vr_1-\vR_{N_s}|g\ran\\
  \vdots &\ddots &\vdots \\
  \lan\vr_{\scriptscriptstyle{N_s}}-\vR_1|g\ran & \cdots &  \lan\vr_{\scriptscriptstyle{N_s}}-\vR_{N_s}|g\ran
  \end{vmatrix}
  \ee
  which  definitely is far heavier to handle than the operator form given above, let alone all the tricky minus signs that would have to be followed up carefully when calculating scalar products involving such Slater determinants. We will come back to these Slater determinants in Sec.~\ref{Sota}.

 \noindent $\bullet$ In the absence of Coulomb interactions between the electron of a  given ion and the other $({N_s}-1)$ ions, and between the electrons themselves, the $|\Phi_g\ran$ energy reduces to
 \be\label{25}
 E_g={N_s} \,   \va_g
 \ee  
With these Coulomb interactions treated at first order, the ground-state energy reads as
\be\label{26}
E'_g= \lan \Phi_g|\hat{H} |\Phi_g\ran
\ee 

\subsubsection{Lowest set of excited states $ |\Phi_{\vR_\ell}\ran$}

% \subsubsection{ In the absence of Coulomb interaction}
 
\noindent $\bullet$ The lowest excited states for ${N_s}$ electrons have one electron  excited from the ground level $g$ of the $\vR_\ell$ ion, to the lowest excited level $e$ of the \textit{same} ion. Indeed, as previously said, exciting the electron onto another ion site would induce a charge separation with an electrostatic energy cost that would lead to a higher excited-state subspace. So, the lowest excited states follow from $|\Phi_g\ran$ with $\hat{a}^\dag_{g,\ell}$ replaced by $\hat{a}^\dag_{e,\ell}$, namely
 \bea
 |\Phi_{\vR_\ell}\ran&=& \hat{a}^\dag_{g,1}\hat{a}^\dag_{g,2}\cdots \hat{a}^\dag_{g,\ell-1}\left(\hat{a}^\dag_{e,\ell}\right)\hat{a}^\dag_{g,\ell+1}\cdots \hat{a}^\dag_{g,N}|v\ran\nn\\
 &=& \hat{a}^\dag_{e,\ell} a_{g,\ell} |\Phi_g\ran\label{38}
 \eea
since a pair of electron operators commutes with different electron operators. By using the anticommutation relations (\ref{8},\ref{9}), and the fact that in $|\Phi_g\ran$, the ground level of all ion sites is occupied by one electron, we can check that these excited states form an orthogonal set, namely
 \be\label{27'}
 \lan \Phi_{\vR_{\ell'}}|\Phi_{\vR_\ell}\ran = \delta_{\ell',\ell}
\ee

\noindent $\bullet$  Since such a jump can occur on any $\vR_\ell$ ion, the energy of the ${N_s}$ states $ |\Phi_{\vR_\ell}\ran$ in the absence of Coulomb interactions between the electron of an ion and the other ions and between the electrons themselves, reads
\be\label{39}
E_e=({N_s}-1)\va_g+\va_e=E_g +(\va_e-\va_g)
\ee
In the presence of these Coulomb interactions, treated at first order, the excited-state energies follow from the diagonalization of the $\hat{H}$ Hamiltonian within the $|\Phi_{\vR_\ell}\ran$ degenerate subspace, that is, the diagonalization of the (${N_s}\times {N_s}$) matrix
\be\label{40}
\lan \Phi_{\vR_{\ell'}}|\hat{H}|\Phi_{\vR_\ell}\ran 
\ee

 \subsection{Relevant parts of the system Hamiltonian}
  
 The relevant parts of $\hat{H}$ in $ \lan \Phi_{\vR_{\ell'}}|\hat{H} |\Phi_{\vR_\ell}\ran$ are the ones that conserve the number of ground-level electrons and the number of excited-level electrons. Let us write them explicitly to better catch their physics.
 
 \subsubsection{One-body parts}
 
 The part of the one-body Hamiltonian $\hat{H}_0$ given in Eq.~(\ref{17}) that conserves the number of ground-level electrons and the number of excited-level electrons, reduces to 
 \be\label{40_1}
 \hat{H}_0\longrightarrow  \hat{H}_{0,g}+\hat{H}_{0,e}
 \ee
 with the ground and excited parts  given by
 \bea
 \hat{H}_{0,g}&=& (\va_g+t_{g,g})\sum_{\ell=1}^{N_s} \hat{a}^\dag_{g,\ell}\hat{a}_{g,\ell}\label{40_2}\\
 \hat{H}_{0,e}&=& (\va_e+t_{e,e})\sum_{\ell=1}^{N_s} \hat{a}^\dag_{e,\ell}\hat{a}_{e,\ell}\label{40_3}
 \eea
 
 \subsubsection{Two-body electron-electron interactions}
 The  part of the $\hat{V}_{e-e}$ interaction given in Eq.~(\ref{22}), that conserves the number of ground-level electrons and the number of excited-level electrons, reduces to
 \be\label{40_4}
 \hat{V}_{e-e}\longrightarrow \hat{V}_{gg}+\hat{V}^{(intra)}_{eg}+\hat{V}^{(inter)}_{eg}
 \ee

\begin{figure}[t]
\centering
\includegraphics[trim=0cm 6cm 0cm 6.5cm,clip,width=5in]{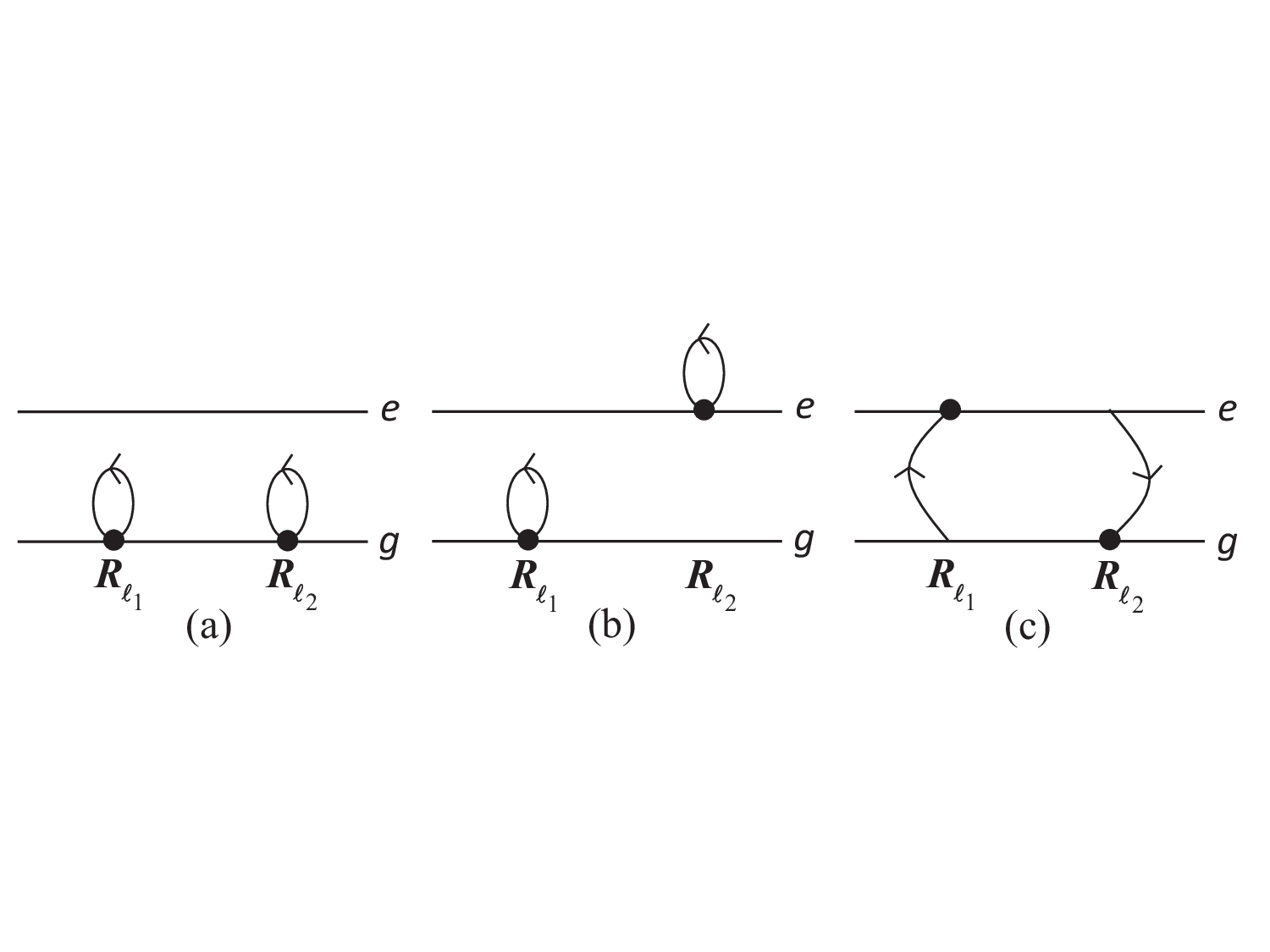}
\vspace{-0.7cm}
\caption{(a) In $\hat{V}_{gg}$ given in Eq.~(\ref{40_5}), the Coulomb interaction acts between two ground-level electrons. (b) In $\hat{V}^{(intra)}_{eg}$ given in Eq.~(\ref{40_6}), the Coulomb interaction acts between one ground-level electron and one excited-level electron, each electron staying in its level. (c) In $\hat{V}^{(inter)}_{eg}$ given in Eq.~(\ref{40_7}), these electrons change level. }
\label{fig1}
\end{figure}

 \noindent $\bullet$ In $\hat{V}_{gg}$, given by
 \be
 \label{40_5}
 \hat{V}_{gg}=\frac{1}{2}\sum_{\ell_1=1}^{N_s} \,\,  \sum_{\ell_2\not=\ell_1}^{N_s} \mathcal{V}_{\vR_{\ell_1}-\vR_{\ell_2}}\left(\begin{smallmatrix}
g& g\\ g& g\end{smallmatrix}\right)\hat{a}^\dag_{g,\ell_1}\hat{a}^\dag_{g,\ell_2}\hat{a}_{g,\ell_2}\hat{a}_{g,\ell_1}
 \ee
 the Coulomb interaction acts on the ground level only (see Fig.~\ref{fig1}(a)), with $\ell_2$  different from $\ell_1$ due to the Pauli exclusion principle.

  \noindent $\bullet$ In the intralevel interaction on the same ion, $\hat{V}^{(intra)}_{eg}$, given by
  \be\label{40_6}
\hat{V}^{(intra)}_{eg}=\sum_{\ell_1=1}^{N_s} \,\, \sum_{\ell_2=1}^{N_s} \mathcal{V}_{\vR_{\ell_1}-\vR_{\ell_2}}\left(\begin{smallmatrix}
e& e\\ g& g\end{smallmatrix}\right)\hat{a}^\dag_{g,\ell_1}\hat{a}^\dag_{e,\ell_2}\hat{a}_{e,\ell_2}\hat{a}_{g,\ell_1}
 \ee
each electron stays in its level (see Fig.~\ref{fig1}(b)).
 
 \noindent $\bullet$ The electron-electron Coulomb interaction also has interlevel processes
    \be\label{40_7}
\hat{V}^{(inter)}_{eg}=\sum_{\ell_1=1}^{N_s} \,\,    \sum_{\ell_2=1}^{N_s} \mathcal{V}_{\vR_{\ell_1}-\vR_{\ell_2}}\left(\begin{smallmatrix}
g& e\\ e& g\end{smallmatrix}\right)\hat{a}^\dag_{e,\ell_1}\hat{a}^\dag_{g,\ell_2}\hat{a}_{e,\ell_2}\hat{a}_{g,\ell_1}
 \ee
 in which each electron changes level (see Fig.~\ref{fig1}(c)). These processes are the crucial ones to produce the Frenkel exciton wave because when $\ell_2\not=\ell_1$,  the excitation moves from the $\vR_{\ell_2}$ site to the $\vR_{\ell_1}$ site.
  
\subsubsection{System Hamiltonian in the  ground and excited  subspaces}

We now have all the tools to calculate the energy  of the ${N_s}$-electron ground state $|\Phi_g\ran$ through $\lan \Phi_g|\hat{H} |\Phi_g\ran$, and the energy of  the lowest excited states  through the diagonalization of the (${N_s}\times {N_s}$) matrix $\lan \Phi_{\vR_{\ell'}}|\hat{H}|\Phi_{\vR_\ell}\ran $.  These calculations are given in \ref{app:M2} and \ref{app:M3}.

Actually, the precise value of the ground-state energy $\lan \Phi_g|\hat{H} |\Phi_g\ran$ does not matter when it comes to determining the Frenkel exciton waves because they just follow from the diagonalization of the $\lan \Phi_{\vR_{\ell'}}|\hat{H}|\Phi_{\vR_\ell}\ran$ matrix. Moreover, the proper way to calculate the $\lan \Phi_{\vR_{\ell'}}|\hat{H}|\Phi_{\vR_\ell}\ran$ matrix elements is not in terms of ground-level and excited-level electrons as done in \ref{app:M3}, but in terms of excitations, that is, electrons and holes. Let us turn to this language.  
 
\subsection{From ground-level and excited-level electrons to electrons and holes}

\subsubsection{Electron and hole operators}

The smart way to perform calculations involving semiconductor excitations is to introduce the concept of hole\cite{ChernyakJOS,Mukamelbook}: the destruction of a ground-level electron on the $\vR_\ell$ lattice site  corresponds to the creation of a hole on the same site. In terms of operators, this reads
\be\label{41}
\hat{a}_{g,\ell}=\hat{b}^\dag_\ell
\ee
without any phase factor in the absence of spin and spatial degeneracy. In this language, an electron in the lowest excited  level is just called  ``electron'', with creation operator
\be\label{41_0}
\hat{a}^\dag_{e,\ell}=\hat{a}^\dag_{\ell}
\ee

 The lowest set of  excited states $|\Phi_{\vR_\ell}\ran$ given in Eq.~(\ref{38}) then appears as
\be
|\Phi_{\vR_\ell}\ran=\hat{a}^\dag_{\ell} \hat{b}^\dag_{\ell} |\Phi_g\ran 
\equiv \hat{a}^\dag_{\ell} \hat{b}^\dag_{\ell} |0\ran
\label{42}
\ee
as the $|\Phi_g\ran$ ground state contains zero hole and zero ``electron'' in the sense of Eq.~(\ref{41_0}).

To  fully  exploit the advantage of describing the system in terms of electrons and holes, we also have to write the relevant parts of the $\hat{H}$ Hamiltonian in terms of these operators.

 \subsubsection{Hamiltonian in terms of electrons and holes}

\noindent $\bullet$  The one-body ground part of the $\hat{H}$ Hamiltonian given in Eq.~(\ref{40_2})  becomes, when using hole operators,
\bea
\hat{H}_{0,g}=N(\va_g+t_{g,g})-(\va_g+t_{g,g})\sum_{\ell=1}^N \, \hat{b}^\dag_{\ell}\hat{b}_{\ell}\label{44}
\eea
since $\hat{a}^\dag_{g,\ell}\hat{a}_{g,\ell}=1-\hat{a}_{g,\ell}\hat{a}^\dag_{g,\ell}$. The first term is just equal to $\lan \Phi_g|\hat{H}_0|\Phi_g\ran$ because  the $|\Phi_g\ran$ ground state does not have  hole nor electron in the sense of ``excited electron''.  
 
 The one-body excited part of the $\hat{H}$ Hamiltonian given in Eq.~(\ref{40_3}) simply leads to
\bea
 \hat{H}_{0,e}= (\va_e+t_{e,e}) \sum_{\ell=1}^N  \hat{a}^\dag_{\ell}\hat{a}_{\ell}
\eea 

%+\sum_n \Big(t_{e,g}\hat{a}^\dag_{n}\hat{b}^\dag_n+h.c.\Big)

\noindent $\bullet$ Next, we consider the $\hat{V}_{gg}$ part of the $\hat{V}_{e-e}$ Coulomb interaction between ground-level electrons only, given in Eq.~(\ref{40_5}). 
From  $\hat{a}^\dag_{g,\ell_1}\hat{a}^\dag_{g,\ell_2}\hat{a}_{g,\ell_2}\hat{a}_{g,\ell_1}=1-\hat{a}_{g,\ell_1}\hat{a}^\dag_{g,\ell_1}-\hat{a}_{g,\ell_2}\hat{a}^\dag_{g,\ell_2}+\hat{a}_{g,\ell_1}\hat{a}_{g,\ell_2}\hat{a}^\dag_{g,\ell_2}\hat{a}^\dag_{g,\ell_1}$ for $\ell_2\not=\ell_1$, the $\hat{V}_{gg}$ interaction splits into three terms when written with electron and hole operators, namely
\bea
\label{46}
\hat{V}_{gg}=
\frac{1}{2}\sum_{\ell_1}\!\!\sum_{\ell_2\not=\ell_1}\!\!\mathcal{V}_{\vR_{\ell_1}-\vR_{\ell_2}}\!\left(\begin{smallmatrix}
g& g\\ g& g\end{smallmatrix}\right)
-2\,\frac{1}{2}\sum_{\ell_1}\hat{b}^\dag_{\ell_1}\hat{b}_{\ell_1}\!\!\left(\sum_{\ell_2\not=\ell_1}\mathcal{V}_{\vR_{\ell_1}-\vR_{\ell_2}}\!\left(\begin{smallmatrix}
g& g\\ g& g\end{smallmatrix}\right)\right)+\hat{V}^{(intra)}_{hh}\nn
\eea
The first $\hat{V}_{gg}$ term is nothing but $\lan \Phi_g|\hat{V}_{gg} |\Phi_g\ran$ because the other terms of $\hat{V}_{gg}$ require states having one hole at least to produce a nonzero contribution while $|\Phi_g\ran$ has no hole. The second term brings a constant shift to the hole energy since for a periodic crystal, the prefactor of $\hat{b}^\dag_{\ell_1}\hat{b}_{\ell_1}$ does not depend on $\ell_1$.
The last $\hat{V}_{gg}$ term corresponds to the Coulomb repulsion between two holes
\be
\hat{V}^{(intra)}_{hh}=\frac{1}{2}\sum_{\ell_1}\sum_{\ell_2\not=\ell_1}\mathcal{V}_{\vR_{\ell_1}-\vR_{\ell_2}}\!\left(\begin{smallmatrix}
g& g\\ g& g\end{smallmatrix}\right)\hat{b}^\dag_{\ell_1}\hat{b}^\dag_{\ell_2}\hat{b}_{\ell_2}\hat{b}_{\ell_1}
\ee

 \begin{figure}[t]
\centering
\includegraphics[trim=2.5cm 5cm 5cm 6.5cm,clip,width=4in]{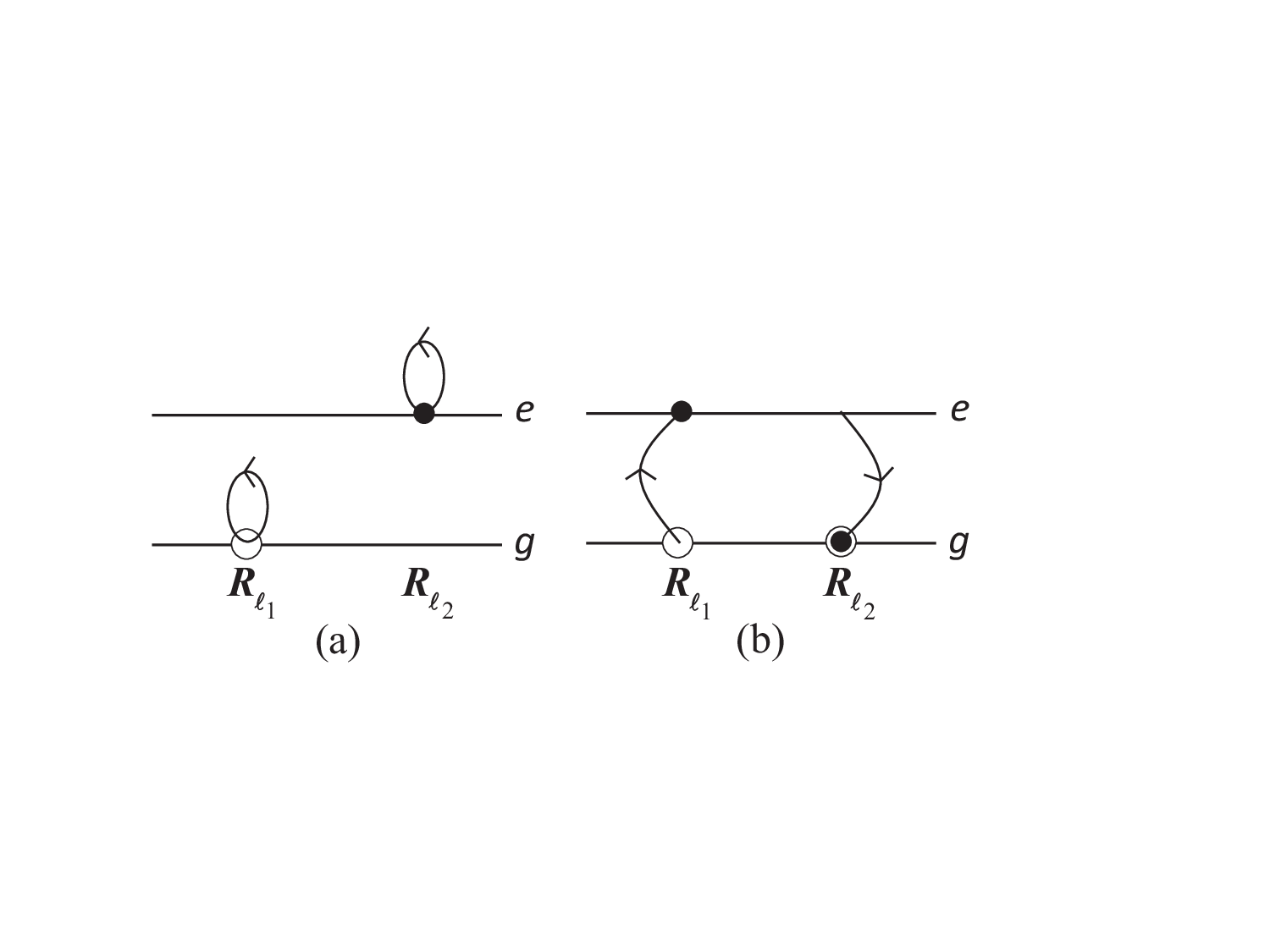}
\vspace{-0.7cm}
\caption{(a) In $\hat{V}^{(intra)}_{eh}$ given in Eq.~(\ref{49_0}), the Coulomb interaction acts between one excited-level electron and one ground-level electron absence, that is, one hole,  each carrier staying in their level. (b) In  $\hat{V}^{(inter)}_{eh}$ given in Eq.~(\ref{49_1}), the electron and the hole of the $\vR_{\ell_2}$ site recombine, while an electron-hole pair is created on the $\vR_{\ell_1}$ site, possibly different from $\vR_{\ell_2}$; this then leads to an excitation transfer from the $\ell_2$ site to the $\ell_1$ site. }
\label{fig2}
\end{figure}

\noindent $\bullet$ The intralevel part (\ref{40_6}) of the Coulomb interaction, $\hat{V}^{(intra)}_{eg}$ between  ground-level and excited-level electrons, leads to two terms since $\hat{b}_{\ell}\hat{b}^\dag_{\ell}=1-\hat{b}^\dag_{\ell}\hat{b}_{\ell}$, namely
\bea
\hat{V}^{(intra)}_{eg}&=& \sum_{\ell_1}\sum_{\ell_2}\mathcal{V}_{\vR_{\ell_1}-\vR_{\ell_2}}\!\left(\begin{smallmatrix}
e& e\\ g& g\end{smallmatrix}\right)\hat{b}_{\ell_1}\hat{a}^\dag_{\ell_2}\hat{a}_{\ell_2} \hat{b}^\dag_{\ell_1}
\nn\\
&=& \sum_\vR \mathcal{V}_{\vR}\left(\begin{smallmatrix}
e& e\\ g& g\end{smallmatrix}\right)\sum_\ell \hat{a}^\dag_\ell \hat{a}_\ell +\hat{V}^{(intra)}_{eh}
\eea
The first $\hat{V}^{(intra)}_{eg}$ term brings a constant shift to the electron energy, while $\hat{V}^{(intra)}_{eh}$ corresponds to a Coulomb attraction in which the electron stays electron and the hole stays hole (see Fig.~\ref{fig2}(a)), namely
\be
\hat{V}^{(intra)}_{eh}=- \sum_{\ell_1}\sum_{\ell_2}\mathcal{V}_{\vR_{\ell_1}-\vR_{\ell_2}}\!\left(\begin{smallmatrix}
e& e\\ g& g\end{smallmatrix}\right)\hat{a}^\dag_{\ell_2}\hat{b}^\dag_{\ell_1}\hat{b}_{\ell_1}\hat{a}_{\ell_2}\label{49_0}
\ee

\noindent $\bullet$ The Coulomb interaction also has an interlevel part given in Eq.~(\ref{40_7}), which  becomes, since $\hat{b}_{\ell_2}\hat{b}^\dag_{\ell_1}=\delta_{\ell_1,\ell_2}- \hat{b}^\dag_{\ell_1}\hat{b}_{\ell_2}$,
\bea
\hat{V}^{(inter)}_{eg}&=& \sum_{\ell_1}\sum_{\ell_2}\mathcal{V}_{\vR_{\ell_1}-\vR_{\ell_2}}\!\left(\begin{smallmatrix}
g& e\\ e& g\end{smallmatrix}\right)\hat{a}^\dag_{\ell_1}\hat{b}_{\ell_2}\hat{a}_{\ell_2}\hat{b}^\dag_{\ell_1}
\nn\\
&=&-  \mathcal{V}_{\vR=\bf0}\left(\begin{smallmatrix}
g& e\\ e& g\end{smallmatrix}\right)\sum_\ell \hat{a}^\dag_\ell \hat{a}_\ell +\hat{V}^{(inter)}_{eh} \label{49}
\eea
where $\hat{V}^{(inter)}_{eh}$ corresponds to an interaction in which one electron-hole pair recombines on the $\vR_{\ell_2}$ site, while another pair is created on the $\vR_{\ell_1}$ site, possibly different from  $\vR_{\ell_2}$ (see Fig.~\ref{fig2}(b)), namely
\be
\hat{V}^{(inter)}_{eh}=\sum_{\ell_1}\sum_{\ell_2}\mathcal{V}_{\vR_{\ell_1}-\vR_{\ell_2}}\!\left(\begin{smallmatrix}
g& e\\ e& g\end{smallmatrix}\right)\hat{a}^\dag_{\ell_1}\hat{b}^\dag_{\ell_1}\hat{b}_{\ell_2}\hat{a}_{\ell_2}\label{49_1}
\ee
Note the positive sign in front of this important term for the Frenkel exciton formation.

 \subsubsection{Relevant Hamiltonian for Frenkel excitons}

  Since $\hat{V}^{(intra)}_{hh}$ acts between two holes, the relevant parts of the Coulomb interaction in the one-hole subspace  reduce to  $\hat{V}^{(intra)}_{eh}$ given in Eq.~(\ref{49_0}) and $\hat{V}^{(inter)}_{eh}$ given in Eq.~(\ref{49_1}). 
As a result, the relevant parts of the total Hamiltonian $\hat{H}$  in the electron-hole subspace $|\Phi_{\vR_\ell}\ran$ reduces to 
\be\label{49_2}
\hat{H}_{eh}=\lan \Phi_g| \hat{H}|\Phi_g\ran+\hat{H}_{e}+\hat{H}_{h}+\hat{V}^{(intra)}_{eh}+\hat{V}^{(inter)}_{eh}
\ee

With the help of Eq.~(\ref{49_2}), we find that the first term of $\hat{H}_{eh}$ is just
\be\label{49_7}
\lan \Phi_g| \hat{H}_{eh}|\Phi_g\ran=\lan \Phi_g| \hat{H}|\Phi_g\ran=E'_g
\ee
the other terms of $\hat{H}$ requiring states with one electron or one hole to contribute.

 Note that the one-body part of $\hat{H}_{eh}$ also contains Coulomb contributions, as seen from its electron part
\bea
\hat{H}_{e}&=&\tilde{\va}_e \sum_\ell \hat{a}^\dag_\ell\hat{a}_\ell\label{49_3}\\
\tilde{\va}_e&=& \va_e+t_{e,e}+\sum_\vR\mathcal{V}_{\vR}\!\left(\begin{smallmatrix}
e& e\\ g& g\end{smallmatrix}\right)-\mathcal{V}_{\vR=\bf0}\!\left(\begin{smallmatrix}
g& e\\ e& g\end{smallmatrix}\right)\label{49_4}
\eea
and its hole part
\bea
\hat{H}_{h}&=&\tilde{\va}_h \sum_\ell \hat{b}^\dag_\ell\hat{b}_\ell\label{49_5}\\
\tilde{\va}_h&=& -\va_g-t_{g,g}-\sum_{\vR\not=\bf0}\mathcal{V}_{\vR}\!\left(\begin{smallmatrix}
g& g\\ g& g\end{smallmatrix}\right)\label{49_6}
\eea

%In the following, we will drop this physically irrelevant constant  from $\hat{H}_{eh}$.[Monique, je prefere le garder]

 \subsubsection{Electron-hole Hamiltonian in the $|\Phi_{\vR_\ell}\ran$ excited subspace}

\noindent $\bullet$ By using Eqs.~(\ref{42},\ref{49_3},\ref{49_5}), we get the contribution from the one-body part of the Hamiltonian as
\bea\label{50}
\left(\hat{H}_{e}+\hat{H}_{h}\right)|\Phi_{\vR_\ell}\ran=\left(\hat{H}_{e}+\hat{H}_{h}\right)\,\hat{a}^\dag_\ell\hat{b}^\dag_\ell|0\ran=\left(\tilde{\va}_e+\tilde{\va}_h\right)|\Phi_{\vR_\ell}\ran
\eea

Turning to the Coulomb interaction, we find that the intralevel part of Eq.~(\ref{49_0}) simply gives
\bea
\label{51}
\hat{V}^{(intra)}_{eh}|\Phi_{\vR_\ell}\ran&=&
-\sum_{\ell_1,\ell_2}\mathcal{V}_{\vR_{\ell_1}-\vR_{\ell_2}}\!\left(\begin{smallmatrix}
e& e\\ g& g\end{smallmatrix}\right)\left(\hat{a}^\dag_{\ell_2}\hat{b}^\dag_{\ell_1}\hat{b}_{\ell_1}\hat{a}_{\ell_2}\right)\hat{a}^\dag_\ell\hat{b}^\dag_\ell|0\ran\nn\\
&=&-\mathcal{V}_{\vR=\bf0}\!\left(\begin{smallmatrix}
e& e\\ g& g\end{smallmatrix}\right)|\Phi_{\vR_\ell}\ran \hspace{1cm}
\eea
for $\vR=\bf0$ because $\hat{b}_{\ell_1}\hat{a}_{\ell_2}\hat{a}^\dag_\ell\hat{b}^\dag_\ell|\Phi_g\ran$ differs from zero for $\ell_2=\ell=\ell_1$ only.

  In the same way, the interlevel part of Eq.~(\ref{49_1}) leads to
\bea
\label{52}
\hat{V}^{(intra)}_{eh}|\Phi_{\vR_\ell}\ran&=&
\sum_{\ell_1,\ell_2}\mathcal{V}_{\vR_{\ell_1}-\vR_{\ell_2}}\!\left(\begin{smallmatrix}
g& e\\ e& g\end{smallmatrix}\right)\left(\hat{a}^\dag_{\ell_1}\hat{b}^\dag_{\ell_1}\hat{b}_{\ell_2}\hat{a}_{\ell_2}\right)\hat{a}^\dag_\ell\hat{b}^\dag_\ell|0\ran\nn\\
&=&\sum_{\ell_1}\mathcal{V}_{\vR_{\ell_1}-\vR_\ell}\!\left(\begin{smallmatrix}
g& e\\ e& g\end{smallmatrix}\right)|\Phi_{\vR_{\ell_1}}\ran 
\eea

\noindent $\bullet$ So, the diagonal term of the $\hat{H}_{eh}$ operator in the $|\Phi_{\vR_\ell}\ran$ subspace reduces to
\be\label{53}
\lan \Phi_{\vR_\ell}|\hat{H}_{eh}|\Phi_{\vR_\ell}\ran=E'_g+\va_e-\va_g+v_{eg}\equiv E'_e
\ee
with $v_{eg}$ given  by
\bea\label{54_1}
v_{eg}=\iint_{L^3}d^3rd^3r'
\sum_{\vR\not=\bf0}\left(\frac{e^2}{|\vR+\vr-\vr'|}-\frac{e^2}{|\vR-\vr'|}\right)
 \bigg(|\lan\vr'|e\ran|^2-|\lan\vr'|g\ran|^2\bigg)|\lan\vr|g\ran|^2
\eea 
 while the nondiagonal terms  simply read
\be\label{53_0}
\lan \Phi_{\vR_{\ell'\not=\ell}}|\hat{H}_{eh}|\Phi_{\vR_\ell}\ran=\mathcal{V}_{\vR_{\ell'}-\vR_\ell}\!\left(\begin{smallmatrix}
g& e\\ e& g\end{smallmatrix}\right) 
\ee
These nondiagonal terms correspond to Coulomb processes associated with interlevel processes on different lattice sites (see Fig.~\ref{fig2}(b)). These processes are the ones responsible for the Frenkel exciton formation.

 \subsubsection{Diagonalization of the corresponding matrix}

\noindent $\bullet$ The linear combinations of $|\Phi_{\vR_\ell}\ran$ states that render diagonal the $\hat{H}_{eh}$ operator in the one-electron-hole-pair subspace, read as
\be\label{56}
|\Phi_{\vK_n}\ran=\frac{1}{\sqrt{N_s}}\sum_{\ell=1}^{N_s} e^{i\vK_n\cdot\vR_\ell}|\Phi_{\vR_\ell}\ran
\ee
for the ${N_s}$ vectors $\vK_n$ of the first Brillouin zone,  quantized in $2\pi/L$. These linear combinations give rise to the so-called Frenkel excitons.

\noindent $\bullet$ To show it, we start with
\be\label{57}
\lan \Phi_{\vK_{n'}}|\hat{H}_{eh}|\Phi_{\vK_n}\ran=\frac{1}{{N_s}}\sum_{\ell'}\sum_\ell e^{i(-\vK_{n'}\cdot\vR_{\ell'}+\vK_n\cdot\vR_\ell)}\lan \Phi_{\vR_{\ell'}}|\hat{H}_{eh}|\Phi_{\vR_\ell}\ran
\ee
For $\ell'=\ell$, the $\lan \Phi_{\vR_{\ell'}}|\hat{H}_{eh}|\Phi_{\vR_\ell}\ran$ matrix element is equal to $E'_e$ given in Eq.~(\ref{53}), while for $\ell'\neq \ell$, it is given by Eq.~(\ref{53_0}). So, the above equation leads to
\bea\label{58}
\lan \Phi_{\vK_{n'}}|\hat{H}_{eh}|\Phi_{\vK_n}\ran &=& \frac{E'_e}{{N_s}}\sum_\ell e^{i(-\vK_{n'}+\vK_n)\cdot\vR_\ell}\\
&&+\frac{1}{{N_s}}\sum_{\ell'}\sum_{\ell\not=\ell'} \mathcal{V}_{\vR_{\ell'}-\vR_\ell}\!\left(\begin{smallmatrix}
g& e\\ e& g\end{smallmatrix}\right) e^{i(-\vK_{n'}\cdot\vR_{\ell'}+\vK_n\cdot\vR_\ell)}\nn
\eea
For $\vK_n$ quantized in $2\pi/L$, the first sum is equal to ${N_s}$ when $n'=n$ and to zero otherwise. To derive the second sum, we write $\vR_\ell$ as $\vR_{\ell'}+\vR_{\ell_1}$. This second term then gives
\bea\label{59}
\frac{1}{{N_s}}\sum_{\vR_{\ell_1}\not=\bf0} \mathcal{V}_{\vR_{\ell_1}}\!\left(\begin{smallmatrix}
g& e\\ e& g\end{smallmatrix}\right)e^{i\vK_n\cdot\vR_{\ell_1}}\sum_{\ell'} e^{i(\vK_n-\vK_{n'})\cdot\vR_{\ell'}}\hspace{2cm}\nn\\
=\delta_ {n',n}   \sum_{\vR_{\ell_1}\not=\bf0} \mathcal{V}_{\vR_{\ell_1}}\!\left(\begin{smallmatrix}
g& e\\ e& g\end{smallmatrix}\right)e^{i\vK_n\cdot\vR_{\ell_1}}\equiv\delta_ {n',n} \,\, v_{_{\vK_n}}
\eea
since the $\ell'$ sum is equal to ${N_s}$ when $n'=n$ and to zero otherwise. 

So, we end with
\be\label{60}
\lan \Phi_{\vK_{n'}}|\hat{H}_{eh}|\Phi_{\vK_n}\ran=\delta_{n',n}(E'_e+v_{_{\vK_n}})\equiv\delta_{n',n}\,\,E_{\vK_n}
\ee
which  proves that the $|\Phi_{\vK_n}\ran$ states render the $\hat{H}_{eh}$ Hamiltonian diagonal.

 \subsubsection{Frenkel exciton energy in the small  $\vK_n$ limit}
 
\begin{figure}[t]
\centering
\includegraphics[trim=5.5cm 6cm 7cm 7.8cm,clip,width=3.2in]{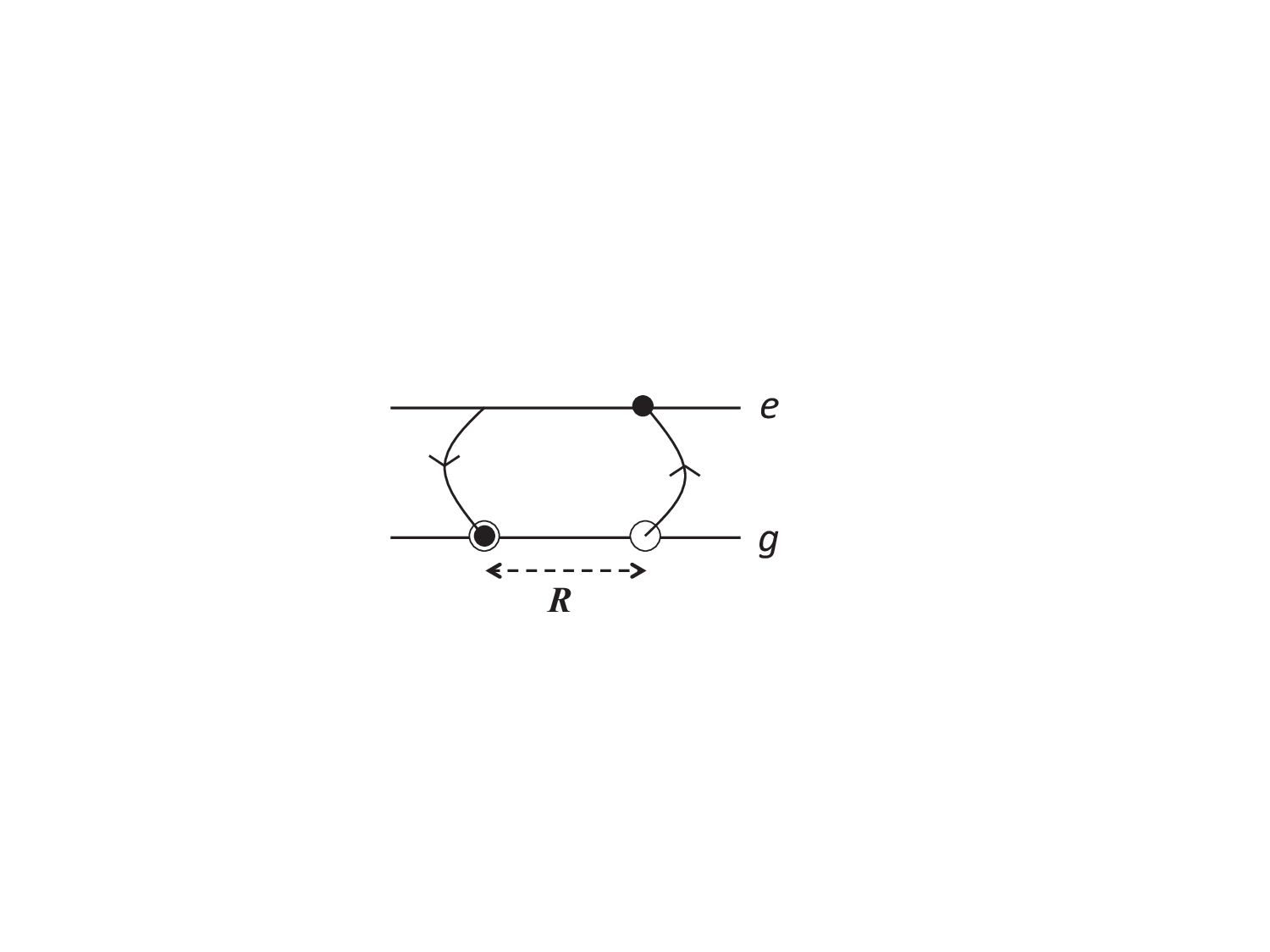}
\vspace{-0.7cm}
\caption{Coulomb processes in which one electron is excited from the ground level $g$ to the excited level $e$, while another electron located on another lattice site at a finite distance $R$ is de-excited from $e$ to $g$. These processes are the ones responsible for the energy splitting of the $|\Phi_{\vR_\ell}\ran$ subspace into $N_s$ Frenkel excitons $|\Phi_{\vK_n}\ran$ (see Eqs.~(\ref{60}) and (\ref{61})). }
\label{fig3}
\end{figure}

 Each $\vK_n$ vector of the first Brillouin zone is associated with  a  Frenkel exciton $|\Phi_{\vK_n}\ran$ that diagonalizes the $\hat{H}_{eh}$ Hamiltonian. Equation ($\ref{59}$)
gives the $\vK_n$ dependence of its energy through 
\be\label{61}
v_{_{\vK_n}}=\sum_{\vR\not=\bf0} \mathcal{V}_{\vR}\!\left(\begin{smallmatrix}
g& e\\ e& g\end{smallmatrix}\right)e^{i\vK_n\cdot\vR}
\ee
It comes from interlevel Coulomb transitions between $g$ and $e$, as given in Eq.~(\ref{49_1}).
The $v_{\scriptscriptstyle{\vK_n}}$ energy  comes from  all possible Coulomb processes in which a ground-level electron jumps to the excited level of the same ion, while the reverse occurs on ions $\vR$ apart (see Fig.~\ref{fig3})\cite{Forster,Jones}.

By writing $v_{_{\vK_n}}$ as
\be\label{62}
v_{_{\vK_n}}=v_{_{\bf0}} + \sum_{\vR\not=\bf0} \mathcal{V}_{\vR}\!\left(\begin{smallmatrix}
g& e\\ e& g\end{smallmatrix}\right)\left(e^{i\vK_n\cdot\vR}-1\right)
\ee
we see that the small $\vK_n$ behavior of  $v_{_{\vK_n}}$ is controlled by the large-$\vR$ behavior of the Coulomb scattering $\mathcal{V}_{\vR}\!\left(\begin{smallmatrix}
g& e\\ e& g\end{smallmatrix}\right)$, which, according to Eq.~(\ref{21_0}), is given by
 \bea
 \label{63}
\mathcal{V}_{\vR}\left(\begin{smallmatrix}
g& e\\ e& g\end{smallmatrix}\right)=
\iint_{L^3}d^3r d^3r'  \,\,   \lan e|\vr\ran\lan\vr|g\ran\frac{e^2}{|\vR+\vr-\vr'|}\lan g|\vr'\ran\lan\vr'|e\ran
 \eea
 Since the $g$ and $e$ electron wave functions confine the $(\vr,\vr')$ variables at a distance small compared to the lattice cell size, the large-$\vR$ behavior of $\mathcal{V}_{\vR}\left(\begin{smallmatrix}
g& e\\ e& g\end{smallmatrix}\right)$ follows from the  $1/R$ expansion of $1/|\vR+\vr-\vr'|$. By using
\bea\label{64}
\frac{1}{\sqrt{R^2+\vr^2+2\vR\cdot\vr}}
=\frac{1}{R}\left(1+\frac{2\vR}{R}\cdot\frac{\vr}{R}+\frac{\vr^2}{R^2}\right)^{-1/2}\nn\hspace{2cm}\\
\simeq \frac{1}{R}\left(1-\frac{1}{2}\left(\frac{2\vR}{R}\cdot\frac{\vr}{R}+\frac{\vr^2}{R^2}\right)+\frac{3}{8}\left(\frac{2\vR}{R}\cdot\frac{\vr}{R}\right)^2 + \cdots  \right )
\eea
 into  Eq.~(\ref{63}), we see that the prefactor of the $1/R$ term reduces to zero because the $e$ and $g$ wave functions, eigenstates of a single ion charge, are orthogonal, $\lan e|g\ran=0$. The terms in $\vr$, $\vr^2$, $\vr'$, $\vr'^2$ reduce to zero for the same reason. As a result, the dominant contribution to $\mathcal{V}_{\vR}\left(\begin{smallmatrix}
g& e\\ e& g\end{smallmatrix}\right)$ in the large $R$ limit, comes from the parts of the above equation in ($\vr, \vr'$); they precisely read\cite{Born,Cohen}
 \be\label{65} 
 \frac{1}{R^3}\left[\vr\cdot\vr'-3\left(\frac{\vR}{R}\cdot\vr\right)\,\,\left(\frac{\vR}{R}\cdot\vr'\right)\right]
 \ee

When inserted into Eq.~(\ref{63}), $\vr$ appears along with  wave function overlaps which physically corresponds to the dipole moment of the excited-electron distribution, namely
\be\label{66}
{\bf d}_{eg}=e\int_{L^3}d^3r\,\vr\Big( \lan e|\vr\ran\, \lan\vr|g\ran\Big )
\ee
We then note that, in order for the above integral to differ from zero, the ground and excited states must have different parities. So, to go further and derive the Frenkel energy dispersion relation which requires ${\bf d}_{eg}$ to differs from zero, it is necessary to reconsider the assumption that the electronic states $|g,\vR\ran$ and $|e,\vR\ran$ have no spatial degeneracy. Rare-gas crystals\cite{Knox_rg} like solid neon\cite{neon1,neon2} and solid argon\cite{argon1,argon2}, which host Frenkel excitons, do have a threefold atomic ground level and a nondegenerate excited level. 

To properly control this spatial degeneracy, it is necessary to redo the calculation all over again, with the state degeneracy introduced from the very first line.

Yet, before doing it, we are going to introduce the electron degeneracy associated with spin because the spin degree of freedom erects a similar but simpler difficulty that ensues from the ground-level degeneracy: when the electronic ground level is degenerate, either due to spin or to spatial degeneracy, the charge neutrality of the semiconductor crystal imposes the ion charge to differ from $|e|$. One then has to question using the eigenstates of Eq.~(\ref{7}) that are the ones for a $|e|$ ion, as a good one-body basis for the second quantization formulation of the Frenkel exciton problem.

\section{Frenkel exciton with spin but no spatial degeneracy\label{sec4}}

\subsection{Quantum formulation}

\subsubsection{System Hamiltonian}

\begin{figure}[t]
\centering
\includegraphics[trim=2.5cm 4.5cm 8cm 7.9cm,clip,width=3in]{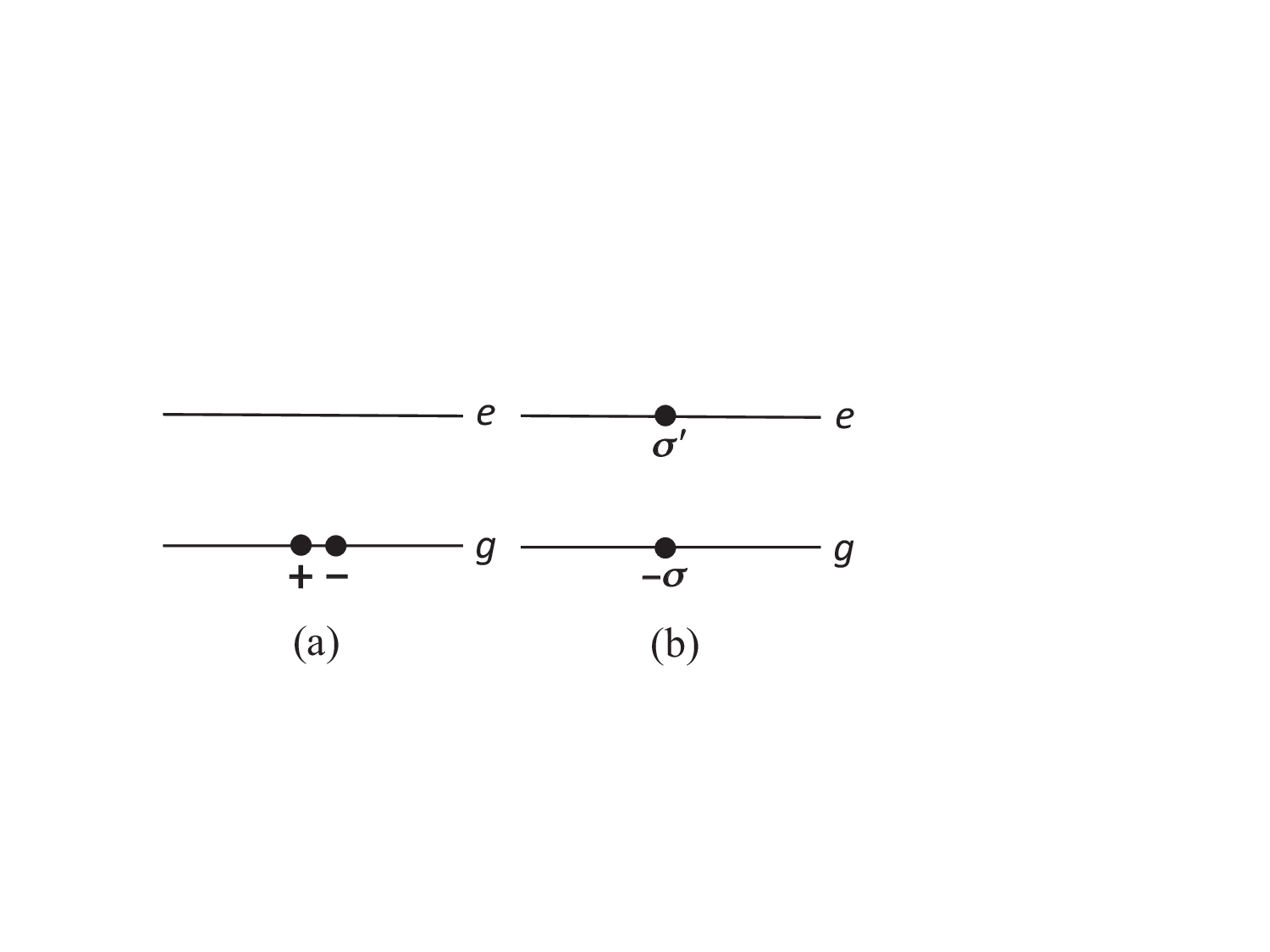}
\vspace{-0.7cm}
\caption{(a) In the presence of spin, each ion site can be occupied by two ground-level electrons, with up and down spins. (b) In the lowest set of excited states, one site has  a ground-level electron with spin $-\sigma$, and an excited-level electron with spin $\sigma'=\sigma$ if we include the fact that for excitation induced by Coulomb interaction or electron-photon interaction, the electron spin is conserved.  }
\label{fig8}
\end{figure}

We now consider electrons with spin $s=\pm1/2$, in electronic levels with no spatial degeneracy. Each lattice site can be occupied by \textit{two} opposite-spin electrons (see Fig.~\ref{fig8}(a)); so, the ion charge must be equal to $2|e|$ in order to ensure the crystal neutrality. The system we then have to consider is made of $2N_s$ electrons with mass $m_0$, charge $-|e|$ and spin $s=\pm1/2$, located at $\textbf{r}_{s,j}$ for $j=(1,\cdots, N_s)$, and $N_s$ ions with infinite mass, charge $2|e|$, located at the $\vR_\ell$ nodes  of a periodic lattice, for $\ell=(1,\cdots,N_s$). In first quantization, the Hamiltonian of this system reads as
\bea
\label{70'}
H_{2N_s}&=&\sum_{s=\pm 1/2}\,\,    \sum_{j=1}^{N_s}\frac{\vp_{s,j}^2}{2m_0}+ \sum_{s=\pm 1/2} \,\,\sum_{j=1}^{N_s}\sum_{\ell=1}^{N_s}\frac{-2e^2}{|\vr_{s,j}-\vR_\ell|}\nn\\
&&+   \frac{1}{2}  \sum_{s=\pm 1/2}\,\,  \sum_{j=1}^{N_s}\left(\sum_{j'\not=j}^{N_s}\frac{e^2}{|\vr_{s,j}-\vr_{s,j'}|}+\sum_{j'=1}^{N_s}\frac{e^2}{|\vr_{s,j}-\vr_{-s,j'}|} \right)\nn\\
&&+\frac{1}{2}\sum_{\ell=1}^{N_s}\sum_{\ell'\not=\ell}^{N_s}\frac{(2e)^2}{|\vR_\ell-\vR_{\ell'}|}
\eea

Here again, this Hamiltonian corresponds to point-charge ions located at each $\vR_\ell$ lattice site, for simplicity. When spin is included, the atom or molecule located in the $\vR_\ell$ cell should be visualized as a core plus two electrons. The ``core'' then includes the  nucleus plus the cloud of the remaining electrons, the total core charge being $2|e|$ due to neutrality. This would lead us to replace the point-charge potential $-2e^2/|\vr_{s,j}-\vR_\ell|$ in Eq.~(\ref{70'}) by a $v_{2|e|}(\vr_{s,j}-\vR_\ell)$ potential with the same $2|e|$ charge somewhat broadened over the $\vR_\ell$ cell. However, here again, the precise shape of this potential is not going to affect the Coulomb physics we study for it entirely comes from the electron-electron Coulomb interaction.

\subsubsection{Good one-electron basis}

To determine the good one-electron basis for  second quantization, we first have to perform the physical analysis of the problem. In spite of the fact that the ion charge now is $2|e|$ instead of $|e|$, the electronic states for one electron in the presence of a $|e|$ charge located at $\vR_\ell$,
\be\label{100}
|s,\nu,\vR_\ell\ran= \hat{a}^\dag_{s,\nu,\ell}|v\ran \qquad \qquad \nu=(g,e)
\ee 
with energies $\va_\nu$, still form the relevant one-electron basis to describe this $2{N_s}$-electron system. The reason is that when one electron jumps from the ground level to the excited level of the same ion, the electron with opposite spin, stays on this ion. So, the excited electron sees the $2|e|$ ion shielded by the $-|e|$ cloud of the remaining electron, which looks more as a $|e|$ ion than as a $2|e|$ ion. By extending this understanding to the ground level, we are led to conclude that the relevant states to describe the initial and final states for the excited electron are the electronic ground state $g$ of a $2|e|$ ion surrounded by a $-|e|$ electron cloud, and the excited state $e$ of this $|e|$ effective ion. A way to represent the system ground state then is
\be
\label{101}
|\Phi_g\ran=( \hat{a}^\dag_{+,g,1} \hat{a}^\dag_{-,g,1})(\hat{a}^\dag_{+,g,2} \hat{a}^\dag_{-,g,2})\cdots \cdots (\hat{a}^\dag_{+,g,{N_s}} \hat{a}^\dag_{-,g,{N_s}})|v\ran
\ee 
from which the set of system excited states simply follows from replacing one of the ground levels $g$ by the excited level $e$, the energy difference between the excited and ground states of $2N_s$ electrons then being $(\va_e-\va_g)$, if we neglect Coulomb processes between the electron and the other ions and between the electrons themselves.

Indeed, the lowest set of excited states we are going to consider has one electron with spin $\sigma$ in the excited level, and the second electron of the same ion, with spin $-\sigma$, in the ground level (see  Fig.~\ref{fig8}(b)).  
The reason is that the physical interactions responsible for this jump, like Coulomb interaction or electron-photon interaction, conserve the spin; so, the electron keeps its spin when jumping to the excited level. It is possible to formally include a spin-flip along the electron excitation. We would then have to consider all spin-triplet and spin-singlet electron-hole states, not just the two $S_z=0$ combinations, that is, four states instead of two, without any more physical insight on the understanding of the energy splitting between triplet and singlet states because both singlet and triplet states exist for $S_z=0$. This is why, to render the restriction of the interlevel Coulomb coupling to spin-singlet states  more transparent, we have decided to restrict the possible excited states to the physically relevant configuration, $\sigma'=\sigma$. The resulting states, shown in Fig.~\ref{fig8}(b), then read
\be\label{102}
|\Phi_{\sigma;\vR_\ell}\ran=\hat{a}^\dag_{\sigma,e,\ell}\hat{a}_{\sigma,g,\ell}|\Phi_g\ran
\quad\quad\quad \quad  \sigma=\pm \frac 1 2
\ee
 with energy close to $E_g+\va_e-\va_g$ if we neglect additional Coulomb interactions. These states form a $(2\times N_s)$ degenerate subspace, with
 \be\label{102'}
\lan \Phi_{\sigma';\vR_{\ell'}}|\Phi_{\sigma;\vR_\ell}\ran = \delta_ {\sigma',\sigma} \,\,\delta_ {\ell',\ell}
\ee 
due to the orthogonality of the $|s,\nu,\vR_\ell\ran$ states in  the tight-binding limit (\ref{5}).
  
 The Frenkel excitons follow from the diagonalization of the system Hamiltonian $H_{2N_s}$ in this subspace, namely, the diagonalization of the (${2N_s}\times{2N_s}$) matrix 
 \be\label{43'}
\lan \Phi_{\sigma';\vR_{\ell'}}|H_{2N_s}|\Phi_{\sigma;\vR_\ell}\ran 
\ee 
To derive these matrix elements, the first step is to derive  the operator  $\hat{H}$ that represents the $H_{2N_s}$ Hamiltonian in the basis  $|s,\nu,\vR_\ell\ran$ with $\nu=(g,e)$ and $\ell=(1,\cdots,{N_s})$, taken in the tight-binding limit (\ref{5}).

 \subsubsection{Hamiltonian in terms of  $(\nu,\ell)$ electron operators}
 
 $\bullet$ The one-body part of the $H_{2N_s}$ Hamiltonian,
 \be\label{75'}
H_{0,{2N_s}}= \sum_{s=\pm 1/2} \,\,\sum_{j=1}^{N_s}  \left(\frac{\vp^2_{s,j}}{2m_0}+ \sum_{\ell=1}^{N_s}\frac{-2e^2}{|\vr_{s,j}-\vR_{\ell}|} \right)  
\ee
appears in terms of electron  operators $\hat{a}^\dag_{s,\nu,\ell}$ as
\be\label{13'}
\hat{H}_0=\sum_{s=\pm 1/2}\,\, \sum_{\nu', \ell'}\sum_{\nu, \ell}h_{\nu', \ell';\nu, \ell}\,\,\hat{a}^\dag_{s,\nu',\ell'}\hat{a}_{s,\nu,\ell}
\ee
with $\hat{H}_0$ diagonal with respect to spin because $H_{0,{2N_s}}$ does not act on spin. 
 The second quantization procedure gives the prefactor as
 \bea
 \label{14_2}
h_{\nu', \ell';\nu, \ell}=
\int_{L^3}d^3r\, \lan\nu',\vR_{\ell'}|\vr \ran \left(\frac{\hat{\vp}^2}{2m_0}{+}\sum_{\ell''=1}^{N_s}\frac{-2e^2}{|\vr{-}\vR_{\ell''}|}\right)\lan\vr|\nu,\vR_{\ell}\ran
\eea

To calculate it, we first extract the $-e^2/|\vr{-}\vR_{\ell}|$ potential from the $\ell''$ sum to make appear the Hamiltonian of one electron in the presence of a $|e|$ charge located at $\vR_{\ell}$, that brings the  energy $\va_\nu$. The remaining part of the $\ell''$ sum corresponds to the electron interaction with all the other $2|e|$ ions and  the electron interaction with 
the $|e|$ effective ion located at $\vR_{\ell}$.
Within the  tight-binding limit which enforces $\ell'=\ell$ in Eq.~(\ref{14_2}), this leads us to split $h_{\nu', \ell';\nu, \ell}$ as 
\be\label{15'}
h_{\nu', \ell';\nu, \ell}=\delta_{\ell',\ell}\Big(\va_\nu\delta_{\nu',\nu}+t_{\nu',\nu}+u_{\nu',\nu} \Big)
\equiv
\delta_{\ell',\ell}\Big(\va_\nu\delta_{\nu',\nu}+t'_{\nu',\nu}\Big)
\ee

The $t_{\nu',\nu}$ contribution, that comes from the electron interaction with all the other  $2|e|$ ions, is given, for $\lan \vr|\nu,\vR_\ell\ran$ in Eq.~(\ref{3}) and $\vr-\vR_{\ell}=\vr_\ell$, by 
\bea
t_{\nu',\nu}=t^\ast_{\nu,\nu'}
%\\
%&=&
&=& \int_{L^3}d^3r_\ell \,\lan\nu'|\vr_\ell\ran\lan\vr_\ell|\nu\ran \sum_{\ell''\not=\ell}^{N_s}\frac{-2e^2}{|\vr_\ell+\vR_\ell-\vR_{\ell''}|}\nn\\
&=&\int_{L^3}d^3r\, \lan\nu'|\vr\ran\lan\vr|\nu\ran \sum_{\vR\not=\bf0}\frac{-2e^2}{|\vr-\vR|}\label{16_2}
\eea
the $\ell''$ sum being $\ell$-independent due to the Born-von Karman boundary condition used to extend the lattice periodicity to a size $L$ crystal, that is,  $f(\vr)=f(\vr+{\bf L})$ for a sample volume $L^3$. 

The electron interaction with the $|e|$ effective ion located at $\vR_{\ell}$ brings another contribution that reads
\be
u_{\nu',\nu}=\int_{L^3}d^3r \,\lan\nu',\vR_{\ell}|\vr \ran
 \frac{-e^2}{|\vr{-}\vR_{\ell}|}
\lan\vr|\nu,\vR_{\ell}\ran
=\int_{L^3}d^3r\, \lan\nu'|\vr \ran
 \frac{-e^2}{|\vr|}
\lan\vr|\nu\ran
\ee
This additional term comes from the fact that the ion charge of this system is $2|e|$, but the electronic states that are used for the second quantization procedure is defined for a $|e|$ ion. This $u_{\nu',\nu}$ term  brings  additional transitions between the electronic levels of the same ion, just like the $t_{\nu',\nu}$ terms. 

The above results lead us to split the one-body Hamiltonian $\hat{H}_0$ in a way quite similar to its expression (\ref{17}) in the absence of spin, namely
\be\label{17'}
\hat{H}_0=\sum_{s,\nu,\ell} \va_\nu\hat{a}^\dag_{s,\nu,\ell}\hat{a}_{s,\nu,\ell}+
\sum_{s,\nu',\nu,\ell} t'_{\nu',\nu} \hat{a}^\dag_{s,\nu',\ell}\hat{a}_{s,\nu,\ell}
\ee

 $\bullet$ The electron-electron interaction has the same form as in the absence of spin, because this interaction is not affected by the ion-charge change from $2|e|$ to $|e|$. It simply follows from Eq.~(\ref{22}) as
  \bea\label{22'}
 \hat{V}_{e-e}=\frac{1}{2}\sum_{\{s,\nu\}}\sum_{\ell_1,\ell_2}\mathcal{V}_{\vR_{\ell_1}-\vR_{\ell_2}}
 \left(\begin{smallmatrix}
\nu'_2&\hspace{0.1cm} \nu_2\\ \hspace{0.1cm}\nu'_1& \nu_1\end{smallmatrix}\right)
\hat{a}^\dag_{s_1,\nu'_1,\ell_1}\hat{a}^\dag_{s_2,\nu'_2,\ell_2}\hat{a}_{s_2,\nu_2,\ell_2}\hat{a}_{s_1,\nu_1,\ell_1}
 \eea
   with the same scattering as the one given in Eq.~(\ref{21_0}), since the Coulomb interaction does not act on spin. Note that the Pauli exclusion principle imposes $(s,j)\not=(s',j')$ in the first quantization expression (\ref{70'}) of the electron-electron interaction, while the commutation relations for fermionic operators take care of this exclusion automatically, in the above expression.

\subsubsection{Relevant parts of the Hamiltonian}
The parts of the Hamiltonian that play a role in the derivation of Frenkel excitons, are the ones that keep constant the number of ground-level electrons and the number of excited-level electrons, as the other states have a higher energy.

  \noindent $\bullet$ The relevant parts of $\hat{H}_{0}$ given in Eq.~(\ref{17'}) reduce to $\hat{H}_{0,g}+\hat{H}_{0,e}$, which formally read as the ones given in Eqs.~(\ref{40_2},\ref{40_3}), namely
 \bea
 \hat{H}_{0,g}&=&(\va_g+t'_{g,g})\sum_{s,\ell} \hat{a}^\dag_{s,g,\ell}\hat{a}_{s,g,\ell}\label{103}\\
 \hat{H}_{0,e}&=&(\va_e+t'_{e,e})\sum_{s,\ell} \hat{a}^\dag_{s,e,\ell}\hat{a}_{s,e,\ell}\label{104}
 \eea

\begin{figure}[t]
\centering
\includegraphics[trim=3cm 4.5cm 4.5cm 7.5cm,clip,width=3.5in]{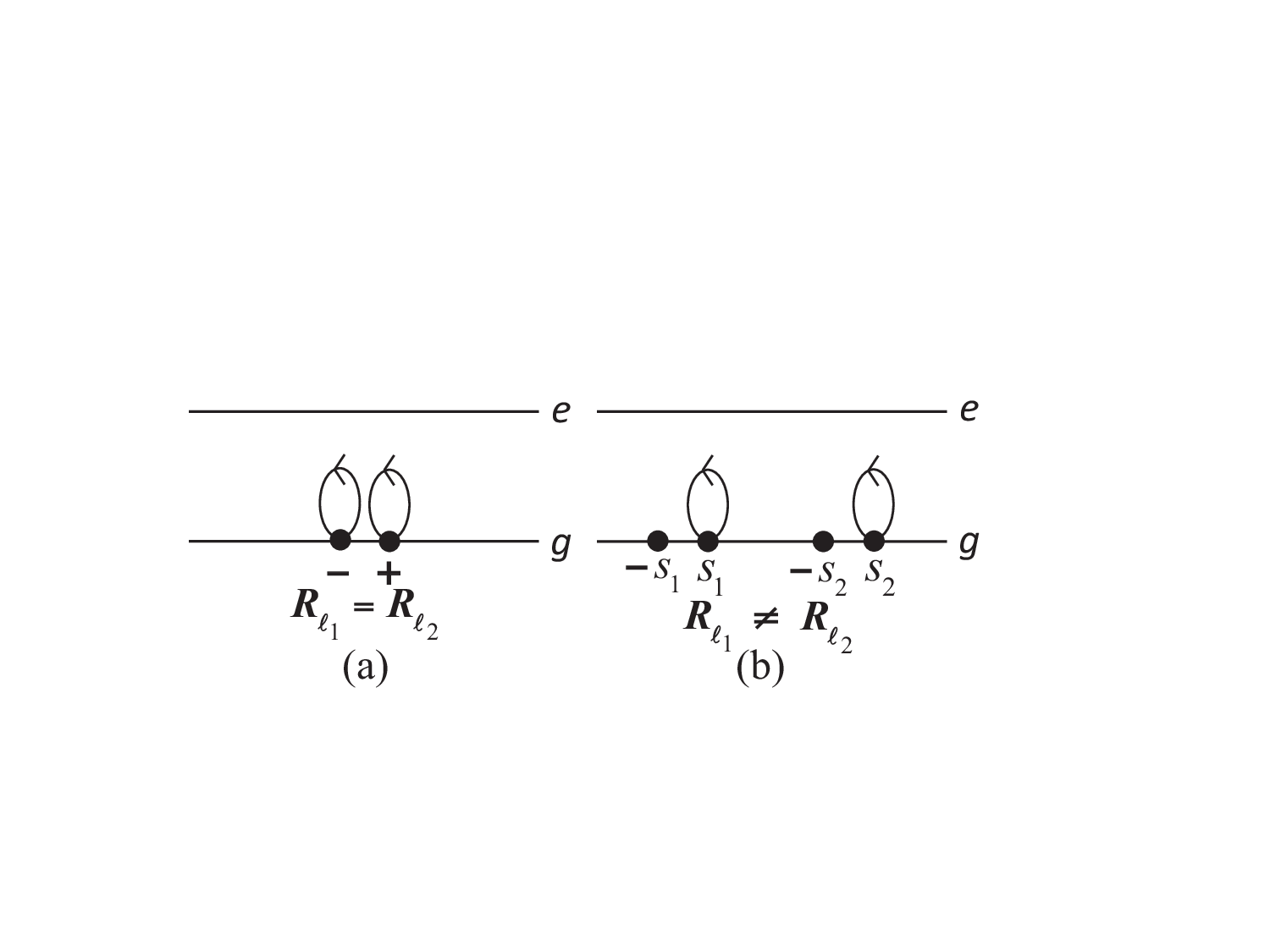}
\vspace{-0.7cm}
\caption{The Coulomb interaction among ground-level electrons can take place on the same ion site, provided that the two electrons have different spins as in (a), or on different ion sites, whatever the electron spins, as in (b).}
\label{fig9}
\end{figure}

  \noindent $\bullet$ Relevant parts of the Hamiltonian also come from the Coulomb interaction (\ref{22'}). The part between ground levels reads
 \bea\label{105}
 \hat{V}_{gg}=\frac{1}{2}\sum_{\ell_1,s_1}\sum_{\ell_2,s_2}\mathcal{V}_{\vR_{\ell_1}-\vR_{\ell_2}}\left(\begin{smallmatrix}
 g& g\\ g& g\end{smallmatrix}\right)
 \hat{a}^\dag_{s_1,g,\ell_1}\hat{a}^\dag_{s_2,g,\ell_2}\hat{a}_{s_2,g,\ell_2}\hat{a}_{s_1,g,\ell_1}
 \eea
 Note that $(s_1,\ell_1)$ must differ from $(s_2,\ell_2)$ in order for $\hat{a}_{s_2,g,\ell_2}\hat{a}_{s_1,g,\ell_1}$ to differ from zero, due to the Pauli exclusion principle. This allows intersite processes, $\ell_1\not=\ell_2$, for arbitrary spin $(s_1,s_2)$, as in Fig.~\ref{fig9}(b), but also on-site processes, $\ell_2=\ell_1$, provided that $s_1\not=s_2$, as in Fig.~\ref{fig9}(a).

\begin{figure}[t]
\centering
\includegraphics[trim=3cm 4.5cm 5.5cm 6.5cm,clip,width=3.5in]{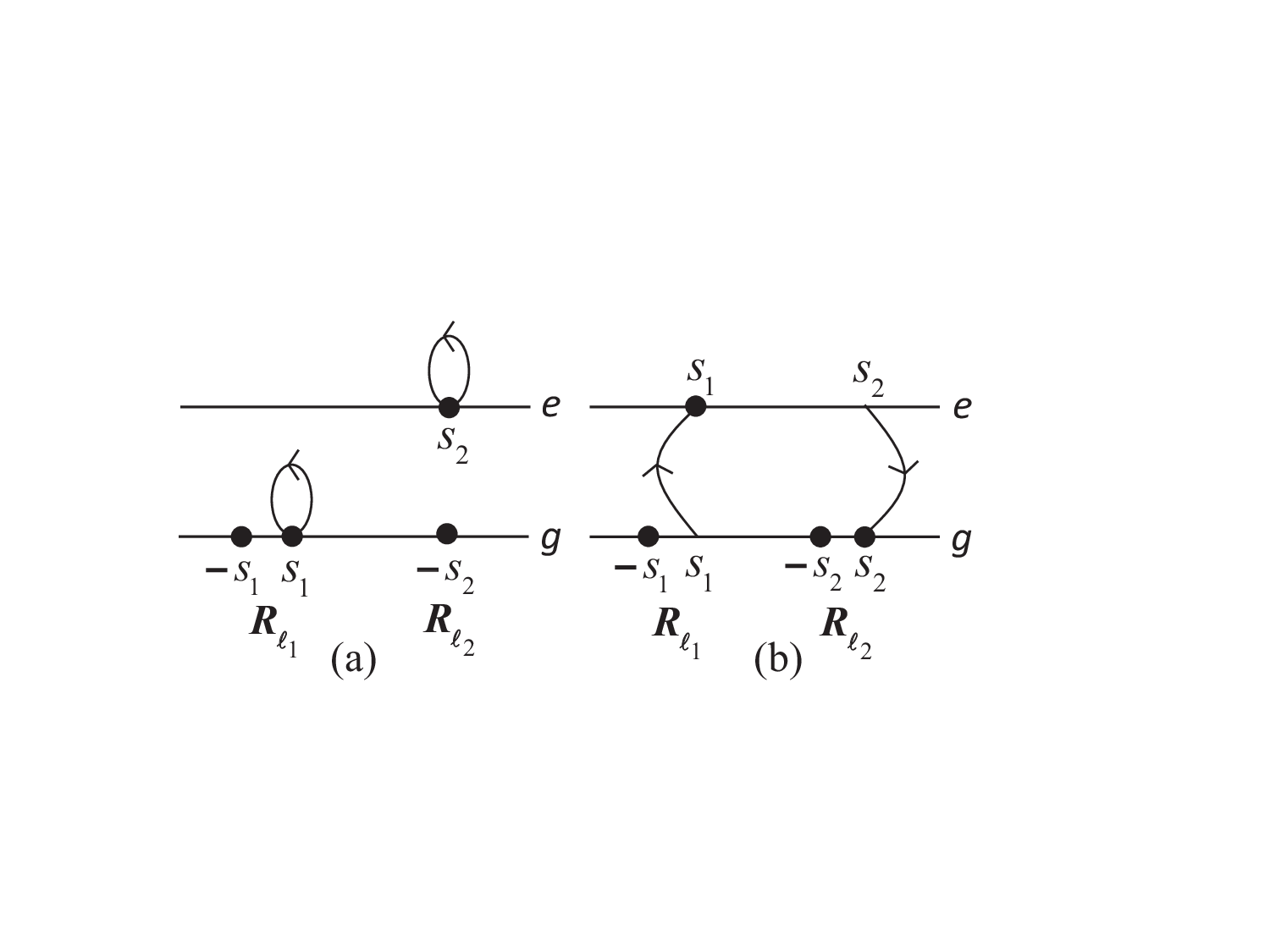}
\vspace{-0.7cm}
\caption{Coulomb interaction between electrons in different levels, as given in Eqs.~(\ref{106},\ref{107}). The electrons can stay in their level as in (a) or change  level as in (b). In the latter case, they still keep their spin because the Coulomb interaction does not act on spin.  }
\label{fig10}
\end{figure}

 \noindent $\bullet$ Relevant parts of the Coulomb interaction also involve the ground and excited levels. The intralevel Coulomb interaction between ground and excited levels given in Eq.~(\ref{40_6}) reads, when spin is included, as (see Fig.~\ref{fig10}(a))
  \bea\label{106}
 \hat{V}^{(intra)}_{eg}=\sum_{\ell_1,s_1}\sum_{\ell_2,s_2}\mathcal{V}_{\vR_{\ell_1}-\vR_{\ell_2}}\left(\begin{smallmatrix}
 e& e\\ g& g\end{smallmatrix}\right)
\hat{a}^\dag_{s_1,g,\ell_1}\hat{a}^\dag_{s_2,e,\ell_2}\hat{a}_{s_2,e,\ell_2}\hat{a}_{s_1,g,\ell_1}
 \eea
 In the same way, the interlevel Coulomb interaction, given in Eq.~(\ref{40_7}), becomes (see Fig.~\ref{fig10}(b))
  \bea\label{107}
 \hat{V}^{(inter)}_{eg}=\sum_{\ell_1,s_1}\sum_{\ell_2,s_2}\mathcal{V}_{\vR_{\ell_1}-\vR_{\ell_2}}\left(\begin{smallmatrix}
 g& e\\ e& g\end{smallmatrix}\right)
\hat{a}^\dag_{s_1,e,\ell_1}\hat{a}^\dag_{s_2,g,\ell_2}\hat{a}_{s_2,e,\ell_2}\hat{a}_{s_1,g,\ell_1}
 \eea

 From the above results, it is possible to derive the ground-state energy $\lan \Phi_g|\hat{H}|\Phi_g\ran $, and the Hamiltonian matrix $\lan \Phi_{\sigma';\vR_{\ell'}}|\hat{H}|\Phi_{\sigma;\vR_\ell}\ran $ in the excited subspace, (see \ref{app:D}). However, as mentioned earlier, the appropriate formulation of the exciton problem is not in terms of electron states but in terms of electron excitations, that is, electrons and holes. When dealing with spin, this formulation makes even more sense because the differentiation between spin-singlet and spin-triplet configurations is better seen with electron-hole pairs.

 \subsubsection{Formulation in terms of electrons and holes}
 
 \noindent $\bullet$ The sign difference between spin-singlet an spin-triplet configurations follows from the phase factor that appears when turning from electron-absence to hole operator\cite{SS2021,JPCM2021}, namely
 \be\label{115}
\hat{a}_{s,g,\ell}=(-1)^{1/2-s}\hat{b}^\dag_{-s,\ell} 
  \ee
  while  the electron operator simply follows from the excited state operator as
  \be\label{115'}
 \hat{a}^\dag_{s,e,\ell}=\hat{a}^\dag_{s,\ell} 
   \ee

  \noindent $\bullet$ The above relations can be readily used to  rewrite the various parts of the Hamiltonian given above. The one-body parts (\ref{103},\ref{104}) appear in terms of electrons and holes as
 \bea\label{116}
 \hat{H}_{0,g}&=&2N(\va_g+t'_{g,g})-\left(\va_g+t'_{g,g}\right)\sum_{s,\ell}\hat{b}^\dag_{s,\ell}\hat{b}_{s,\ell}
 \\
 \hat{H}_{0,e}&=&\left(\va_e+t'_{e,e}\right)\sum_{s,\ell}\hat{a}^\dag_{s,\ell}\hat{a}_{s,\ell}
 \eea

Turning to the Coulomb parts, the $\hat{V}_{gg}$ interaction given in Eq.~(\ref{105}) becomes
 \be
 \hat{V}_{gg}=\lan \Phi_g|\hat{V}_{gg}|\Phi_g\ran\label{116_1}
 -\bigg(\mathcal{V}_{\vR=\bf0}\left(\begin{smallmatrix}
 g& g\\ g& g\end{smallmatrix}\right)+2\sum_{\vR\not=\bf0} \mathcal{V}_{\vR}\left(\begin{smallmatrix}
 g& g\\ g& g\end{smallmatrix}\right) \bigg) \sum_{s,\ell}\hat{b}^\dag_{s,\ell}\hat{b}_{s,\ell}+\cdots
 \ee
within an additional term that acts on two holes and thus  plays no role in the excited subspace that contains one hole only. So, we forget it.

With regard to the Coulomb interaction between ground and excited levels,  the intralevel part (\ref{106}) becomes
  \bea
 \hat{V}^{(intra)}_{eg}&=& \sum_{s_1,\ell_1}\sum_{s_2,\ell_2}\mathcal{V}_{\vR_{\ell_1}-\vR_{\ell_2}}\left(\begin{smallmatrix}
 e& e\\ g& g\end{smallmatrix}\right)
\hat{b}_{-s_1,\ell_1}\hat{a}^\dag_{s_2,\ell_2}\hat{a}_{s_2,\ell_2}\hat{b}^\dag_{-s_1,g,\ell_1}
  \nn
   \\
 &=& 2\sum_{\vR} \mathcal{V}_{\vR}\left(\begin{smallmatrix}
 e& e\\ g& g\end{smallmatrix}\right)\sum_{s,\ell}\hat{a}^\dag_{s,\ell}\hat{a}_{s,\ell}+ \hat{V}^{(intra)}_{eh}\label{116_2}
  \eea
while the interlevel part (\ref{107}) appears as
 \bea
 \hat{V}^{(inter)}_{eg}&=&
 \sum_{s_1,\ell_1}\sum_{s_2,\ell_2}(-1)^{1-s_1-s_2}\,\mathcal{V}_{\vR_{\ell_1}-\vR_{\ell_2}}\left(\begin{smallmatrix}
 g& e\\ e& g\end{smallmatrix}\right)\,
\hat{a}^\dag_{s_1,\ell_1}\hat{b}_{-s_2,\ell_2}\hat{a}_{s_2,\ell_2}\hat{b}^\dag_{-s_1,\ell_1}
  \nn
   \\
  &=& -\mathcal{V}_{\vR=\bf0}\left(\begin{smallmatrix}
 g& e\\ e& g\end{smallmatrix}\right)\sum_{s,\ell}\hat{a}^\dag_{s,\ell}\hat{a}_{s,\ell}+ \hat{V}^{(inter)}_{eh}\label{116_3}
 \eea

\begin{figure}[t]
\centering
\includegraphics[trim=4.5cm 4.5cm 3cm 6.5cm,clip,width=3.7in]{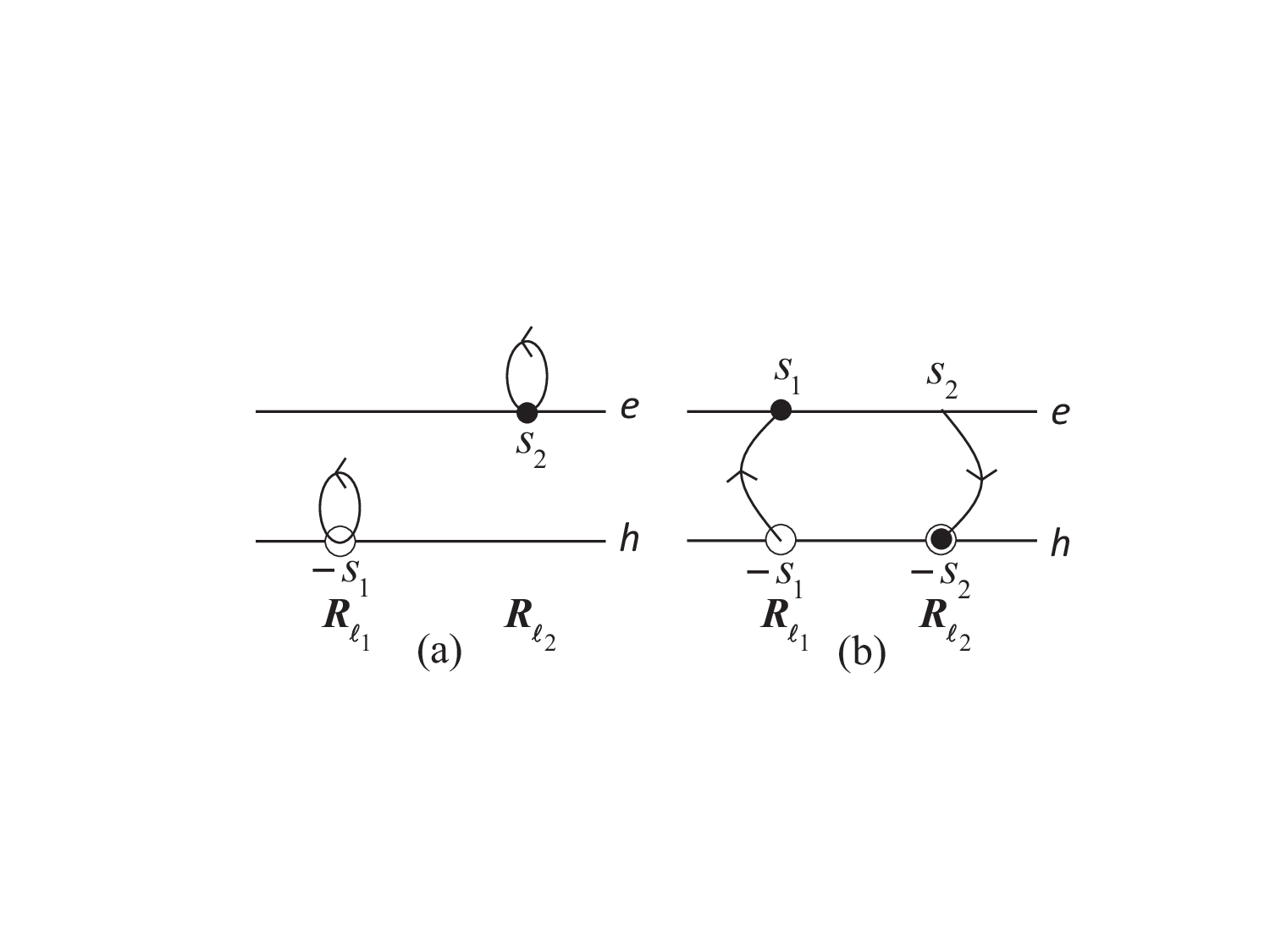}
\vspace{-0.7cm}
\caption{Same as Fig.~\ref{fig10}, but in terms of electron and hole: (a) intralevel processes given in Eq.~(\ref{116_4}); (b) interlevel processes given in Eq.~(\ref{117}). The fact that the interlevel Coulomb processes take place between electron-hole pairs in spin-singlet states and not in ($S_z=0$) triplet states, follows from the phase factor between electron-absence and hole operators (see Eq.~(\ref{115})).   }
\label{fig13}
\end{figure}

The Coulomb interaction between electrons and holes that corresponds to the Coulomb processes shown in Fig.~\ref{fig13}(a), with each carrier staying in its level, is given by
\bea\label{116_4}
\hat{V}^{(intra)}_{eh}=-\sum_{s_1,\ell_1}\sum_{s_2,\ell_2}\mathcal{V}_{\vR_{\ell_1}-\vR_{\ell_2}}\left(\begin{smallmatrix}
 e& e\\ g& g\end{smallmatrix}\right)\,
 \hat{a}^\dag_{s_2,\ell_2}\hat{b}^\dag_{-s_1,\ell_1}\hat{b}_{-s_1,\ell_1}\hat{a}_{s_2,\ell_2}
\eea

  The Coulomb interaction in which each electron changes electronic level leads, in terms of electrons and holes, to an interaction in which one electron-hole pair recombines while another pair is excited, within a phase factor that comes from Eq.~(\ref{115}). It reads (see Fig.~\ref{fig13}(b))
\bea\label{117}
\hat{V}^{(inter)}_{eh}=\sum_{s_1,\ell_1}\sum_{s_2,\ell_2}\mathcal{V}_{\vR_{\ell_1}-\vR_{\ell_2}}\left(\begin{smallmatrix}
 g& e\\ e& g\end{smallmatrix}\right)
  (-1)^{1-s_1-s_2}\,\hat{a}^\dag_{s_1,\ell_1}\hat{b}^\dag_{-s_1,\ell_1}\hat{b}_{-s_2,\ell_2}\hat{a}_{s_2,\ell_2}
\eea
 or, by writing the spin part explicitly, 
 \bea\label{118}
\hat{V}^{(inter)}_{eh}&=&\sum_{\ell_1,\ell_2}\mathcal{V}_{\vR_{\ell_1}-\vR_{\ell_2}}\left(\begin{smallmatrix}
 g& e\\ e& g\end{smallmatrix}\right)\\
 &&
 \Big(\hat{a}^\dag_{\frac{1}{2},\ell_1}\hat{b}^\dag_{-\frac{1}{2},\ell_1}-\hat{a}^\dag_{-\frac{1}{2},\ell_1} \hat{b}^\dag_{\frac{1}{2},\ell_1} \Big)\Big(\hat{b}_{-\frac{1}{2},\ell_2}\hat{a}_{\frac{1}{2},\ell_2}-
 \hat{b}_{\frac{1}{2},\ell_2}\hat{a}_{-\frac{1}{2},\ell_2}\Big)
 \nn
\eea
This evidences that the pairs involved in the interlevel Coulomb processes are in a spin-singlet state $(S=0,S_z=0)$ with creation operator
\be\label{125'}
\hat{B}^\dag_{0,\ell}=\frac{\hat{a}^\dag_{\frac{1}{2},\ell}\hat{b}^\dag_{-\frac{1}{2},\ell}-\hat{a}^\dag_{-\frac{1}{2},\ell}\hat{b}^\dag_{\frac{1}{2},\ell}}{\sqrt{2}}
\ee
in contrast to pairs that also have a total spin equal to zero, but that are in a spin-triplet state $(S=1,S_z=0)$, their creation operator reading as
\be\label{124'}
\hat{B}^\dag_{1,\ell}=\frac{\hat{a}^\dag_{\frac{1}{2},\ell}\hat{b}^\dag_{-\frac{1}{2},\ell}+\hat{a}^\dag_{-\frac{1}{2},\ell}\hat{b}^\dag_{\frac{1}{2},\ell}}{\sqrt{2}}
\ee

 \noindent $\bullet$ As a result, the relevant parts of the $\hat{H}$ Hamiltonian in the $|\Phi_{\sigma;\vR_\ell}\ran$
subspace reduce, in terms of electron and hole operators, to 
\bea
\label{121a}
\hat{H}_{eh}=\lan \Phi_g|\hat{H}|\Phi_g\ran+
\hat{H}_{e}+\hat{H}_{h}+\hat{V}^{(intra)}_{eh}+\hat{V}^{(inter)}_{eh}
\eea
The two-body Coulomb parts are given in Eqs.~(\ref{116_4}) and (\ref{118}), while the one-body parts now read
\bea
\label{121b}
\hat{H}_{e}&=&\tilde{\va}_e\sum_{s,\ell}\hat{a}^\dag_{s,\ell}\hat{a}_{s,\ell}\label{119}\\
\hat{H}_{h}&=&\tilde{\va}_h\sum_{s,\ell}\hat{b}^\dag_{s,\ell}\hat{b}_{s,\ell}\label{120}
\eea
with the electron and hole energies given by 
\bea\label{121}
\tilde{\va}_e&=&\va_e+t'_{e,e}+2\sum_{\vR}\mathcal{V}_{\vR}\left(\begin{smallmatrix}
 e& e\\ g& g\end{smallmatrix}\right)- \mathcal{V}_{\vR=\bf0}\left(\begin{smallmatrix}
 g& e\\ e& g\end{smallmatrix}\right)
 \\
\tilde{\va}_h &=& -\va_g-t'_{g,g}-2\sum_{\vR\not=\bf0}\mathcal{V}_{\vR}\left(\begin{smallmatrix}
 g& g\\ g& g\end{smallmatrix}\right)- \mathcal{V}_{\vR=\bf0}\left(\begin{smallmatrix}
 g& g\\ g& g\end{smallmatrix}\right)
\eea
%due to contributions coming from the various Coulomb interactions.

\subsection{Spin-singlet and spin-triplet subspaces $|\Phi_{S;\vR_\ell}\ran$ }

\noindent $\bullet$ The lowest set of excited states $|\Phi_{\sigma;\vR_\ell}\ran$, given in Eq.~(\ref{102}), appears in terms of electron and hole operators as
\be\label{96'}
|\Phi_{\sigma;\vR_\ell}\ran=\hat{a}^\dag_{\sigma,e,\ell}\hat{a}_{\sigma,g,\ell}|\Phi_g\ran
=(-1)^{1/2-\sigma} \,\hat{a}^\dag_{\sigma,\ell}\hat{b}^\dag_{-\sigma,\ell}|0\ran
\ee
They are made of electron-hole pairs located on the ion site $\textbf{R}_\ell$ with total spin $S_z=0$. Yet, in view of Eq.~(\ref{118}), it is clear that the physically relevant electron-hole pair states are not these $|\Phi_{\sigma;\vR_\ell}\ran$ states with $\sigma=\pm1/2$, but their symmetric and antisymmetric combinations, namely the spin-triplet and spin-singlet states, which read in terms of the operators defined in Eqs.~(\ref{125'}) and (\ref{124'}), as
\be\label{96''}
|\Phi_{S;\vR_\ell}\ran= \hat{B}^\dag_{S,\ell}  |0\ran
   \quad\quad\quad  \textrm{for} \quad\quad\quad    S=(1,0)
\ee
These states also form an orthogonal set, as possible to check from Eq.~(\ref{102'}),
 \be\label{102'_0}
\lan \Phi_{S';\vR_{\ell'}}|\Phi_{S;\vR_\ell}\ran = \delta_ {S',S} \,\,\delta_ {\ell',\ell}
\ee 

\noindent $\bullet$ When acting on these singlet and triplet states, the one-body parts of the system Hamiltonian give
\be\label{127}
\Big(\hat{H}_{e}+\hat{H}_{h}\Big)|\Phi_{S;\vR_\ell}\ran=(\tilde{\va}_e+\tilde{\va}_h)|\Phi_{S;\vR_\ell}\ran
\ee
 while the intralevel part of the Coulomb interaction leads to
\bea\label{127'}
\hat{V}^{(intra)}_{eh} |\Phi_{S;\vR_\ell}\ran=-\mathcal{V}_{\vR=\v0}\left(\begin{smallmatrix}
 e& e\\ g& g\end{smallmatrix}\right)
  |\Phi_{S;\vR_\ell}\ran
\eea
If we now consider the interlevel part of the Coulomb interaction given in Eq.~(\ref{118}) which only acts on spin-singlet states, we readily find
\bea\label{127''}                   
\hat{V}^{(inter)}_{eh}|\Phi_{S;\vR_\ell}\ran=2\delta_ {S,0} \sum_{\ell_1} 
\mathcal{V}_{\vR_\ell-\vR_{\ell_1}}\left(\begin{smallmatrix}
 g& e\\ e& g\end{smallmatrix}\right)
 |\Phi_{S;\vR_{\ell_1}}\ran
\eea

By combining the above equations, we end with
\bea\label{127a}
\hat{H}_{eh}|\Phi_{S;\vR_\ell}\ran=E^{(pair)}_S |\Phi_{S;\vR_\ell}\ran
+ 2\delta_ {S,0} \sum_{\ell_1\neq \ell} 
\mathcal{V}_{\vR_\ell-\vR_{\ell_1}}\left(\begin{smallmatrix}
 g& e\\ e& g\end{smallmatrix}\right)
 |\Phi_{S;\vR_{\ell_1}}\ran
\eea
 with the pair energy given by  
\bea
E^{(pair)}_S\!\!\!&=&\!\!\! \lan \Phi_g|\hat{H}|\Phi_g\ran+(\va_e-\va_g) +(t'_{e,e}-t'_{g,g})+2\sum_{\vR\not=\bf0}\Big(\mathcal{V}_{\vR}\left(\begin{smallmatrix}
 e& e\\ g& g\end{smallmatrix}\right)- \mathcal{V}_{\vR}\left(\begin{smallmatrix}
 g& g\\ g& g\end{smallmatrix}\right) \Big)\nn\\
&&\!\!\!+\Big(\mathcal{V}_{\vR=\bf0}\left(\begin{smallmatrix}
 e& e\\ g& g\end{smallmatrix}\right)-\mathcal{V}_{\vR=\bf0}\left(\begin{smallmatrix}
 g& g\\ g& g\end{smallmatrix}\right)\Big)
+(-1)^S \,\mathcal{V}_{\vR=\bf0}\left(\begin{smallmatrix}
 g& e\\ e& g\end{smallmatrix}\right)\label{133}
\eea
This energy  depends on the spin state, triplet or singlet, through its last term: the on-site $(\ell_1=\ell)$ interlevel Coulomb interaction in Eq.~(\ref{127''}) differentiates the spin-triplet state $(S=1,S_z=0)$ from the spin-singlet state $(S=0,S_z=0)$.

\subsection{Frenkel excitons with spin}

 \noindent $\bullet$ To go further, we first note that
\bea
\hat{B}_{S',\ell'}\hat{B}^\dag_{S,\ell}|0\ran=\delta_{\ell',\ell}\delta_{S',S}|0\ran
\eea
It then becomes trivial to show that the ($2N_s\times 2N_s$) matrix for $\hat{H}_{eh}$ in the $|\Phi_{S;\vR_\ell}\ran$ excited subspace, splits into a ($N_s\times N_s$) diagonal submatrix in the spin-triplet subspace $(S=1,S_z=0)$, with diagonal energy $E^{(pair)}_{S=1}$, and a non-diagonal ($N_s\times N_s$) submatrix in the spin-singlet subspace $(S=0,S_z=0)$, that remains to be diagonalized,
\be\label{135}
\left(\begin{matrix}
E^{(pair)}_{S=1}  &\cdots & 0 &0 & \cdots& 0  \\
%0 & E^{(pair)}_{S=1} & \cdots  & 0& \cdots&  \cdots&\vdots \\
 \vdots   & \ddots & 0 & \vdots &\ddots & \vdots  \\
 0  & 0& E^{(pair)}_{S=1}   &0 & \cdots& 0 \\
 0  &\cdots  &0 & E^{(pair)}_{S=0} & \cdots &  \mathcal{V}^{(e,g)}_{\ell',\ell}  \\
 \vdots   &\ddots   &\vdots  & \vdots& \ddots &\vdots \\
  0  & \cdots &0  & \left(\mathcal{V}^{(e,g)}_{\ell',\ell}\right)^\ast & \cdots &  E^{(pair)}_{S=0}  \end{matrix}\right)
\ee 
with $\mathcal{V}^{(e,g)}_{\ell',\ell} =2\mathcal{V}_{\vR_{\ell'}-\vR_{\ell}}\left(\begin{smallmatrix}
 g& e\\ e& g\end{smallmatrix}\right)$ for $\mathcal{V}_{\vR}\left(\begin{smallmatrix}
 g& e\\ e& g\end{smallmatrix}\right)$ given in Eq.~(\ref{63}).

\noindent $\bullet$ To diagonalize this ($N_s\times N_s$) submatrix, we use the same phase factor as the one given in Eq.~(\ref{0}), to transform one-site excitations into wave excitations, namely
\be\label{136}
|\Phi_{0;\vK_n}\ran=\frac{1}{\sqrt{{N_s}}}\sum_{\ell=1}^{N_s} e^{i\vK_n\cdot\vR_\ell}|\Phi_{0;\vR_\ell}\ran
\ee
Indeed, for the same reason as the one leading to Eq.~(\ref{60}), we find that the $\hat{H}_{eh}$ Hamiltonian is diagonal in the $|\Phi_{0;\vK_n}\ran$ subspace 
\be
 \lan \Phi_{0;\vK_{n'}}|  \hat{H}_{eh}|\Phi_{0;\vK_n}\ran=\delta_{n',n}\left(E^{(pair)}_{S=0}+v_{_{\vK_n}}\right)
\ee
with $v_{_{\vK_n}}$ now given by
\be\label{61_1}
v_{_{\vK_n}}=2\sum_{\vR\not=\bf0} \mathcal{V}_{\vR}\!\left(\begin{smallmatrix}
g& e\\ e& g\end{smallmatrix}\right)e^{i\vK_n\cdot\vR}
\ee
the extra factor $2$, compared to Eq.~(\ref{61}), coming from spin. The degeneracy of the spin-singlet subspace is lifted by these linear combinations, thanks to the interlevel Coulomb processes between different ion sites, $\mathcal{V}_{\vR\not=\bf0}\!\left(\begin{smallmatrix}
g& e\\ e& g\end{smallmatrix}\right)$. By contrast, the spin-triplet subspace, already diagonal, remains degenerate.

The major result of this section is the fact that spin-triplet pairs $(S=1,S_z=0)$ do not form excitonic waves. The Frenkel excitons that result from interlevel Coulomb processes through the delocalization of on-site electronic excitations, are only made of spin-singlet pairs. Since these spin-singlet pairs are coupled to photons, the Frenkel excitons that are formed are bright. We wish to stress that this strong conclusion is derived within the strict tight-binding limit that we have used to possibly formulate the problem in second quantization: once accepted, the delocalization of on-site excitations to waves, can only come from interlevel Coulomb processes, \textit{i.e.}, processes which are fundamentally similar to the interlevel processes that occur in photon absorption or emission: an electron-hole pair recombination on one site and a pair excitation on another site.

\section{Frenkel exciton with spatial degeneracy but no spin\label{sec5}}

We now  introduce the spatial degeneracy of the electronic states associated with one ion and we forget the spin, \textit{i.e.}, we  perform the same calculations as the ones in Sec.~\ref{sec3} but we take into account this degeneracy from the very first line. 

\subsection{Threefold excited  level}

\subsubsection{Lowest excited subspace $| \Phi_{\mu;\vR_\ell}\ran$}

We start with the simplest case: $N_s$ electrons with charge $-|e|$ and $N_s$ ions with charge $|e|$ located at the  $\vR_\ell$ lattice sites. The system  Hamiltonian is the same as the one given in Eq.~(\ref{2'}), so is the good one-electron basis for second quantization: this basis is made of the eigenstates of one electron $-|e|$ in the presence of one ion $|e|$ located at $\textbf{R}_\ell$. We consider that the ground level, with energy $\va_g$, is nondegenerate, but that the lowest excited level, with energy $\va_e$, has a threefold degeneracy. We will label these three excited states as $\mu=(x,y,z)$ along axes that can be chosen at will: indeed, the basis we here consider follows from the eigenstates of a single electron in the presence of a single ion;  the other ions play no role, along with the axes of the crystal to which they belong. We will come back to the importance of properly choosing these $(x,y,z)$ axes. 

The relevant electronic states $|\nu,\vR_\ell\ran$ now read
\begin{subeqnarray}\label{67}
|g,\vR_\ell\ran&=& \hat{a}^\dag_{g,\ell}|v\ran\slabel{67a}\\
|\mu,e,\vR_\ell\ran&=& \hat{a}^\dag_{\mu,e,\ell}|v\ran\slabel{67b}
\end{subeqnarray}
The system ground state is still given by
 \be\label{68}
| \Phi_g\ran=\hat{a}^\dag_{g,1}\hat{a}^\dag_{g,2}\cdots \hat{a}^\dag_{g,N}|v\ran
 \ee 
 with energy $E_g=N\va_g$ in the absence of additional Coulomb processes.  The lowest excited subspace is made of states in which the ground-state electron of a $\vR_\ell$ ion is replaced by one of the three excited states of the same ion
 \be\label{69}
 | \Phi_{\mu;\vR_\ell}\ran=\hat{a}^\dag_{\mu,e,\ell}\hat{a}_{g,\ell}| \Phi_g\ran
 \ee
 This subspace now has a $3N_s$ degeneracy, its energy being  $E_e=E_g+\va_e- \va_g $ in the absence of Coulomb processes other than the ones of the electron with ``its'' ion.

 \subsubsection{Hamiltonian in terms of electron operators}
\noindent $\bullet$ The one-body part of the system Hamiltonian $H_{N_s}$ given in Eq.~(\ref{1}) has a ground-level  term still given by Eq.~(\ref{40_2}) 
% namely \be\label{70}
% \hat{H}_{0,g}=\left(\va_g+t_{gg}\right)\sum_{m=1}^N\hat{a}^\dag_{g,m}\hat{a}_{g,m}
% \ee
and an excited-level term slightly different from the one given in Eq.~(\ref{40_3}), due to the spatial degeneracy of the excited level
 \be\label{71}
 \hat{H}_{0,e}= \sum_{(\mu',\mu)=(x,y,z)} \,\, \sum_{\ell=1}^{N_s}
 \left(\va_e\delta_{\mu',\mu}+t_{\mu',\mu}\right)\hat{a}^\dag_{\mu',e,\ell}\hat{a}_{\mu,e,\ell}
 \ee
the Coulomb term being now given by
\be\label{72}
t_{\mu',\mu}= \int_{L^3}d^3r \lan \mu',e|\vr\ran\lan\vr|\mu,e\ran \sum_{\vR\not={\bf0}}\frac{-e^2}{|\vr-\vR|}
\ee 
instead of Eq.~(\ref{16}). This term comes from interactions between one excited electron and all the other ions. Through it, a change of excited-level index from $\mu$ to $\mu'\not=\mu$ can take place.

\noindent $\bullet$ The relevant parts of the Coulomb interaction $\hat{V}_{e-e}$ given in Eq.~(\ref{18}) also reduce to $\hat{V}_{gg}+\hat{V}^{(intra)}_{eg}+\hat{V}^{(inter)}_{eg}$. The $\hat{V}_{gg}$ interaction still reads as  Eq.~(\ref{40_5}), while the other two interactions now contain additional excited-level indices $\mu$. The intralevel part, given in Eq.~(\ref{40_6}), now appears as
\bea
\label{73}
\hat{V}^{(intra)}_{eg}=
\sum_{\mu',\mu}\,\,
\sum_{\ell_1,\ell_2}\mathcal{V}_{\vR_{\ell_1}-\vR_{\ell_2}}\left(\begin{smallmatrix}
\mu',e& \,\,\mu,e\\ g& \,\,g\end{smallmatrix}\right)\hat{a}^\dag_{g,\ell_1}\hat{a}^\dag_{\mu',e,\ell_2}\hat{a}_{\mu,e,\ell_2}\hat{a}_{g,\ell_1}
\eea
 and   the interlevel part, given in Eq.~(\ref{40_7}), now appears as
 \bea
 \label{74}
\hat{V}^{(inter)}_{eg}=
\sum_{\mu',\mu} \,\,   \sum_{\ell_1,\ell_2}\mathcal{V}_{\vR_{\ell_1}-\vR_{\ell_2}}\left(\begin{smallmatrix}
g&\, \mu,e\\ \mu',e&\, g\end{smallmatrix}\right)\hat{a}^\dag_{\mu',e,\ell_1}\hat{a}^\dag_{g,\ell_2}\hat{a}_{\mu,e,\ell_2}\hat{a}_{g,\ell_1}
\eea

 \subsubsection{Hamiltonian in terms of electrons and holes}
\noindent $\bullet$ Next, we turn from electron absence to hole and from excited electron to ``electron'' in the same way as in Eq.~(\ref{41},\ref{41_0}), namely
 \be\label{75}
 \hat{a}_{g,\ell}=\hat{b}^\dag_\ell
  \qquad\qquad\qquad
 \hat{a}^\dag_{\mu,e,\ell}=\hat{a}^\dag_{\mu,\ell}
 \ee
 This allows us to rewrite the lowest set of excited states, defined in Eq.~(\ref{69}), as
 \be\label{76'}
 |\Phi_{\mu;\vR_{\ell}}\ran=\hat{a}^\dag_{\mu,\ell}\hat{b}^\dag_\ell|0\ran 
 \ee

 \noindent $\bullet$ The parts of the Hamiltonian that are relevant in this subspace reduce to 
 \be\label{76}
 \hat{H}_{eh}=E'_g+\hat{H}_{e}+\hat{H}_{h}+\hat{V}^{(intra)}_{eh}+\hat{V}^{(inter)}_{eh}
 \ee
 The constant term $E'_g$ is equal to $\lan \Phi_g|\hat{H}|\Phi_g\ran$, while the $\hat{H}_{h}$ part is just the one given in Eqs.~(\ref{49_5},\ref{49_6}).
 The other terms can be readily obtained from their expression in the absence of excited-level degeneracy, by simply adding some $\mu$ indices, namely
 \be\label{77}
 \hat{H}_{e}=  \sum_{\mu',\mu}   \,\, \sum_\ell  \tilde{\va}_e(\mu',\mu)   \,\,   \hat{a}^\dag_{\mu',\ell}\hat{a}_{\mu,\ell}
 \ee
 for the one-body part, with
 \be\label{78}
\tilde{\va}_e(\mu',\mu)=\va_e\delta_{\mu',\mu}+t_{\mu',\mu}+\sum_\vR \mathcal{V}_\vR\left(\begin{smallmatrix}
\mu',e&\, \mu,e \\g& \,g \end{smallmatrix}\right)-\mathcal{V}_{\vR=\bf0}\left(\begin{smallmatrix}
g& \mu,e\\ \mu',e& g\end{smallmatrix}\right)
 \ee
 The third term in the above equation comes from the intralevel interaction between the excited-level electron and \textit{all} the ground-level electrons, while the last term comes from the interlevel interaction on a single ion site.

   \begin{figure}[t]
\centering
\includegraphics[trim=2.5cm 5cm 5.5cm 6.5cm,clip,width=3.5in]{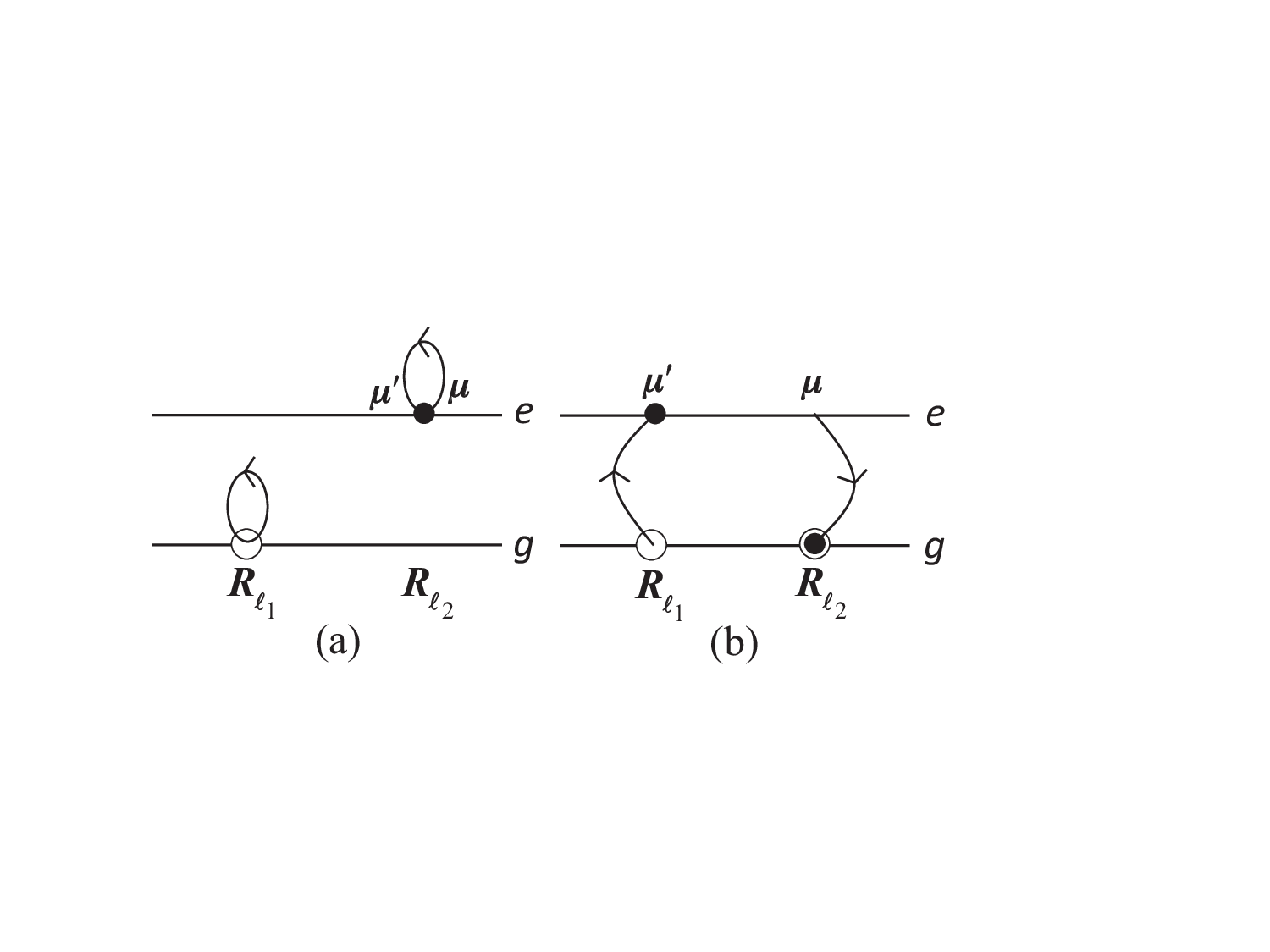}
\vspace{-0.7cm}
\caption{Same as Fig.~\ref{fig2}, when the excited level has a spatial degeneracy (see Eqs.~(\ref{79},\ref{80})).  (a) The electron can change its spatial index $\mu$. (b) The excited pair can have a spatial index $\mu'$ different from the $\mu$ pair that recombines.  }
\label{fig4}
\end{figure}
 
 Finally, the intralevel part of the electron-hole interaction, that now reads
 \bea
 \label{79}
\hat{V}^{(intra)}_{eh}=
-   \sum_{\mu',\mu} \,\,  \sum_{\ell_1,\ell_2}\mathcal{V}_{\vR_{\ell_1}-\vR_{\ell_2}}\left(\begin{smallmatrix}
 \mu',e& \mu,e \\g& g \end{smallmatrix}\right)\hat{a}^\dag_{\mu',\ell_2}\hat{b}^\dag_{\ell_1}\hat{b}_{\ell_1}\hat{a}_{\mu,\ell_2}
\eea
allows transition between different excited levels of the same lattice site (see Fig.~\ref{fig4}(a)), while the interlevel part  
 \bea
 \label{80}
\hat{V}^{(inter)}_{eh}=
\sum_{\mu',\mu}  \,\, \sum_{\ell_1,\ell_2}  \mathcal{V}_{\vR_{\ell_1}-\vR_{\ell_2}}\left(\begin{smallmatrix}
g& \mu,e\\ \mu',e& g\end{smallmatrix}\right)\hat{a}^\dag_{\mu',\ell_1}\hat{b}^\dag_{\ell_1}\hat{b}_{\ell_2}\hat{a}_{\mu,\ell_2}
\eea
allows for excitations of electron-hole pairs in a different $\mu'$ state of a different lattice site (see Fig.~\ref{fig4}(b)).

\subsubsection{Electron-hole Hamiltonian in the $| \Phi_{\mu;\vR_{\ell}} \ran$ excited subspace}

\noindent $\bullet$ We start with the $|\Phi_{\mu;\vR_{\ell}}\ran=\hat{a}^\dag_{\mu,\ell}\hat{b}^\dag_\ell|0\ran$ state of the $3N_s$-degenerate excited subspace and we calculate the relevant parts of the system Hamiltonian acting on this state. From the one-body parts
\begin{subeqnarray}\label{81}
\hat{H}_{h}|\Phi_{\mu;\vR_{\ell}}\ran&=&\tilde{\va}_h |\Phi_{\mu;\vR_{\ell}}\ran\slabel{81a}\\
\hat{H}_{e}|\Phi_{\mu;\vR_{\ell}}\ran&=&\sum_{\mu'}\tilde{\va}_e (\mu',\mu) |\Phi_{\mu';\vR_{\ell}}\ran\slabel{81b}
\end{subeqnarray} 
 we already see that the existence of a degenerate excited level keeps the $\hat{H}_{e}$ operator diagonal with respect to lattice sites, but not diagonal with respect to  excited-level indices $\mu$.

\begin{figure}[t]
\centering
\includegraphics[trim=5cm 5.5cm 4cm 7cm,clip,width=3.5in]{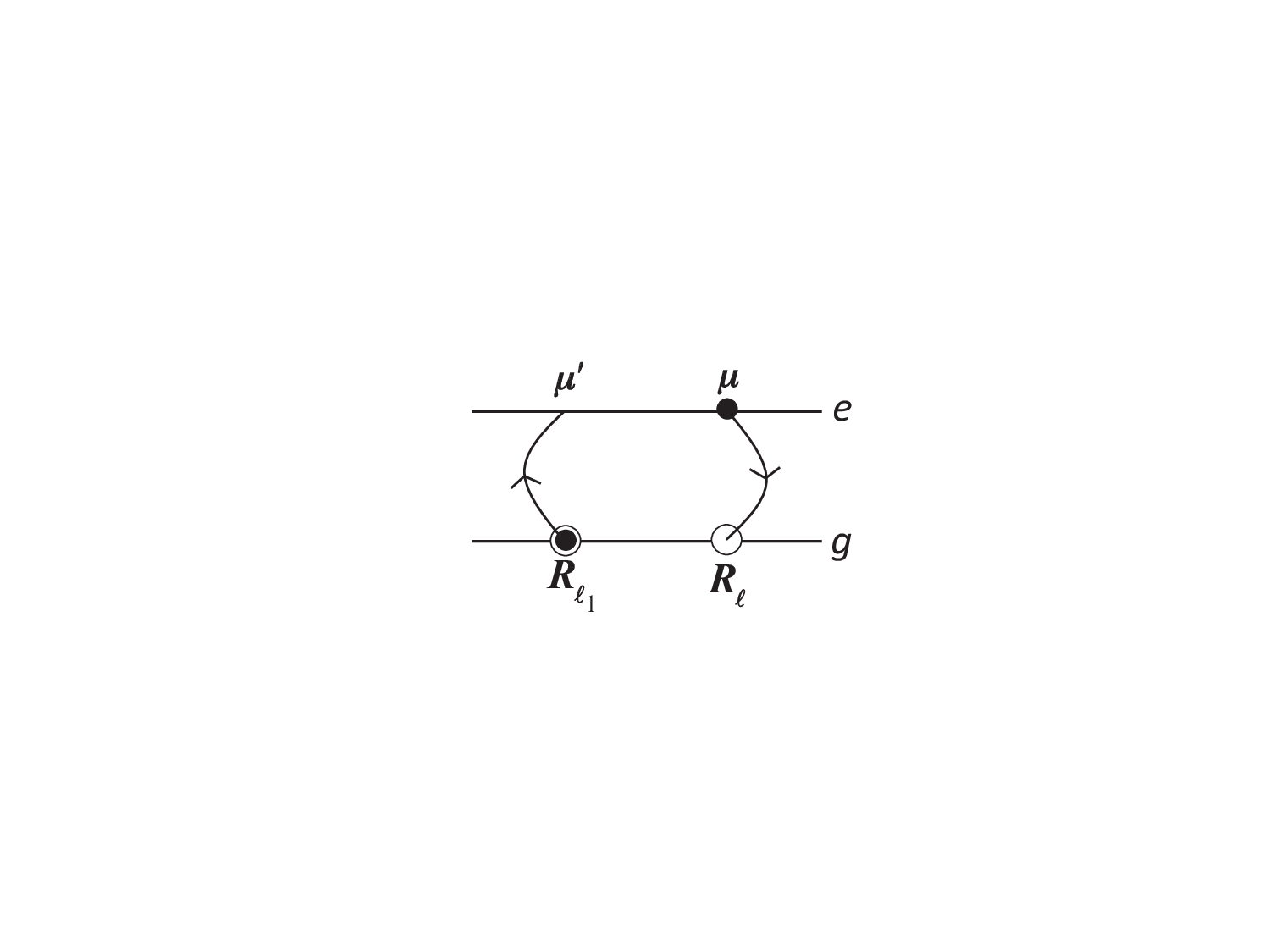}
\vspace{-0.7cm}
\caption{Effect of  $\hat{V}^{(inter)}_{eh}$ on the $N_s$-electron excited state $|\Phi_{\mu;\vR_{\ell}}\ran$, as given in Eq.~(\ref{83}). Under Coulomb interaction, the excited level can change its spatial index from $\mu$ to $\mu'$, with a possible  ion site change from $\ell$ to $\ell_1\not=\ell$.   }
\label{fig5}
\end{figure}

 In the same way, instead of Eq.~(\ref{51}), the intralevel Coulomb interaction now gives 
 \be\label{82}
 \hat{V}^{(intra)}_{eh}|\Phi_{\mu;\vR_{\ell}}\ran=-\sum_{\mu'}\mathcal{V}_{\vR=\bf0}\left(\begin{smallmatrix}
\mu',e& \mu,e\\ g& g \end{smallmatrix}\right)|\Phi_{\mu';\vR_{\ell}}\ran
 \ee  
while instead of Eq.~(\ref{52}), the interlevel Coulomb interaction now gives
(see Fig.~\ref{fig5})
 \bea\label{83}
\hat{V}^{(inter)}_{eh}|\Phi_{\mu;\vR_{\ell}}\ran=
 \sum_{\mu'}\sum_{\ell_1}\mathcal{V}_{\vR_{\ell_1}-\vR_{\ell}}\left(\begin{smallmatrix}
g& \mu,e\\ \mu',e& g\end{smallmatrix}\right)|\Phi_{\mu';\vR_{\ell_1}}\ran
 \eea

 The resulting ($3N_s\times 3N_s$) matrix $\lan \Phi_{\mu';\vR_{\ell'}}|\hat{H}_{eh}| \Phi_{\mu;\vR_{\ell}} \ran$ is made of a set of identical $3\times 3$ submatrices $h$  on the main diagonal, that reads
 \be\label{86}
 h^{(\ell'=\ell)}=\big(E'_g+\va_e-\va_g\big)\rm{I}^{(3)}+\left(\begin{matrix}
 v_{x,x}& v_{x,y}&v_{x,z}\\ v_{y,x}&   v_{y,y}& v_{y,z}\\
v_{z,x}&v_{z,y}&  v_{z,z}\end{matrix}\right)
 \ee
  where $\rm{I}^{(3)}$ is the $3\times 3$ identity matrix, while $v_{\mu',\mu}$ is given by
 \bea
 v_{\mu',\mu}&=& \iint_{L^3}d^3rd^3r' \,|\lan \vr|g\ran|^2
\Big(\lan \mu',e|\vr'\ran \lan \vr'| \mu,e\ran- \delta_{\mu',\mu} |\lan \vr'|g\ran|^2 \Big)\nn\\
&&\times
\sum_{\vR\not=\bf0}\left[\frac{e^2}{|\vR+\vr-\vr'|}-\frac{e^2}{|\vR-\vr'|}\right]\label{87}
\eea
The ($3N_s\times 3N_s$) matrix $\lan \Phi_{\mu';\vR_{\ell'}}|\hat{H}_{eh}| \Phi_{\mu;\vR_{\ell}} \ran$ also has nondiagonal $(\ell'\not=\ell)$ terms that come from interlevel excitations that read  
 \be\label{89}
 h_{\mu',\mu}^{(\ell'\not=\ell)}=\mathcal{V}_{\vR_{\ell'}-\vR_{\ell}}\left(\begin{smallmatrix}
g&\, \mu,e\\ \mu',e&\, g\end{smallmatrix}\right)
 \ee 
 and that allow coupling between lattice sites.

 \subsubsection{Diagonalization of the corresponding matrix}
 
 The diagonalization of this matrix is performed in two steps:
 
 \noindent $\bullet$ In a first step, we diagonalize the ($3N_s\times 3N_s$) matrix with respect to the lattice sites. This amounts to delocalizing the excitation on the $\vR_\ell$ site into a $\vK_n$ wave, as done through the same phase factor as the one given in Eq.~(\ref{0}). By introducing the linear combination 
 \be\label{90}
 | \Phi_{\mu;\vK_n} \ran=\frac{1}{\sqrt{N_s}}\sum_{\ell=1}^{N_s} e^{i\vK_n\cdot \vR_\ell} | \Phi_{\mu;\vR_{\ell}} \ran
 \ee
  we get
 \be\label{91}
\lan \Phi_{\mu';\vK_{n'}}  |\hat{H}_{eh}| \Phi_{\mu;\vK_n} \ran=\delta_{n',n}\Big(h_{\mu',\mu}+v_{_{\vK_n}}(\mu',\mu)\Big)
 \ee
 the $\vK_n$ dependence of this matrix element coming from the interlevel Coulomb interaction between lattice sites
 \be\label{92}
 v_{_{\vK_n}}(\mu',\mu)=\sum_{\vR\not=\bf0} e^{i\vK_n\cdot \vR}  \, \mathcal{V}_{\vR}\left(\begin{smallmatrix}
g& \mu,e\\ \mu',e& g\end{smallmatrix}\right)
 \ee
 
 This renders the ($3N_s\times 3N_s$) matrix block-diagonal with respect to the $N_s$ wave vectors $\vK_n$. The matrix elements of the remaining ($3\times 3$) submatrices, associated with a given $\vK_n$, are equal to  $h_{\mu',\mu}+v_{_{\vK_n}}(\mu',\mu)$.

 \noindent $\bullet$  In a second step, we diagonalize these ($3\times 3$) submatrices. A close look at $ v_{_{\vK_n}}(\mu',\mu)$ defined above, which also reads, with the help of Eq.~(\ref{63}), as 
\be
 v_{_{\vK_n}}(\mu',\mu)= \iint_{L^3} d^3r'd^3r \,\Big(\lan  g|\vr'\ran \lan \vr'|\mu,e\ran\Big)\Big(\lan \mu',e|\vr\ran \lan \vr|g \ran\Big)
 \sum_{\vR\not=\bf0}e^{i\vK_n \cdot \vR}\frac{e^2}{|\vR+\vr-\vr'|}\label{93}
 \ee
shows that, in the small $\vK_n$ limit, this coupling is controlled by large $\vR$'s, for the same reason as the one given in Eq.~(\ref{62}). By noting that the electronic wave functions keep $(\vr,\vr')$ small compared to $\vR\not=\bf0$,  we are led to perform a large-$\vR$ expansion of $1/|\vR+\vr-\vr'|$. Since the ground and excited levels now have a different parity, the dominant nonzero terms in the sum are just the ones in Eq.~(\ref{65}).
%linear in $\vr$ and $\vr'$. As previously shown, they read
 %\be\label{94}
 %\frac{e^2}{R^3}\left[\vr\cdot \vr'-3\left(\frac{\vR}{R}\cdot\vr\right)\left(\frac{\vR}{R}\cdot\vr'\right)+\mathcal{O}(R^{-4})\right]
 %\ee 
 When inserted into Eq.~(\ref{93}), this gives
 \be\label{95}
  v_{_{\vK_n}}(\mu',\mu)=|d_{ge}|^2 S_{\vK_n} (\mu',\mu)
 \ee
the excited-electron dipole moment
 \be\label{96}
 d_{ge}=e\int_{L^3}d^3r  \,\lan g|\vr\ran\lan \vr|\mu,e\ran  \,r_\mu
 \ee
 being $\mu$ independent due to cyclic invariance. The $S_{\vK}(\mu',\mu)$ sum, defined as
 \be\label{97}
 S_{\vK}(\mu',\mu)=\sum_{\vR\not=\bf0}\frac{e^{i\vK\cdot \vR}}{R^3}\left(\delta_{\mu',\mu}-3 \frac{R_{\mu'}R_\mu}{R^2}\right)
 \ee
 is singular in the small $\vK$ limit: as explicitly shown in  \ref{app:B}, it reads in this limit as\cite{Born,Cohen}
 \be\label{98}
 \lim_{\vK\rightarrow \bf0}S_\vK(\mu',\mu)=-\frac{4\pi}{3a_c^3}\left(\delta_{\mu',\mu}-3 \frac{K_{\mu'}K_\mu}{K^2}\right)
 \ee
 where $a_c$ is the distance between two adjacent lattice sites.
 
\begin{figure}[t]
\centering
\includegraphics[trim=8cm 5.5cm 9cm 5.6cm,clip,width=2in]{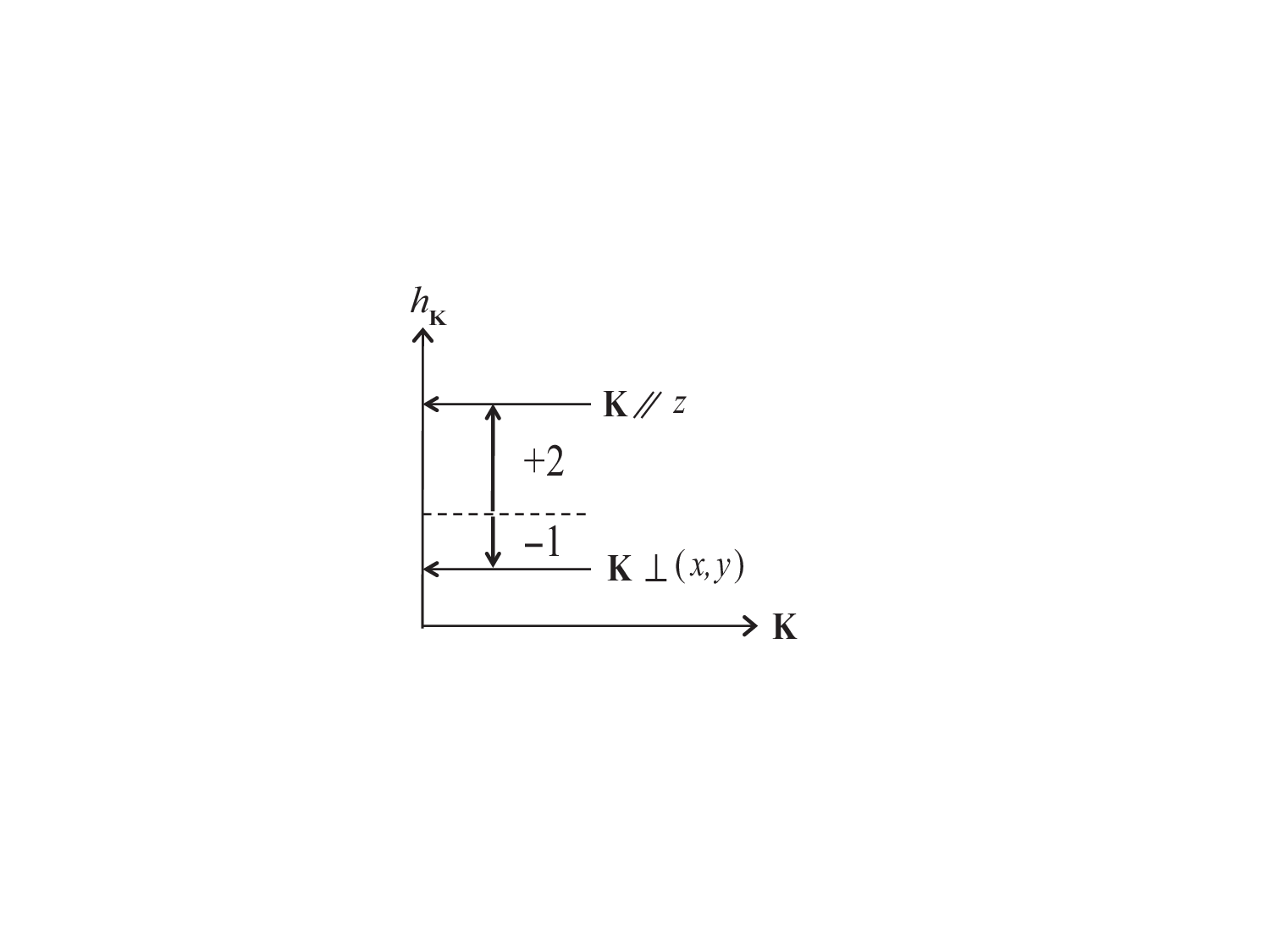}
\vspace{-0.7cm}
\caption{Frenkel exciton energy in the small $\vK$ limit, as given in Eq.~(\ref{99}). The splitting results from the singular behavior of $S_\vK(\mu',\mu)$ in the small $\vK$ limit (see Eqs.~(\ref{97},\ref{98})).}
\label{fig6}
\end{figure}

 By taking the $z$ axis of the arbitrary  $(x,y,z)$ set along $\vK$, the above equation gives  $\lim_{\vK\rightarrow \bf0}S_\vK(\mu',\mu)=0$ for $\mu'\not=\mu$, this limit being equal to $-4\pi/3a_c^3$ for $\mu'=\mu=(x,y)$, and to $8\pi/3a_c^3$ for $\mu'=\mu=z$ (see Fig.~\ref{fig6}). This induces to the Frenkel exciton energy a singular splitting in the small wave-vector limit that depends on the $\vK_n$ direction, the barycenter of this splitting reducing to zero, as usual. 
 
As a last step, we note (see \ref{app:A}) that the Coulomb terms in the $\vR$ sum of $v_{\mu',\mu}$ defined in Eq.~(\ref{87}), scales as $\mathcal{O}(R^{-4})$ in the large $\vR$ limit, while the ones in the $v_{_\vK}(\mu',\mu)$ sum scale as $\mathcal{O}(R^{-3})$. Consequently, we can neglect the $v_{\mu',\mu}$ nondiagonal part of the ($3\times 3$) submatrix $h$ in the large sample limit. As a result, by choosing $z$ along $\vK_n$, we render diagonal the ($3\times 3$) submatrix $h$ given in Eq.~(\ref{86}), in the small Frenkel exciton wave vector limit: this submatrix then appears as\cite{Heller,Fano}  
  \be\label{99}
\big(E'_g+\va_e-\va_g\big){\rm{I}}^{(3)}+\frac{4\pi}{3a_c^3}|d_{ge}|^2\left(\begin{matrix}
-1& 0&0\\ 0& -1 &0\\
0&0&2\end{matrix}\right)
 \ee

 We wish to stress that, by choosing the $z$ axis for the degenerate electronic basis along $\vK_n$, we render diagonal the $h$ submatrix for exciton with wave vector $\vK_n$. However, since each $3\times 3$ submatrix is associated with a different $\vK_n$, this does not simultaneously diagonalize all $v_{_\vK}$ submatrices. Consequently, it is not possible to diagonalize the whole ($3N_s\times 3N_s)$ matrix  $\lan \Phi_{\mu';\vR_{\ell'}}|\hat{H}| \Phi_{\mu;\vR_{\ell}} \ran$ by properly choosing the $(x,y,z)$ axes once for all.
 
 \begin{figure}[t]
\centering
\includegraphics[trim=1cm 4.5cm 0cm 8cm,clip,width=4.5in]{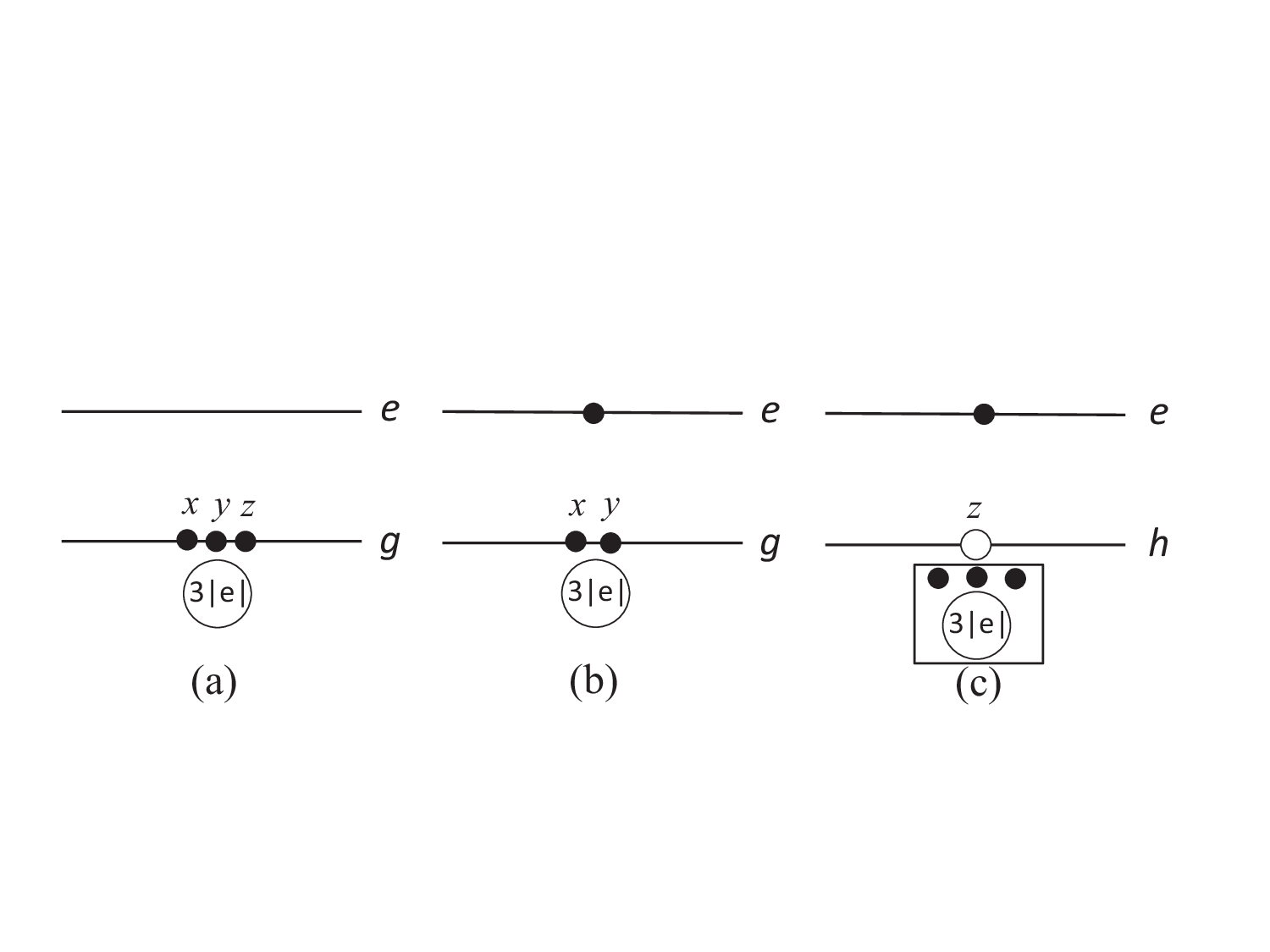}
\vspace{-0.3cm}
\caption{(a) In the presence of a threefold spatial degeneracy in the electronic ground level, the system ground state has three ground-level electrons with spatial symmetry $\mu=(x,y,z)$ and a $3|e|$ ion.  (b) The lowest set of excited states has a $3|e|$ ion and two electrons in the ground  level, let's say $(x,y)$, the third electron being in the nondegenerate excited level. (c) This fundamentally corresponds to the $3|e|$ ion with its three ground-level electrons $(x,y,z)$, plus one excited electron and one ground-level hole with spatial symmetry $\mu=z$. }
\label{fig7}
\end{figure}

 \subsection{Threefold electronic ground level\label{IVB}}
 
 The situation seems at first more complicated when the threefold degenerate level is not the excited level but the ground level: due to charge neutrality, the ion charge then has to be  $3|e|$ (see Fig.~\ref{fig7}(a)) because the ground level has its three $(x,y,z)$ states occupied by an electron. In the lowest set of excited states, one of the three ground electrons of a particular lattice site $\vR_\ell$ is replaced by an excited electron, while the other two electrons stay in the ground level (see Fig.~\ref{fig7}(b)). As a result, the excited electron does not feel the bare $3|e|$ ion, but a $3|e|$ ion surrounded by a  cloud made of the remaining two electrons; so, the charge felt by the excited electron  looks very much like a $|e|$ ion.  This difficulty is fundamentally the same as the one we faced in Sec.~\ref{sec4} for electrons with up and down spins: the ion charge was $2|e|$ instead of $|e|$, but the relevant electronic basis still is the one for one electron in the presence of a $|e|$ ion.

 To handle this problem, we first introduce the appropriate basis for one electron and one $|e|$ ion located at $\textbf{R}_\ell$; the electronic eigenstates we are going to use are threefold for the ground level and nondegenerate for the lowest excited level. In terms of the corresponding electron operators $\hat{a}^\dag_{\mu,g,\ell}$ with $\mu=(x,y,z)$ for the ground level and $\hat{a}^\dag_{e,\ell}$ for the excited level, the system ground state reads
 \begin{align}
 \label{140'}
|\Phi_g\ran=\Big (\hat{a}^\dag_{x,g,1}\hat{a}^\dag_{y,g,1}\hat{a}^\dag_{z,g,1}\Big)\Big (\hat{a}^\dag_{x,g,2}\hat{a}^\dag_{y,g,2}\hat{a}^\dag_{z,g,2}\Big)\cdots \cdots
\Big (\hat{a}^\dag_{x,g,N_s}\hat{a}^\dag_{y,g,N_s}\hat{a}^\dag_{z,g,N_s}\Big)|v\ran
\end{align}
with energy  $E_g=3N_s\va_g$, and the lowest set of excited states corresponds to  (see Fig.~\ref{fig7}(c))
\be\label{141'}
|\Phi_{\mu;\vR_\ell}\ran=\hat{a}^\dag_{e,\ell}\hat{a}_{\mu,g,\ell}|\Phi_g\ran
\ee
with energy $E_e=E_g+\va_e-\va_g$ in the absence of additional Coulomb processes.

The change from ground and excited electron operators to electron and hole operators follows from
\be\label{75a}
 \hat{a}_{\mu,g,\ell}=\hat{b}^\dag_{\mu,\ell}
  \qquad\qquad
 \hat{a}^\dag_{e,\ell}=\hat{a}^\dag_{\ell}
 \ee
 without phase factor, due to cyclic invariance. So, the lowest set of excited states reads in terms of electron and hole as
 \be\label{141b}
|\Phi_{\mu;\vR_\ell}\ran=\hat{a}^\dag_{\ell}\hat{b}^\dag_{\mu,\ell}|0\ran
\ee
 This state  has the same spatial symmetry $\mu$ as the one for a nondegenerate ground level and an electron in the  $\mu$ state of a threefold excited  level. So, we  end with the same Frenkel exciton matrix as for nondegenerate ground level and threefold excited level. 
 
A first diagonalization is performed with respect to the lattice sites through the $e^{i\vK_n\cdot\vR_\ell}$ phase factor. The diagonalization of the resulting $3\times 3$ submatrix associated with a given $\vK_n$ is then performed by choosing the $z$ axis of the electronic states along $\vK_n$; this renders the $3\times 3$ submatrix diagonal in the small $\vK_n$ limit.

\section{Frenkel excitons with spin and spatial degeneracies \label{sec6}}

\subsection{The whole problem}

\noindent $\bullet$ The results obtained in the previous sections provide the keys for analytically solving the whole Frenkel exciton problem, that is, a system made of $6N_s$ electrons with spin $1/2$ or $-1/2$, and $N_s$ ions located on lattice sites, the electronic states for a single ion being nondegenerate for the lowest excited level and threefold degenerate for the ground level with states labeled as $\mu=(x,y,z)$ along axes that can be chosen at will, since these states correspond to a single ion, the other ions of the lattice playing no role in the basis. (The simpler case, with a nondegenerate ground level and a threefold excited level, is easy to solve along the same line\cite{Knox1963}.) The lowest set of system excited states is made of states in which all the electronic ground states of the $N_s$ lattice sites are occupied, except one site for which one of its three ground-level electrons is replaced by an excited-level electron. The corresponding excited subspace then has a ($2\times 3\times N_s$)-fold degeneracy, 2 for the spin of the excited electron, 3 for its spatial degeneracy in the ground level, and $N_s$ for the lattice site on which the electron is excited. As a result, the mathematical problem we face is the diagonalization of a ($6N_s\times 6N_s)$ matrix.

To overcome the formidable task of analytically diagonalizing such a large matrix, we use the physics we have previously learned, namely

\noindent \textit{(i)} we introduce a one-electron basis made of the eigenstates of one electron in the presence of a single $|e|$ charge located at a $\vR_\ell$ lattice site, and we only consider the two lowest electronic levels on all $\vR_\ell$ sites, for which the strict tight-binding limit is valid, in order to possibly perform a second quantization formulation of the problem;

\noindent\textit{(ii)} next, we turn from ground-level and excited-level electron operators to electron and hole operators;

\noindent \textit{(iii)} we further introduce electron-hole pairs in spin-triplet and spin-singlet states. This readily splits the ($6N_s\times6N_s)$ matrix into a diagonal ($3N_s\times3N_s$) submatrix in the spin-triplet subspace and a nondiagonal ($3N_s\times3N_s)$ submatrix in the spin-singlet subspace;

\noindent \textit{(iv)} we diagonalize this nondiagonal matrix with respect to the $N_s$ lattice sites, by constructing exciton waves that are linear combination of spin-singlet pair excitations  on $\vR_\ell$ sites, with a prefactor equal to $e^{i\vK_n\cdot\vR_\ell}$. We are left with $N_s$ submatrices ($3\times3$) in the $\mu=(x,y,z)$ subspace, each submatrix being characterized by a distinct wave vector $\vK_n$;

\noindent \textit{(v)} this ($3\times3$) submatrix can be made diagonal in the small $\vK_n$ limit, by choosing the electronic-state axis $z$ along $\vK_n$; its eigenstates, quantized along $z$ and $(x,y)$, then have a positive and a negative energy shift, the splitting barycenter being equal to zero.  

This set of smart transformations relies on a deep understanding of how Frenkel exciton waves come to be formed out of on-site electronic excitations, and the fact that the interlevel excitations responsible for the excitonic waves occur for spin-singlet electron-hole pairs only. The last step, associated with the singularity of the interlevel scattering in the small center-of-mass wave vector limit,  requires a careful mathematical study of this scattering.

\noindent $\bullet$ The fact that  the system ground state has all the threefold  ground states of each lattice site occupied by an up-spin and a down-spin electron, imposes

\noindent (\textit{i}) the ion to have a $6|e|$ charge in order for the system to be neutral, and

\noindent (\textit{ii}) the electron number to be equal to $3N_s$ for each up and down spin for a system having $N_s$ ions.

 So, the system Hamiltonian considered in Eq.~(\ref{1}) now reads, for these $6N_s$ electrons, as
\bea
\label{137}
H_{6N_s}&=&\sum_{s=\pm 1/2}
\sum_{j=1}^{{3N_s}}\frac{\vp_{s,j}^2}{2m_0}+\sum_{s=\pm 1/2}\sum_{j=1}^{{3N_s}}\sum_{\ell=1}^{N_s}\frac{-6e^2}{|\vr_{s,j}-\vR_\ell|}\nn\\
&&+ \frac{1}{2}  \sum_{s=\pm 1/2}  \sum_{j=1}^{{3N_s}}\left(\sum_{j'\not=j}^{{3N_s}}\frac{e^2}{|\vr_{s,j}-\vr_{s,j'}|}+\sum_{j'=1}^{3N_s}\frac{e^2}{|\vr_{s,j}-\vr_{-s,j'}|} \right)\nn\\
&&+\frac{1}{2}\sum_{\ell=1}^{N_s}\sum_{\ell'\not=\ell}^{N_s}\frac{(6e)^2}{|\vR_\ell-\vR_{\ell'}|}
\eea

\subsection{Good one-electron basis}  

\noindent $\bullet$ As already mentioned in Secs.~\ref{sec4} and \ref{sec5} for degenerate ground level, the physically relevant electronic states to describe these $6{N_s}$ electrons are not the ones of an electron in the presence of a $6|e|$ ion, as the first two terms of Eq.~(\ref{137}) could na\"{i}vely suggest, but rather the ones of an electron in the presence of a $6|e|$ ion surrounded by a cloud made of the five negatively-charged electrons that remain in the ground level. As this ensemble is closer to a $|e|$ charge than to a $6|e|$ charge, the \textit{one-body} Hamiltonian that provides the physically relevant one-electron states for second quantization, cannot be isolated from the $H_{6{N_s}}$ Hamiltonian due to the \textit{two-body} repulsive interaction between one electron and the five-electrons cloud on the same lattice site. One of the beauties of the second quantization formalism is that the  one-body operators can \textit{a priori} be defined from any basis, not necessarily a basis constructed from a part of the system Hamiltonian. Choosing a ``good'' basis requires to first perform a physical analysis of the problem, in order for this basis to include as much physics as possible.

\noindent $\bullet$ This good basis is made of the eigenstates of one electron in the presence of a single $|e|$ ion located at $\vR_\ell$. The associated Hamiltonian reads
\be\label{138}
h_{\scriptscriptstyle{\vR_\ell}}=\frac{\vp^2}{2m_0}-\frac{e^2}{|\vr-{\vR_\ell}|}
\ee
The $h_{\scriptscriptstyle{\vR_\ell}}$ eigenstate wave function $\lan \vr|\nu,\vR_\ell\ran=\lan \vr-\vR_\ell|\nu\ran$ only depends on the distance of the electron to the $\vR_\ell$ site. 

It turns out that the electronic states physically relevant  for the Frenkel exciton problem are highly localized on lattice sites at the lattice size scale. For this reason, the good one-body basis to describe an electron located on any lattice site, can hardly be  the $|\nu,\vR_\ell\ran$ states for a specific $\vR_\ell$ and all $\nu$'s, although these states do form a complete set. The good basis should include  $|\nu,\vR_\ell\ran$ states for all $\vR_\ell$'s. This can be done by restricting $\nu$ to its two lowest states, that are the relevant states to describe $6{N_s}$ electrons  in the case of Frenkel excitons, \textit{and} by enforcing the tight-binding limit (\ref{5}) on these lowest states, namely
\be\label{5'}
\lan\nu'|\vr-\vR_{\ell'}\ran\lan\vr-\vR_{\ell}|\nu\ran= 0  \quad \textrm{for} \quad \ell'\not=\ell
\ee

\noindent $\bullet$ The $|\nu,\vR_{\ell}\ran$ states for the two lowest $\nu$'s, which are the relevant states of the problem, are such that
 \be\label{6'}
\lan \nu',\vR_{\ell'}|\nu,\vR_{\ell}\ran=\delta_{\nu',\nu}\delta_{\ell',\ell}
\ee
 They can thus be used to define fermionic creation operators for spin  $s=\pm1/2$  electron, through 
\be\label{139}
|s,\nu, \vR_\ell\ran= \hat{a}^\dag_{s,\nu,\ell}|v\ran
\ee
with $\nu=e$ for the nondegenerate excited level, and  $\nu=(\mu,g)$  for the threefold ground level, the index  $\mu$ being taken along $(x,y,z)$ axes that can be chosen at will. These axes  do not have to coincide with the $(X,Y,Z)$ cubic axes of the crystal because  $h_{\scriptscriptstyle{\vR_\ell}}$ corresponds to the Hamiltonian of  a single ion, the position of the other ions in the lattice playing no role in this basis.

     \begin{figure}[t]
\centering
\includegraphics[trim=0.5cm 4.8cm 9cm 8cm,clip,width=3.5in]{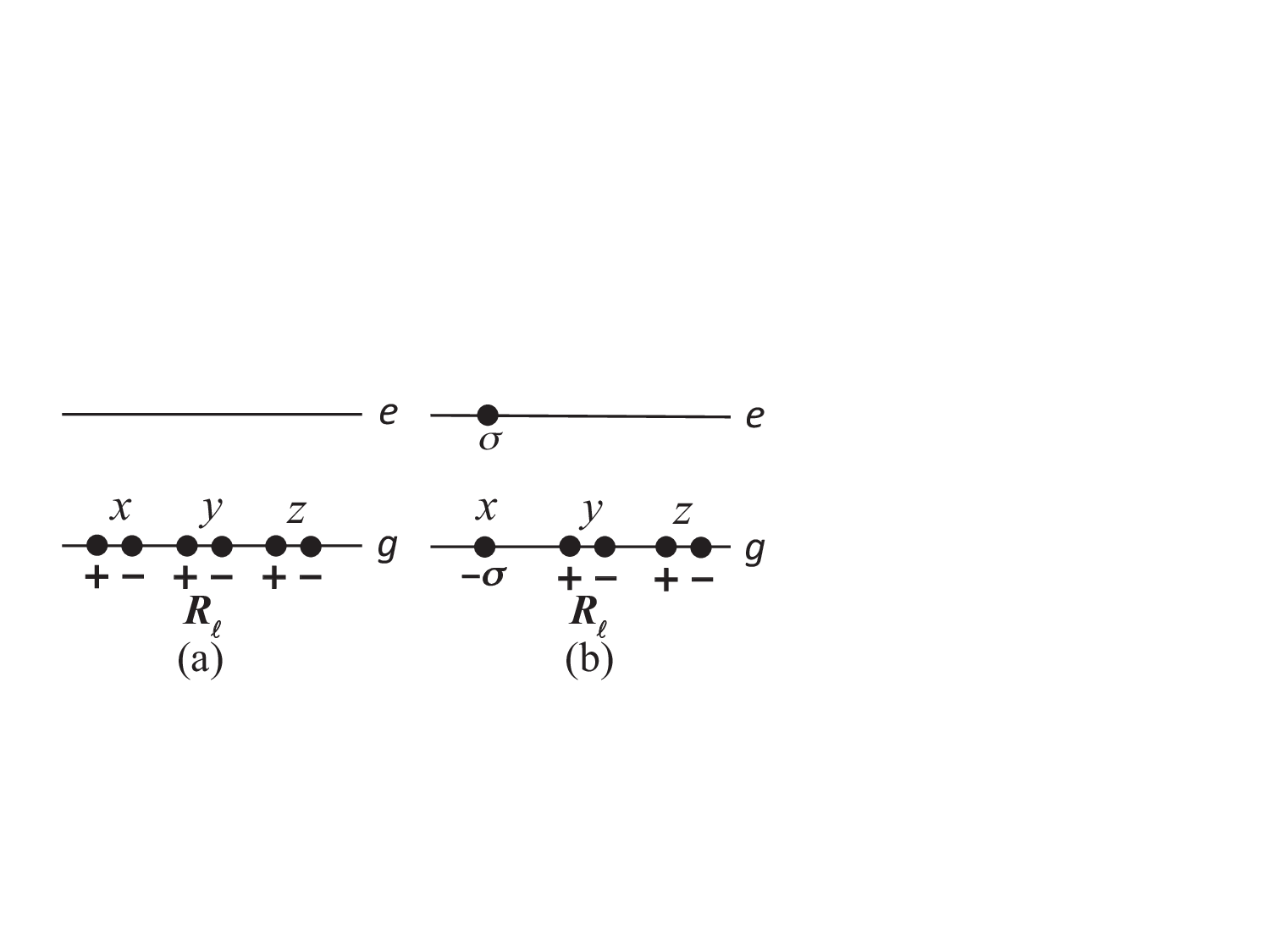}
\vspace{-0.7cm}
\caption{(a) System ground state in which all up-spin and down-spin electron states, with spatial symmetry $\mu=(x,y,z)$ of each $\vR_\ell$ site, are  occupied (see Eq.~(\ref{140})). (b) In the lowest set of excited states, one ground-level electron jumps to the excited level with its $\sigma$ spin, the opposite-spin ($-\sigma$) electron staying in the ground level (see Eq.~(\ref{141})).  }
\label{fig14}
\end{figure}

\noindent $\bullet$ The $(6N_s)$-electron ground state (see Fig.~\ref{fig14}(a)) reads within this electronic basis as   
\bea
 \label{140}
|\Phi_g\ran\!\!\!&=&\!\!\!\Big (\hat{a}^\dag_{+,x,g,1}\hat{a}^\dag_{-,x,g,1}\hat{a}^\dag_{+,y,g,1}\hat{a}^\dag_{-,y,g,1}\hat{a}^\dag_{+,z,g,1}\hat{a}^\dag_{-,z,g,1}\Big)\cdots
\\
&&\,\,\,\,\,\,\,\,\,\,\,\,\,\,\,\, \,\,\,\,\,\,\,\,       \Big(\hat{a}^\dag_{+,x,g,{N_s}}\hat{a}^\dag_{-,x,g,{N_s}}\hat{a}^\dag_{+,y,g,{N_s}}\hat{a}^\dag_{-,y,g,{N_s}}\hat{a}^\dag_{+,z,g,{N_s}}\hat{a}^\dag_{-,z,g,{N_s}}\Big )|v\ran
 \nn
\eea
The energy of the $H_{6{N_s}}$ Hamiltonian (\ref{137}) for this state follows from 
\bea
 \label{140'_6}
  E'_g=  \lan \Phi_g|    H_{6{N_s}}|\Phi_g\ran
 \eea
It is close to $E_g=6{N_s}\va_g$ if we only include Coulomb interaction between electrons and their effective $|e|$ ion. 

\noindent $\bullet$ In the lowest set of $(6N_s)$-electron excited states, one spin-$\sigma$ electron goes from the spatial ground level $(\mu, g)$ of the $\vR_\ell$ site to the excited level $e$ of the same site. The lowest set of system excited states thus reads (see Fig.~\ref{fig14}(b))
\be\label{141}
|\Phi_{\sigma,\mu;\vR_\ell}\ran=\hat{a}^\dag_{\sigma,e,\ell}\hat{a}_{\sigma,\mu,g,\ell}|\Phi_g\ran
\ee
This excited subspace, that  has an excitation energy close to $\va_e-\va_g$, is ($2\times 3\times {N_s}$)-degenerate. The various Coulomb interactions that appear in the $H_{6{N_s}}$ Hamiltonian (\ref{137}) split this degeneracy,  as obtained from the diagonalization of the ($6{N_s}\times6{N_s}$) matrix
\bea
 \label{141'_6}
   \lan \Phi_{\sigma',\mu';\vR_{\ell'}}| H_{6{N_s}}|\Phi_{\sigma,\mu;\vR_\ell}\ran
 \eea

\subsection{Hamiltonian in terms of ground and excited electron operators}

In order to calculate the above ($6{N_s}\times6{N_s}$) matrix for the $|\Phi_{\sigma,\mu;\vR_\ell}\ran$ subspace defined in Eq.~(\ref{141}), we first have to rewrite the $H_{6{N_s}}$ Hamiltonian (\ref{137}) in terms of electronic-state operators $\hat{a}^\dag_{s,\nu,\ell}$. The $H_{6{N_s}}$  relevant parts for calculating this matrix are the ones that keep the number of ground-level electrons and the number of excited-level electrons because the other states are higher in energy.

\subsubsection{One-body part}

\noindent $\bullet$ The first two terms of the $H_{6{N_s}}$ Hamiltonian (\ref{137}), which correspond to the electron kinetic energy and the Coulomb interaction of the electrons with \textit{all} $6|e|$  ions, are a sum of one-body terms. As in Eq.~(\ref{11}), we rewrite them as
\be\label{11_6}
H_{0,6{N_s}}=\sum_{s=\pm1/2}\sum_{j=1}^{3{N_s}}  \left(\frac{\vp^2_{s,j}}{2m_0}+\sum_{\ell=1}^{N_s}\frac{-6e^2}{|\vr_{s,j}-\vR_{\ell}|} \right )  
\equiv \sum_{s,j}^{N_s} h_{s,j}
\ee 
The $h_{s,j}$ Hamiltonian  differs from the $h_{\scriptscriptstyle{\vR_\ell}}$ Hamiltonian used to define the one-body basis for second quantization, because, in $h_{s,j}$, the electron interacts with all $6|e|$ ions. The second quantization procedure gives the $\hat{H}_0$ operator associated with $H_{0,6{N_s}}$ as
\be
\hat{H}_0=\sum_s \sum_{\nu', \ell'}\sum_{\nu, \ell}h_{\nu', \ell';\nu, \ell}\,\hat{a}^\dag_{s,\nu', \ell'}\hat{a}_{s,\nu, \ell}
\ee
since $h_{s,j}$ does not act on spin. 

 The prefactor, that reads
 \bea
 \label{14_6}
h_{\nu', \ell';\nu, \ell}=
\int_{L^3}d^3r\, \lan\nu',\vR_{\ell'}|\vr \ran
 \left(\frac{\hat{\vp}^2}{2m_0}+\sum_{\ell''=1}^{N_s}\frac{-6e^2}{|\vr-\vR_{\ell''}|}\right)\lan\vr|\nu,\vR_{\ell}\ran
\eea
is calculated by first noting that the wave-function product is equal to zero for $\ell'\not=\ell$, due to the  tight-binding limit (\ref{5}). 
Next, we note that $|\nu,\vR_\ell\ran$ is  $h_{\scriptscriptstyle{\vR_\ell}}$ eigenstate. This leads us to add and subtract $-e^2/|\vr-\vR_\ell|$ from the above $\ell''$ sum, in order to bring the  energy $\va_\nu$ into this prefactor.  We end with
\be\label{15_6}
h_{\nu', \ell';\nu,\ell}=\delta_{\ell',\ell}\left(\va_\nu\delta_{\nu',\nu}+t_{\nu',\nu}+u_{\nu',\nu}\right)=\delta_{\ell',\ell}\left(\va_\nu\delta_{\nu',\nu}+t'_{\nu',\nu}\right)
\ee 
The $t_{\nu',\nu}$ term, that comes from  the electron interaction with all the $6|e|$ ions located on other lattice sites, precisely reads, for $\lan \vr|\nu,\vR_\ell\ran$ given in Eq.~(\ref{3}) and $\vr-\vR_{\ell}=\vr_\ell$, as 
\bea
t_{\nu',\nu}&=& \int_{L^3}d^3r_\ell\, \lan\nu'|\vr_\ell\ran\lan\vr_\ell|\nu\ran \sum_{\ell''\not=\ell}^{N_s}\frac{-6e^2}{|\vr_\ell+\vR_\ell-\vR_{\ell''}|}\nn\\
&=&\int_{L^3}d^3r\, \lan\nu'|\vr\ran\lan\vr|\nu\ran \sum_{\vR\not=\bf0}\frac{-6e^2}{|\vr-\vR|}\label{16_6}
\eea
the  sum being $\ell$-independent due to the Born-von Karman boundary condition. The electron interaction with the $6|e|$ charge on the $\vR_\ell$ site, brings another contribution
\be
u_{\nu',\nu}=\int_{L^3}d^3r\, \lan\nu',\vR_\ell|\vr\ran \frac{-5e^2}{|\vr-\vR_\ell|}\lan\vr|\nu,\vR_\ell\ran=\int_{L^3}d^3r\, \lan\nu'|\vr\ran \frac{-5e^2}{|\vr|}\lan\vr|\nu\ran
\ee  
as the $|e|$ part of the $6|e|$ charge is already used in the $h_{\scriptscriptstyle{\vR_\ell}}$ Hamiltonian.

The above results lead us to split the one-body Hamiltonian $\hat{H}_0$ as
\be\label{17_6}
\hat{H}_0=\sum_{s,\nu,\ell} \va_\nu  \, \hat{a}^\dag_{s,\nu,\ell}\hat{a}_{s,\nu,\ell}+
\sum_{s,\nu',\nu,\ell} t'_{\nu',\nu} \, \hat{a}^\dag_{s,\nu',\ell}\hat{a}_{s,\nu,\ell}
\ee 
This Hamiltonian allows transitions between different electronic levels of the same lattice site.

\noindent $\bullet$ The parts of the $\hat{H}_0$ Hamiltonian that are relevant in the calculation of the ($6{N_s}\times 6{N_s}$) matrix defined in Eq.~(\ref{141'_6}) are the ones that keep the number of ground-level electrons and the number of excited-level electrons. They reduce to $\hat{H}_{0,g}+\hat{H}_{0,e}$. The one-body Hamiltonian for  threefold ground-level  electrons  appears as
\be\label{142}
\hat{H}_{0,g}=\sum_{s=\pm1/2}\,\sum_{(\mu',\mu)=(x,y,z)}\big(\va_g\delta_{\mu',\mu}+t'_{\mu',\mu}\big)\,\sum_{\ell=1}^{N_s}\,\hat{a}^\dag_{s,\mu',g,\ell}\hat{a}_{s,\mu,g,\ell}
\ee
This Hamiltonian allows transitions between the different states $\mu$ of the ground level. The part of $\hat{H}_0$ for nondegenerate excited electrons simply reads 
\be\label{144_6}
\hat{H}_{0,e}= (\va_e+t'_{e,e}) \sum_{\sigma=\pm1/2}\,\sum_{\ell=1}^{N_s} \hat{a}^\dag_{s,e,\ell}\hat{a}_{s,e,\ell}
\ee

\subsubsection{Two-body electron-electron interactions}

\noindent $\bullet$ The electron-electron interaction in the $H_{6{N_s}}$ Hamiltonian reads in the $|s,\nu,\vR_\ell\ran$ basis just as the one of $H_{N_s}$ in this basis, except for the spin index that is conserved in Coulomb processes. By using Eq.~(\ref{22}), we simply get
 \bea\label{22_6}
 \hat{V}_{e-e}=\frac{1}{2}\sum_{\{\nu\}}\sum_{s_1,\ell_1}\sum_{s_2,\ell_2}\mathcal{V}_{\vR_{\ell_1}-\vR_{\ell_2}}
 \left(\begin{smallmatrix}
\nu'_2&\hspace{0.1cm} \nu_2\\ \hspace{0.1cm}\nu'_1& \nu_1\end{smallmatrix}\right)
\hat{a}^\dag_{s_1,\nu'_1,\ell_1}\hat{a}^\dag_{s_2,\nu'_2,\ell_2}\hat{a}_{s_2,\nu_2,\ell_2}\hat{a}_{s_1,\nu_1,\ell_1}
 \eea 
 where $\mathcal{V}_{\vR}
 \left(\begin{smallmatrix}
\nu'_2&\hspace{0.1cm} \nu_2\\ \hspace{0.1cm}\nu'_1& \nu_1\end{smallmatrix}\right)$ is the scattering given in Eq.~(\ref{21_0}).

\noindent $\bullet$ The parts of this  electron-electron interaction that are relevant for calculating the $\lan \Phi_{\sigma',\mu';\vR_{\ell'}}| H_{6{N_s}}|\Phi_{\sigma,\mu;\vR_\ell}\ran$ matrix, are the ones that conserve the number of ground-level electrons and the number of excited-level electrons. They reduce to $\hat{V}_{gg}+\hat{V}^{(intra)}_{eg}+\hat{V}^{(inter)}_{eg}$. The interaction involving ground-level electrons only, now contains transitions inside the degenerate $\mu$ subspace. It reads
 \be\label{146}
 \hat{V}_{gg}=\frac{1}{2}\sum\mathcal{V}_{\vR_{\ell_1}-\vR_{\ell_2}}\left(\begin{smallmatrix}
 \mu_2',g& \,\mu_2,g\\ \mu_1',g&\,\mu_1, g\end{smallmatrix}\right)\,\hat{a}^\dag_{s_1,\mu'_1,g,\ell_1}\hat{a}^\dag_{s_2,\mu'_2,g,\ell_2}\hat{a}_{s_2,\mu_2,g,\ell_2}\hat{a}_{s_1,\mu_1,g,\ell_1}
 \ee
the  $(s_1,\mu_1,\ell_1)= (s_2,\mu_2,\ell_2)$ process being equal to zero  due to the Pauli exclusion principle.

\begin{figure}[t]
\centering
\includegraphics[trim=3cm 5.5cm 4.5cm 5cm,clip,width=3.5in]{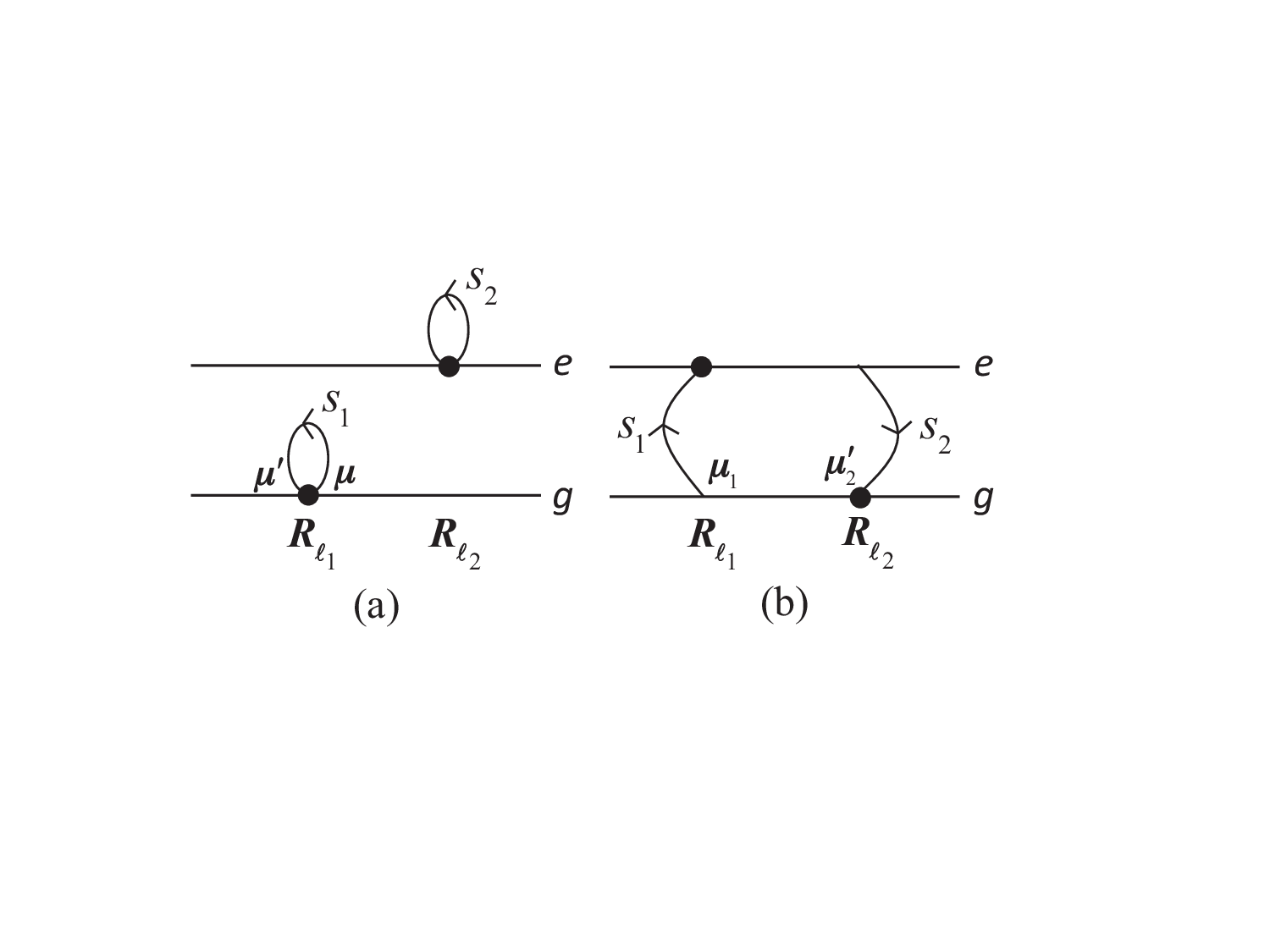}
\vspace{-0.7cm}
\caption{Same as Fig.~\ref{fig10}, the electronic ground states now having a spatial index $\mu$.  }
\label{fig15}
\end{figure}

With regard to the interaction between ground and excited levels, we again have intralevel and interlevel processes. Since the Coulomb interaction conserves the spin, the interaction in which the two scattered electrons stay in their  level, reads  (see Fig.~\ref{fig15}(a))
\bea\label{147}
\hat{V}^{(intra)}_{eg}=\sum \mathcal{V}_{\vR_{\ell_1}-\vR_{\ell_2}}\left(\begin{smallmatrix}
e & \,e\\ \mu',g&\, \mu, g\end{smallmatrix}\right)\hat{a}^\dag_{s_1,\mu',g,\ell_1} \hat{a}^\dag_{s_2,e,\ell_2}\hat{a}_{s_2,e,\ell_2}  \hat{a}_{s_1,\mu,g,\ell_1}
\eea
while the interaction in which they change  level is given by  (see Fig.~\ref{fig15}(b))
\bea\label{148}
\hat{V}^{(inter)}_{eg}=\sum \mathcal{V}_{\vR_{\ell_1}-\vR_{\ell_2}}\left(\begin{smallmatrix}
\mu'_2,g &\, e\\ e &\,\mu_1, g\end{smallmatrix}\right)
\hat{a}^\dag_{s_1,e,\ell_1}\hat{a}^\dag_{s_2,\mu'_2,g,\ell_2} \hat{a}_{s_2,e,\ell_2}  \hat{a}_{s_1,\mu_1,g,\ell_1}
\eea

\noindent $\bullet$ The ion-ion interaction, \textit{i.e.}, the last term of $H_{6{N_s}}$ in Eq.~(\ref{137}), appears as in Eq.~(\ref{32}), but for ion now having a $6|e|$ charge. It reads
\be\label{149}
V_{i-i}=\frac{{N_s}}{2}\sum_{\vR\not=\bf0}\frac{36\,e^2}{|\vR|}
\ee

 \subsection{From ground and excited electrons to electrons and holes}
 
 \noindent $\bullet$ The change from  ground-level electron absence to  hole goes along with a phase factor induced by its spin part, the spatial part bringing no phase factor when the states of the threefold level are labeled as $\mu=(x,y,z)$, due to cyclic invariance \cite{SS2021,JPCM2021}. This change reads
 \be\label{150}
 a_{s,\mu,g,\ell}=(-1)^{1/2-s}   \,    \hat{b}^\dag_{-s,\mu,\ell}
 \ee
The excited-level electron operator simply leads to electron operator as
 \be\label{151}
 a^\dag_{s,e,\ell}= a^\dag_{s,\ell}
 \ee

  \noindent $\bullet$ So, the lowest set of excited states given in Eq.~(\ref{141})  becomes in terms of electron and hole 
 \be\label{141'_7}
|\Phi_{\sigma,\mu;\vR_\ell}\ran= (-1)^{1/2-\sigma} \,   \hat{a}^\dag_{\sigma,\ell}\hat{b}^\dag_{-\sigma,\mu,\ell}|\Phi_g\ran
\ee
Note that these electron-hole pair states have a total spin $S_z$ equal to zero.

\noindent $\bullet$ When written in terms of holes, the one-body ground-level Hamiltonian (\ref{142}) becomes, since $\hat{a}^\dag_{s,\mu',g,\ell}\hat{a}_{s,\mu,g,\ell}=\delta_{\mu',\mu}-\hat{b}^\dag_{-s,\mu,\ell}\hat{b}_{-s,\mu',\ell}$,
\be\label{152}
\hat{H}_{0,g}=2{N_s}\Big(3\va_g+\sum_\mu t'_{\mu,\mu}\Big) 
-\sum_{\mu',\mu}\sum_{s,\ell}(\va_g\delta_{\mu',\mu}+ t'_{\mu',\mu})\, \hat{b}^\dag_{s,\mu,\ell}\hat{b}_{s,\mu',\ell}
\ee

\noindent $\bullet$ In the same way, the  Coulomb interaction $\hat{V}_{gg}$ given in Eq.~(\ref{146}) between ground-level electrons, reads in terms of hole operators as
\be\label{153}
\hat{V}_{gg}=
V_g-\sum \mathcal{V}_g(\mu',\mu)\,\hat{b}^\dag_{s,\mu,\ell}\hat{b}_{s,\mu',\ell}+\cdots
\ee
within a hole-hole interaction that does not act in the ground and lowest-excited subspaces. The precise values of the constant term $V_g$ and the prefactor of the above one-body operator are given in \ref{app:G}.

\noindent $\bullet$ Following the same procedure, we find that the interactions involving ground and excited  levels, given in Eqs.~(\ref{147},\ref{148}), split into a one-body part and two two-body parts
\be\label{156}
\hat{V}^{(intra)}_{eg}+\hat{V}^{(inter)}_{eg}=\mathcal{V}_e\sum_{s,\ell}\hat{a}^\dag_{s,\ell}\hat{a}_{s,\ell}+\hat{V}^{(intra)}_{eh}+\hat{V}^{(inter)}_{eh}
\ee
 with $\mathcal{V}_e$ given by
\be\label{157}
\mathcal{V}_e=2\sum_\mu \sum_\vR \mathcal{V}_{\vR}\left(\begin{smallmatrix}
e&e\\ \mu,g&\mu,g\end{smallmatrix}\right)-\sum_\mu \mathcal{V}_{\vR=\bf0}\left(\begin{smallmatrix}
\mu,g&e\\e&\mu,g \end{smallmatrix}\right)
\ee

Equation (\ref{156}) contains two two-body parts. The one that corresponds to Coulomb processes in which the electron stays  electron reads 
\bea
\hat{V}^{(intra)}_{eh}=-\sum_{s_1,s_2}\sum_{\mu'_1,\mu_1}\sum_{\ell_1,\ell_2}\mathcal{V}_{\vR_{\ell_1}-\vR_{\ell_2}}\left(\begin{smallmatrix}
e & e\\ \mu'_1,g&\mu_1, g\end{smallmatrix}\right)
\hat{a}^\dag_{s_2,\ell_2} \hat{b}^\dag_{s_1,\mu_1,\ell_1}\hat{b}_{s_1,\mu'_1,\ell_1}  \hat{a}_{s_2,\ell_2}\label{158}
\eea
The $\hat{V}^{(inter)}_{eh}$ part of Eq.~(\ref{156}) corresponds  to Coulomb processes in which the electron and the hole of a lattice site recombines, while another electron-hole pair is created on either the same site or a different site. By writing the spin part of these terms explicitly, we find that this interaction appears just as in Eq.~(\ref{118}), with additional $\mu$ indices to label the involved ground states, namely
\bea
\hat{V}^{(inter)}_{eh}&=&\sum_{\mu_1,\mu'_2}\sum_{\ell_1,\ell_2}\mathcal{V}_{\vR_{\ell_1}-\vR_{\ell_2}}\left(\begin{smallmatrix}
\mu'_2,g & e\\ e&\mu_1, g\end{smallmatrix}\right)\label{159}
\\
&& \Big(\hat{a}^\dag_{\frac{1}{2},\ell_1} \hat{b}^\dag_{-\frac{1}{2},\mu_1,\ell_1}- \hat{a}^\dag_{-\frac{1}{2},\ell_1} \hat{b}^\dag_{\frac{1}{2},\mu_1,\ell_1}\Big) 
\Big(\hat{b}_{-\frac{1}{2},\mu'_2,\ell_2}  \hat{a}_{\frac{1}{2},\ell_2} -\hat{b}_{\frac{1}{2},\mu'_2,\ell_2}  \hat{a}_{-\frac{1}{2},\ell_2}  \Big)
\nn
\eea

\noindent $\bullet$ By collecting all these terms, we end with the relevant parts of the Hamiltonian $\hat{H}_{eh}$ for the Frenkel exciton problem, as
\be
\hat{H}_{eh}=E'_g+\hat{H}_{e}+\hat{H}_{h}+\hat{V}^{(intra)}_{eh}+\hat{V}^{(inter)}_{eh}\label{160}
\ee

 The constant term  which precisely reads 
\bea\label{161}
E'_g=\lan \Phi_g|\hat{H}|\Phi_g\ran=6{N_s}\va_g +2{N_s}\sum_\mu t'_{\mu,\mu}+V_g+\frac{{N_s}}{2}\sum_{\vR\not=\bf0}\frac{36\,e^2}{|\vR|}
\eea
corresponds to the ground-state energy. It is possible to show that the overextensive terms that come from the long-range character of the Coulomb potential cancel out exactly, thanks to the ion-ion interaction that appears in the last term of the above equation.

The electron and hole kinetic parts read 
\bea
\widetilde{H}_{e}&=&\tilde{\va}_e\sum_{s,\ell} \hat{a}^\dag_{s,\ell}\hat{a}_{s,\ell}\\
\widetilde{H}_{h}&=&\sum_{s,\ell}\sum_{\mu',\mu}\tilde{\va}_h(\mu,\mu')\,\hat{b}^\dag_{s,\mu',\ell}\hat{b}_{s,\mu,\ell}
\eea
They contain contributions from the $\hat{V}_{e-e}$ Coulomb interaction, as seen from the electron and hole energies given by
\bea
\tilde{\va}_e&=&\va_e+t'_{e,e}+\mathcal{V}_e\label{162}\\
\tilde{\va}_h(\mu,\mu')&=& -\va_g \delta_{\mu,\mu'}-t'_{\mu,\mu'}-\mathcal{V}_g(\mu,\mu')\label{163}
\eea

\subsection{Spin-singlet and spin-triplet subspaces}

\noindent $\bullet$ The last step is to note that as in Sec.~\ref{sec4}, the relevant operators for the interlevel Coulomb interaction are not electron operators and hole operators, but electron-hole pair operators in spin-singlet states (see Eq.~(\ref{159})). This leads us to introduce the symmetric and antisymmetric combinations of electron-hole pair operators that correspond to spin-triplet $(S=1,S_z=0)$ and spin-singlet $(S=0,S_z=0)$ states, defined as
\bea
\label{164}
\hat{B}^\dag_{S,\mu,\ell}=\frac{\hat{a}^\dag_{\frac{1}{2},\ell}\hat{b}^\dag_{-\frac{1}{2},\mu,\ell}-(-1)^S \hat{a}^\dag_{-\frac{1}{2},\ell}\hat{b}^\dag_{\frac{1}{2},\mu,\ell}}{\sqrt{2}}
\quad \quad \textrm{for}\quad \quad
S=(1,0)
\eea  
and to replace the two excited states $|\Phi_{\sigma,\mu;\vR_\ell}\ran$, for $\sigma=\pm 1/2$, by the two states  
\be\label{166}
|\Phi_{S,\mu;\vR_\ell}\ran= \hat{B}^\dag_{S,\mu,\ell} |0\ran 
\ee
for $S=(1,0)$, where $|0\ran$ has zero hole and zero electron in the sense of Eq.~(\ref{151}); so, this state corresponds to the full valence band $|\Phi_g\ran$. We can check that the above states also form an orthogonal set
\be\label{102a}
\lan \Phi_{S',\mu';\vR_{\ell'}}|\Phi_{S,\mu;\vR_\ell}\ran = \delta_ {S',S} \,\,\delta_ {\mu',\mu} \,\,\delta_ {\ell',\ell}
\ee
that readily follows from Eq.~(\ref{164}), so that
\bea
\label{164'}
\left[\hat{B}_{S',\mu',\ell'},\hat{B}^\dag_{S,\mu,\ell} \right]_- =\delta_{S',S} \,  \delta_{\mu',\mu} \,
\delta_{\ell',\ell}
\eea

\noindent $\bullet$ The interlevel interaction given in Eq.~(\ref{159}) then takes a compact form in terms of spin-singlet operators 
\bea
\hat{V}^{(inter)}_{eh}=2\sum_{\mu_1,\mu'_2}\sum_{\ell_1,\ell_2}\mathcal{V}_{\vR_{\ell_1}-\vR_{\ell_2}}\left(\begin{smallmatrix}
\mu'_2,g & e\\ e &\mu_1, g\end{smallmatrix}\right)
\hat{B}^\dag_{0,\mu_1,\ell_1}\hat{B}_{0,\mu'_2,\ell_2}\label{167}
\eea
from which we get, with the help of Eq.~(\ref{164'}),
\bea
\hat{V}^{(inter)}_{eh}|\Phi_{S,\mu;\vR_\ell}\ran=2\delta_{S,0}\sum_{\mu_1,\ell_1}\mathcal{V}_{\vR_{\ell_1}-\vR_{\ell}}\left(\begin{smallmatrix}
\mu,g & e\\ e &\mu_1, g\end{smallmatrix}\right)|\Phi_{0,\mu_1;\vR_{\ell_1}}\ran
\eea
Through this interlevel Coulomb interaction, the excitation of a spin-singlet pair on a $\vR_\ell$ site can be transferred to a different $\vR_{\ell_1}$ site.

By contrast, we note from
\be
\hat{b}_{s_1,\mu'_1,\ell_1}\hat{a}_{s_2,\ell_2} \hat{B}^\dag_{S,\mu,\ell}|0\ran=\delta_{\ell_1,\ell}\delta_{\ell_2,\ell}\delta_{\mu'_1,\mu}\delta_{s_1,-s_2}\frac{\delta_{s_2,\frac{1}{2}}-(-1)^S\delta_{s_2,-\frac{1}{2}}}{\sqrt{2}}|0\ran
\ee
that the intralevel interaction (\ref{158}) acting on any state of the excited subspace 
\be\label{169}
\hat{V}^{(intra)}_{eh}|\Phi_{S,\mu;\vR_\ell}\ran=
-\sum_{\mu_1}\mathcal{V}_{\vR=\bf0}\left(\begin{smallmatrix}
e & e\\ \mu,g &\mu_1, g\end{smallmatrix}\right)|\Phi_{S,\mu_1;\vR_\ell}\ran
\ee
 does not produce an excitation transfer from site to site. 

\noindent $\bullet$ In the same way, the one-body parts of the $\hat{H}_{eh}$ Hamiltonian given in Eq.~(\ref{160}) acting on the excited subspace
\bea
\label{168}
\Big(\hat{H}_{e}+\hat{H}_{h}\Big) |\Phi_{S,\mu;\vR_\ell}\ran=
\sum_{\mu'}\Big(\tilde{\va}_e\delta_{\mu,\mu'}+\tilde{\va}_h(\mu,\mu')\Big)
|\Phi_{S,\mu';\vR_\ell}\ran
\eea
does not delocalize the on-site excitation.

\subsection{Electron-hole Hamiltonian in the $|\Phi_{S,\mu;\vR_\ell}\ran$ excited subspace}

\noindent $\bullet$ By collecting all the above results, we find  that the ($6N_s\times 6N_s$) matrix that represents the $\hat{H}_{eh}$ Hamiltonian in the $|\Phi_{S,\mu;\vR_\ell}\ran$ excited subspace, appears block-diagonal in the spin-triplet subspace and in the spin-singlet  subspace. The ($3N_s\times 3N_s$) submatrix in the spin-triplet subspace $(S=1,S_z=0)$ is diagonal with respect to the $N_s$ lattice sites because the coupling between lattice sites, induced by the electron-electron Coulomb interaction, differs from zero for spin-singlet pairs only. The main diagonal is made of ($3\times3$) submatrices that have $\ell$-independent matrix elements given by
\be\label{170}
E_{S=1}^{(pair)}(\mu,\mu')=\tilde{\va}_e\delta_{\mu,\mu'}+\tilde{\va}_h (\mu,\mu')-\mathcal{V}_{\vR=\bf0}\left(\begin{smallmatrix}
e & e\\ \mu,g &\mu', g\end{smallmatrix}\right)
\ee

By contrast, the $\hat{H}_{eh}$  Hamiltonian in the spin-singlet subspace $(S=0,S_z=0)$, is not diagonal with respect to the lattice sites. The off-diagonal terms, that read
\be\label{172}
\mathcal{V}^{(e,g)}_{\ell'\not=\ell} (\mu,\mu')=2\mathcal{V}_{\vR_{\ell'}-\vR_{\ell}}\left(\begin{smallmatrix}
\mu,g & e\\ e &\mu', g\end{smallmatrix}\right)
\ee
couple different lattice sites through interlevel Coulomb processes. The diagonal part of the ($3N_s\times 3N_s$) submatrix  in the spin-singlet subspace  is also made of ($3\times3$) submatrices, like for the spin-triplet subspace, with $\ell$-independent matrix elements that  read 
\be\label{171}
E_{S=0}^{(pair)}(\mu,\mu')=E_{S=1}^{(pair)}(\mu,\mu')+2\mathcal{V}_{\vR=\bf0}\left(\begin{smallmatrix}
\mu,g & e\\ e &\mu', g\end{smallmatrix}\right)
\ee
They differ from the ones in the spin-triplet subspace due to the existence of on-site interlevel Coulomb processes, that corresponds to the last term of the above equation. 

\noindent $\bullet$ So, just like when the spin enters into play (see Eq.~(\ref{135})), the spin-triplet and spin-singlet states $|\Phi_{S,\mu;\vR_\ell}\ran$ for $S=(1,0)$ taken as a basis, render block-diagonal the ($6N_s\times 6N_s$) matrix that represents the $\hat{H}_{eh}$ Hamiltonian in the lowest-excited subspace: it appears as 
\be\label{135_6}
\left(\begin{matrix}
E^{(pair)}_{1}(\mu,\mu')  &\cdots & 0 &0 & \cdots& 0  \\
%0 & E^{(pair)}_{1}(\mu,\mu') & \cdots  & 0& \cdots&  \cdots&\vdots \\
 \vdots   & \ddots & \vdots & \vdots &\ddots & \vdots  \\
 0  & \cdots& E^{(pair)}_{1} (\mu,\mu')  &0 & \cdots& 0 \\
 0  &\cdots  &0 & E^{(pair)}_{0}(\mu,\mu') & \cdots &  \mathcal{V}^{(e,g)}_{\ell',\ell}(\mu,\mu')  \\
 \vdots   &\ddots   &\vdots  & \vdots& \ddots &\vdots \\
  0  & \cdots &0  & \left(\mathcal{V}^{(e,g)}_{\ell',\ell}(\mu,\mu')\right)^\ast & \cdots &  E^{(pair)}_{0}(\mu,\mu')  \end{matrix}\right)
\ee

The ($3N_s\times 3N_s$) submatrix in the spin-triplet subspace is diagonal with respect to the lattice sites: the Coulomb interaction does not delocalize the on-site excitation because the interlevel interaction only acts on the spin-singlet subspace. We are thus left with diagonalizing the ($3N_s\times 3N_s$) submatrix in the spin-singlet subspace. The Coulomb coupling between lattice sites that exists in this subspace is going to generate the Frenkel exciton waves.

\subsection{Frenkel excitons}

\noindent $\bullet$ The Frenkel excitons, that correspond to wave excitations over the whole sample, are only formed in the spin-singlet  subspace.  To derive them, we first perform a diagonalization with respect to the lattice sites, through the same phase factor as the one given in Eq.~(\ref{0}), that is,  we introduce the following linear combinations of on-site excitations
\be\label{173}
|\Phi_{0,\mu;\vK_n}\ran=\frac{1}{\sqrt{{N_s}}} \sum_{\ell=1}^{N_s} e^{i\vK_n\cdot\vR_\ell} |\Phi_{0,\mu;\vR_\ell}\ran
\ee
A calculation very similar to the one done in the previous sections, leads to  
\be\label{174}
 \lan \Phi_{0,\mu';\vK_{n'}}|\hat{H}_{eh} |\Phi_{0,\mu;\vK_n}\ran=\delta_{n',n}\left(E_0^{(pair)}(\mu,\mu')+v_{_{\vK_n}} (\mu,\mu')\right)
\ee
 where the interlevel Coulomb interaction appears, as in Eq.~(\ref{92}), through
\be\label{175}
v_{_{\vK_n}} (\mu,\mu')=2\sum_{\vR\not=\bf0}e^{i\vK_n\cdot\vR}\mathcal{V}_{\vR}\left(\begin{smallmatrix}
\mu,g & e\\ e &\mu', g\end{smallmatrix}\right)
\ee

The last diagonalization with respect to the $\mu$ spatial degeneracy, is similar to the one we have performed in Sec.~\ref{sec5}. The interlevel Coulomb scattering is singular in the small $\vK$ limit\cite{Born,Cohen}
\be\label{177}
\lim_{\vK\rightarrow \bf0}v_{_{\vK}}(\mu,\mu')= -\frac{8\pi}{3a_c^3}|d_{ge}|^2\left(\delta_{\mu,\mu'}-3\frac{K_\mu K_{\mu'}}{K^2}\right)
\ee
where $d_{ge}$ is the dipole moment between ground and excited distributions defined as in Eq.~(\ref{96}), namely
\be
d_{ge}=e\int_{L^3}d^3r\, \lan \mu,g|\vr\ran \lan \vr|e\ran \,r_\mu
\ee
which does not depend on $\mu$ due to cyclic invariance.

To end the diagonalization with respect to the spatial degeneracy, we choose the $z$ axis of  the arbitrary set $\mu=(x,y,z)$ along $\vK_n$. Just as in  Sec.~\ref{sec5}, this leads to a ``transverse-longitudinal splitting'' of the Frenkel exciton level driven by the direction of its $\vK_n$ wave vector, with two transverse modes having a red shift and one longitudinal mode having a blue shift.

\section{The Frenkel exciton problem formulated  in first quantization \label{Sota}}

The very first book that dealt with excitons was written by Knox\cite{Knox1963} in 1963. The exciton problem is presented within the first quantization framework, common at that time. This approach has been followed by numerous authors afterward, in spite of the fact that the first quantization formalism is surely not a convenient way to approach a many-body problem. In Knox's book, that covers Frenkel and Wannier excitons, the many-electron states are written through  wave function Slater determinants. These determinants render all calculations not only cumbersome but also tricky when the signs play a key role, as for problems in which spin-singlet and spin-triplet states have different properties. Most results in Knox's book are presented without detailed derivations: this is why we find it  useful to also present pedestrian derivations of some key results on Frenkel excitons when written in terms of Slater determinants, to be compared to the more transparent operator formulation given in the present manuscript.

The very first step to approach the exciton problem, that fundamentally deals with semiconductor excitations, is to properly approximate the relevant ground and lowest-excited semiconductor states. As for all problems on solid state physics, the major difficulty comes from the two-body electron-electron Coulomb interaction which is known to generate many-body effects that in most cases, cannot be handled exactly. This forces us to start with approximate semiconductor states written in terms of the most suitable one-body states and to derive the consequences of the electron-electron Coulomb interaction on these approximate states. As the spin plays a key role in the Frenkel exciton physics through differentiating spin-singlet from spin-triplet states, its introduction is far more tricky than just adding a few 2 prefactors to some quantities, we will, here also, start by considering the exciton problem in the absence of spin.

\subsubsection{Slater determinants for semiconductor states without spin}

\noindent $\bullet$ Since semiconductors are nonmetallic crystals, the natural idea is to approximate their ground state by an antisymmetrized product of one-electron ground-state wave functions localized on \textit{all} the $N_s$ lattice sites of the crystal: indeed, in the absence of empty sites, the system cannot change under a small electric field, as required for a semiconductor. A simple idea for these one-electron wave functions is to take them as the eigenfunctions     
\be\label{ssa_1}
\left(\frac{\vp^2}{2m_0}+V_e(\vr-\vR_\ell)\right)\lan \vr|\nu,\vR_\ell\ran=\va_\nu \lan \vr|\nu,\vR_\ell\ran
\ee        
where $V_e$ stands for the Coulomb attraction to an effective charge $|e|$ made of the atomic or molecular nucleus with charge $Z|e|$ and a cloud of $(Z-1)$ electrons. In the present manuscript, this potential is taken as the one of a point charge, for simplicity. The approximate $N_s$-electron ground state then reads as the Slater determinant
  \be\label{ssa_2}
 \lan \vr_{\scriptscriptstyle{N_s}},\cdots, \vr_1|\Phi_g\ran= \frac{1}{\sqrt{{N_s}!}}\begin{vmatrix}
  \lan\vr_1|g,\vR_1\ran & \cdots & \lan\vr_1|g,\vR_\ell\ran & \cdots & \lan\vr_1|g,\vR_{N_s}\ran\\
  \vdots &\ddots &\vdots&\ddots &\vdots \\
  \lan\vr_{\scriptscriptstyle{N_s}}|g,\vR_1\ran & \cdots & \lan\vr_{\scriptscriptstyle{N_s}}|g,\vR_\ell\ran & \cdots & \lan\vr_{\scriptscriptstyle{N_s}}|g,\vR_{N_s}\ran
  \end{vmatrix}
  \ee
where $\nu=g$ stands for the ground level. 

\noindent $\bullet$ In the case of semiconductors hosting Frenkel excitons, the $\lan\vr|g,\vR_\ell\ran$ function is highly localized on the $\vR_\ell$ lattice site, at the cell size scale; so, the $|\Phi_g\ran$ state is made of states highly localized on the $\vR_\ell$ sites. This state can look at first very different from what should be taken for the ground state of semiconductors hosting Wannier excitons because their physically relevant one-electron  states,  known as Bloch states, correspond to itinerant electrons characterized by a wave vector. As a result, the natural way to write the corresponding semiconductor ground state is as Eq.~(\ref{ssa_2}), with  $\lan\vr|g,\vR_\ell\ran$ for $\ell=(1,\cdots,{N_s})$ replaced by the Bloch wave functions $\lan\vr|g,\vk_n\ran$ for $n=(1,\cdots,{N_s})$.

Wannier noted\cite{Wannier1937} that these two ground states actually are identical by introducing the linear combinations of $|g,\vk_n\ran$ states, known as Wannier states,
\be\label{ssa_3}
|g,\vR_\ell\ran=\frac{1}{\sqrt{{N_s}}}\sum_{n=1}^{N_s} e^{-i\vk_n\cdot\vR_\ell}|g,\vk_n\ran
\ee   
 When inserted into Eq.~(\ref{ssa_2}), the resulting Slater determinant can only contain columns made of different $(g,\vk_n)$'s, in order to differ from zero. So, Eq.~(\ref{ssa_2}) must also read
  \be\label{ssa_4}
 \lan \vr_{\scriptscriptstyle{N_s}},\cdots, \vr_1|\Phi_g\ran= \frac{1}{\sqrt{{N_s}!}}\begin{vmatrix}
  \lan\vr_1|g,\vk_1\ran & \cdots & \lan\vr_1|g,\vk_n\ran & \cdots & \lan\vr_1|g,\vk_{N_s}\ran\\
  \vdots &\ddots &\vdots&\ddots &\vdots \\
  \lan\vr_{\scriptscriptstyle{N_s}}|g,\vk_1\ran & \cdots & \lan\vr_{\scriptscriptstyle{N_s}}|g,\vk_n\ran & \cdots & \lan\vr_{\scriptscriptstyle{N_s}}|g,\vk_{N_s}\ran
  \end{vmatrix}
  \ee
  within a  global phase factor, for the two formulations are normalized.

  The equivalence between these two expressions holds when\textit{ all} one-electron states labeled either by $\vR_\ell$ or by $\vk_n$ are occupied.  A pedestrian way to be convinced of this equivalence is to consider ${N_s}=2$. Equation (\ref{ssa_2}) which then reads
 \be\label{ssa_5}
 \lan \vr_2,\vr_1|\Phi_g\ran=\frac{1}{\sqrt{2!}}\Big( \lan\vr_1|g,\vR_1\ran\lan\vr_2|g,\vR_2\ran-\lan\vr_1|g,\vR_2\ran\lan\vr_2|g,\vR_1\ran\Big)
 \ee 
  reduces, with the help of Eq.~(\ref{ssa_3}), to
  \bea\label{ssa_6}
 \lan \vr_2,\vr_1|\Phi_g\ran&=&\frac{1}{\sqrt{2!}}\Big( \lan\vr_1|g,\vk_1\ran\lan\vr_2|g,\vk_2\ran-\lan\vr_1|g,\vk_2\ran\lan\vr_2|g,\vk_1\ran\Big)\nn\\
 &&\,\,\, \,\,\,\,\,\,\,\,\,\,\,\,\,\,\,    \frac{e^{-i\vk_1\cdot\vR_1}e^{-i\vk_2\cdot\vR_2}-e^{-i\vk_2\cdot\vR_1}e^{-i\vk_1\cdot\vR_2}}{2}
 \eea    
The fraction is nothing but an overall phase factor which depends on the chosen origins for $\vR_\ell$ and $\vk_n$: in the case of a one-dimensional sample with length $L$, this fraction is equal to 1 for the $(\vR,\vk)$'s taken as $(\vR_1=0,\vR_2=L/2)$ and $(\vk_1=2\pi/L,\vk_2=4\pi/L)$.

A general  proof of the equivalence between Eq.~(\ref{ssa_2}) and Eq.~(\ref{ssa_4}) can be obtained by noting that the matrix
\bea
\begin{pmatrix}
 \lan\vr_1|g,\vR_1\ran & \cdots & \lan\vr_1|g,\vR_\ell\ran & \cdots & \lan\vr_1|g,\vR_{N_s}\ran\\
  \vdots &\ddots &\vdots&\ddots &\vdots \\
  \lan\vr_{\scriptscriptstyle{N_s}}|g,\vR_1\ran & \cdots & \lan\vr_{\scriptscriptstyle{N_s}}|g,\vR_\ell\ran & \cdots & \lan\vr_{\scriptscriptstyle{N_s}}|g,\vR_{N_s}\ran
  \end{pmatrix}
  \eea
   can be rewritten as the product
 \bea 
 \begin{pmatrix}
  \lan\vr_1|g,\vk_1\ran & \cdots & \lan\vr_1|g,\vk_n\ran & \cdots & \lan\vr_1|g,\vk_{N_s}\ran\\
  \vdots &\ddots &\vdots&\ddots &\vdots \\
  \lan\vr_{\scriptscriptstyle{N_s}}|g,\vk_1\ran & \cdots & \lan\vr_{\scriptscriptstyle{N_s}}|g,\vk_n\ran & \cdots & \lan\vr_{\scriptscriptstyle{N_s}}|g,\vk_{N_s}\ran
  \end{pmatrix}  \,\,\mathcal{U}
\eea
with $\mathcal{U}$ given by
\be
\mathcal{U}=\begin{pmatrix}\displaystyle
  \frac{e^{-i\vk_1\cdot\vR_1}}{\sqrt{{N_s}}} & \cdots & \displaystyle\frac{e^{-i\vk_1\cdot\vR_\ell}}{\sqrt{{N_s}}} & \cdots & \displaystyle\frac{e^{-i\vk_1\cdot\vR_{N_s}}}{\sqrt{{N_s}}}\\
  \vdots &\ddots & \vdots & \ddots &\vdots \\
 \displaystyle \frac{e^{-i\vk_{\scriptscriptstyle{N_s}}\cdot\vR_1}}{\sqrt{{N_s}}} & \cdots & \displaystyle\frac{e^{-i\vk_{\scriptscriptstyle{N_s}}\cdot\vR_\ell}}{\sqrt{{N_s}}} & \cdots &  \displaystyle\frac{e^{-i\vk_{\scriptscriptstyle{N_s}}\cdot\vR_{N_s}}}{\sqrt{{N_s}}}
  \end{pmatrix}
\ee
We then note that $\det(AB)=\det(A)\det(B)$, while ($\mathcal{U}^\dag\mathcal{U})=\textrm{I}$, so that $|\det\mathcal{U}|=1$. As a result, we readily conclude that Eq.~(\ref{ssa_2}) and Eq.~(\ref{ssa_4}) are equal, within a phase.

These Wannier states are commonly used to rewrite the ground state for materials hosting Wannier excitons, originally written in terms of delocalized Bloch functions labeled by a $\textbf{k}_n$ wave vector, in terms of functions labeled by a lattice site $\vR_\ell$. These Wannier functions are definitively more localized on lattice sites than the fully delocalized Bloch functions; however, they are far from being localized enough for the tight-binding approximation to be valid --- as evidenced from the fact that excitations on different lattice sites are accepted when forming the lowest Wannier exciton subspace (see below). Actually, besides nicely relating the two forms of the crystal ground state, these Wannier states are not really useful for semiconductors: indeed, calculations related to Wannier excitons are far easier to perform with Bloch states than with Wannier states, even for the so-called long-range and short-range parts of the ``electron-hole exchange''\cite{M_Wannier_eh}.

\noindent $\bullet$ Starting from the $|\Phi_g\ran$ ground state written as in Eq.~(\ref{ssa_2}), it appears  natural to write the lowest-excited subspace in the same way, with the $|g,\vR_\ell\ran$ column replaced by the $|e,\vR_\ell\ran$ column, where $\nu=e$ stands for the lowest-excited electronic level, namely    
 \be\label{ssa_7}
 \lan \vr_{\scriptscriptstyle{N_s}},\cdots, \vr_1|\Phi_{\vR_\ell}\ran= \frac{1}{\sqrt{{N_s}!}}\begin{vmatrix}
  \lan\vr_1|g,\vR_1\ran & \cdots & \lan\vr_1|e,\vR_\ell\ran & \cdots & \lan\vr_1|g,\vR_{N_s}\ran\\
  \vdots &\ddots &\vdots&\ddots &\vdots \\
  \lan\vr_{\scriptscriptstyle{N_s}}|g,\vR_1\ran & \cdots & \lan\vr_{\scriptscriptstyle{N_s}}|e,\vR_\ell\ran & \cdots & \lan\vr_{\scriptscriptstyle{N_s}}|g,\vR_{N_s}\ran
  \end{vmatrix}
  \ee
  In doing so, we accept that one of the ${N_s}$ electrons of the $|\Phi_g\ran$ state jumps to the lowest excited state of the \textit{same} lattice site. This restriction physically imposes the $|e,\vR_\ell\ran$ states to be highly localized on their lattice site. Indeed, a jump to $\vR_{\ell'}\not=\vR_\ell$ would cost an electrostatic energy because the $|e|$ charge that produces the $V_e(\vr-\vR_\ell)$ potential, would then be well apart from the $-|e|$ charge of the excited electron.

It is of importance to stress the major difference between semiconductors hosting Wannier excitons and semiconductors hosting Frenkel excitons, on that respect. Indeed, even if we can formally write the crystal ground state in the same way, this is not so for the lowest excited states. In the case of Wannier excitons, the $\lan\vr|g,\vR_\ell\ran$ column is replaced by a $\lan\vr|e,\vR_{\ell'}\ran$ column with $\vR_{\ell'}$ possibly different from $\vR_\ell$. So, the lowest set of excited states, then written as $|\Phi_{\vR_{\ell'},\overline{\vR}_\ell}\ran$, depends on two lattice sites, $\vR_{\ell'}$ that denotes where the electron is and $\vR_\ell$ that denotes from where it is removed. The physical reason for not enforcing $\vR_\ell=\vR_{\ell'}$ is that the Wannier states formed on Bloch states through Eq.~(\ref{ssa_3}), are not that much localized on the $\vR_\ell$ sites; so, the subspaces in which the electron jumps to the excited level of a different site, are not so far above in energy. This imposes to include all the  $|\Phi_{\vR_{\ell'},\overline{\vR}_{\ell}}\ran$ states into the excited subspace. By contrast, for semiconductors hosting Frenkel excitons, the  $|\Phi_{\vR_{\ell'}\not=\vR_\ell,\overline{\vR}_\ell}\ran$ subspaces are different enough in energy to allow handling them separately.

The strong restriction $\vR_\ell=\vR_{\ell'}$ that is accepted when separating the $|\Phi_{\vR_\ell,\overline{\vR}_\ell}\ran$ subspace from the $|\Phi_{\vR_{\ell'}\not=\vR_\ell,\overline{\vR}_\ell}\ran$ subspaces, is physically associated with the fact that the electronic wave functions on the $\vR_\ell$ and $\vR_{\ell'\not=\ell}$ sites have a very small wave function overlap. This supports using the tight-binding limit 
\be\label{ssa_8}
\lan \nu',\vR_{\ell'}|\vr\ran\lan \vr|\nu,\vR_\ell\ran= 0 \quad \quad \textrm{for}\quad\quad \ell'\not=\ell
\ee
for the lowest electronic states $\nu=(g,e)$ that are the relevant states for Frenkel excitons. Note that this condition is stronger than 
\be\label{ssa_9}
\lan \nu',\vR_{\ell'\not=\ell}|\nu,\vR_\ell\ran=0
\ee

We wish to  mention that the absence of overlap between the wave functions of different lattice sites, eliminates \textit{intralevel} Coulomb processes between different sites because these overlaps enter their associated scatterings. Through it, this absence also prohibits the formation of Frenkel exciton waves from electronic excitations in spin-triplet states. We will come back to this important point.

\subsubsection{Slater determinants for semiconductor states with spin}

%\noindent \textbf{(a) Ground state}

\noindent \textbf{(a)} By following what can be done for valence Bloch states $|v,\vk_n\ran$, that can be occupied by an up-spin or down-spin electron, we are led to take the possible one-electron ground states as $|\pm,g,\vR_\ell\ran$, where $|g,\vR_\ell\ran$ is the electronic ground state introduced in Eq.~(\ref{ssa_1}). So, the approximate ground-state wave function $\lan \vr_{\scriptscriptstyle{2N_s}},\cdots, \vr_1|\Phi_g\ran$ for a ${N_s}$-site semiconductor crystal occupied by $2{N_s}$ up-spin and down-spin electrons reads, following Eq.~(\ref{ssa_2}), as
  \be\label{ssa_10}
 \frac{1}{\sqrt{(2{N_s})!}}\begin{vmatrix}
  \lan\vr_1|+,g,\vR_1\ran & \lan\vr_1|-,g,\vR_1\ran & \cdots & \lan\vr_1|+,g,\vR_{N_s}\ran& \lan\vr_1|-,g,\vR_{N_s}\ran\\
  \vdots &\vdots &\ddots&\vdots &\vdots \\
   \lan\vr_{\scriptscriptstyle{2N_s}}|+,g,\vR_1\ran & \lan\vr_{\scriptscriptstyle{2N_s}}|-,g,\vR_1\ran & \cdots & \lan\vr_{2{N_s}}|+,g,\vR_{N_s}\ran& \lan\vr_{\scriptscriptstyle{2N_s}}|-,g,\vR_{N_s}\ran
  \end{vmatrix}
  \ee

The above ground state is made of spin-singlet pairs on each lattice site. A simple way to catch it, is to note that for ${N_s}=1$, this ground state reduces to
 \bea\label{ssa_11}
 \lan \vr_2,\vr_1|\Phi_g\ran=\frac{1}{\sqrt{2!}}\begin{vmatrix}
  \lan\vr_1|+,g,\vR_1\ran & \lan\vr_1|-,g,\vR_1\ran \\
   \lan\vr_2|+,g,\vR_1\ran & \lan\vr_2|-,g,\vR_1\ran 
  \end{vmatrix}\hspace{4cm}\\
  =\frac{1}{\sqrt{2}}\Big( \lan\vr_1|g,\vR_1\ran|1_+\ran\otimes \lan\vr_2|g,\vR_1\ran|2_-\ran -\lan\vr_1|g,\vR_1\ran|1_-\ran\otimes  \lan\vr_2|g,\vR_1\ran|2_+\ran \Big)\hspace{-0.5cm}\nn
 \eea 
which corresponds to each electron in the ground level of the $\vR_1$ site, with the electron pair in the spin-singlet state, namely
 \bea\label{ssa_12}
 \lan \vr_2,\vr_1|\Phi_g\ran&=&\lan\vr_1|g,\vR_1\ran \lan\vr_2|g,\vR_1\ran\frac{|1_+,2_-\ran- |1_-,2_+\ran}{\sqrt{2}}\nn\\
 &\equiv&S_{\vR_1}(\vr_1,\vr_2)=-S_{\vR_1}(\vr_2,\vr_1)
 \eea
The above result is physically obvious because, as the $|\Phi_g\ran$ wave function must be antisymmetric with respect to the $(\vr_1,\vr_2)$ electron exchange, $\lan \vr_2,\vr_1|\Phi_g\ran=-\lan \vr_1,\vr_2|\Phi_g\ran$, two electrons with identical spatial wave functions must be in the antisymmetric spin-singlet state.

A pedestrian way to catch that this remains true for ${N_s}>1$, is to consider ${N_s}=2$. We then have 
   \be\label{ssa_13}
\lan \vr_4,\cdots,\vr_1|\Phi_g\ran= \frac{1}{\sqrt{4!}}\begin{vmatrix}
  \lan\vr_1|+,g,\vR_1\ran & \lan\vr_1|-,g,\vR_1\ran & \cdot & \cdot\\
  \lan\vr_2|+,g,\vR_1\ran & \lan\vr_2|-,g,\vR_1\ran & \cdot & \cdot\\
 \cdot & \cdot & \lan\vr_3|+,g,\vR_2\ran& \lan\vr_3|-,g,\vR_2\ran\\
 \cdot & \cdot& \lan\vr_4|+,g,\vR_2\ran& \lan\vr_4|-,g,\vR_2\ran
  \end{vmatrix}
  \ee
that we calculate by first isolating the two diagonal submatrices, namely
\be\label{ssa_14}
\lan \vr_4,\cdots,\vr_1|\Phi_g\ran=\frac{(\sqrt{2})^2}{\sqrt{4!}}\Big(S_{\vR_1}(\vr_1,\vr_2)S_{\vR_2}(\vr_3,\vr_4)+\textrm{other terms}\Big)
\ee  
 These other terms are easy to obtain by noting that the Slater determinant (\ref{ssa_13}) is antisymmetric under $(\vr_i,\vr_j)$ exchanges. So, the six terms that compose $\lan \vr_4,\cdots,\vr_1|\Phi_g\ran$ must read
\bea
\lan \vr_4,\cdots,\vr_1|\Phi_g\ran\!\!\!&=&\!\!\!\frac{1}{\sqrt{3!}}\Big( S_{\vR_1}(\vr_1,\vr_2)S_{\vR_2}(\vr_3,\vr_4)-S_{\vR_1}(\vr_1,\vr_3)S_{\vR_2}(\vr_2,\vr_4)\nn\\
\!\!\!&&\!\!\!-S_{\vR_1}(\vr_1,\vr_4)S_{\vR_2}(\vr_3,\vr_2)-S_{\vR_1}(\vr_3,\vr_2)S_{\vR_2}(\vr_1,\vr_4)\nn\\
\!\!\!&&\!\!\!-S_{\vR_1}(\vr_4,\vr_2)S_{\vR_2}(\vr_3,\vr_1)+S_{\vR_1}(\vr_3,\vr_4)S_{\vR_2}(\vr_1,\vr_2) \Big)\label{ssa_15}
\eea  
which proves that the $|\Phi_g\ran$ ground state is indeed made of spin-singlet pairs on each lattice site.

\noindent \textbf{(b)} In the same  way that a valence Bloch state is replaced by a conduction Bloch state with same spin to get the excited states leading to Wannier excitons, the lowest set of excited states for semiconductors hosting Frenkel excitons in the presence of spin, follows by replacing a $|\sigma,g,\vR_\ell\ran$ column with $\sigma=\pm$, by a $|\sigma,e,\vR_\ell\ran$ column, the other columns staying unchanged. This leads us to write the wave functions for the lowest set of excited states as
  \bea\label{ssa_16}
&&\lan \vr_{\scriptscriptstyle{2N_s}},\cdots,\vr_1|\Phi_{\sigma;\vR_\ell}\ran=
  \\
 && \,\,\,\,\,\,\,\,\,\,\,\,\,\,\,\,\,\,
 \frac{1}{\sqrt{(2N_s)!}}\begin{vmatrix}
  \lan\vr_1|+,g,\vR_1\ran &  \cdots &\lan\vr_1|\sigma,e,\vR_\ell\ran & \cdots & \lan\vr_1|-,g,\vR_{N_s}\ran\\
  \vdots &\ddots &\vdots&\ddots &\vdots \\
   \lan\vr_{\scriptscriptstyle{2N_s}}|+,g,\vR_1\ran &  \cdots &\lan\vr_{\scriptscriptstyle{2N_s}}|\sigma,e,\vR_\ell\ran & \cdots& \lan\vr_{\scriptscriptstyle{2N_s}}|-,g,\vR_{N_s}\ran
  \end{vmatrix}
  \nn
  \eea

It can be of interest to note that the $\vR_\ell$ site of the $|\Phi_{\sigma;\vR_\ell}\ran$ state now contains a linear combination of  spin-singlet and  spin-triplet pairs. Indeed, $|\Phi_{+;\vR_\ell}\ran$ for ${N_s}=1$ reads
\bea\label{ssa_17}
 \lan \vr_2,\vr_1|\Phi_{+;\vR_\ell}\ran=\frac{1}{\sqrt{2!}}\begin{vmatrix}
  \lan\vr_1|+,e,\vR_1\ran & \lan\vr_1|-,g,\vR_1\ran \\
   \lan\vr_2|+,e,\vR_1\ran & \lan\vr_2|-,g,\vR_1\ran 
  \end{vmatrix}\hspace{4cm}\\
  =\frac{1}{\sqrt{2}}\Big( \lan\vr_1|e,\vR_1\ran \lan\vr_2|g,\vR_1\ran\otimes |1_+\ran|2_-\ran -\lan\vr_1|g,\vR_1\ran  \lan\vr_2|e,\vR_1\ran \otimes|1_-\ran|2_+\ran \Big)\hspace{-0.5cm}\nn
 \eea 
 that we can rewrite in terms of spin-singlet and spin-triplet pairs as
 \bea\label{ssa_18}
 \lan \vr_2,\vr_1|\Phi_{+;\vR_\ell}\ran=\hspace{9cm}\\
 \frac{1}{\sqrt{2}}\bigg( \frac{\lan\vr_1|e,\vR_1\ran \lan\vr_2|g,\vR_1\ran +\lan\vr_1|g,\vR_1\ran \lan\vr_2|e,\vR_1\ran}{\sqrt{2}}\otimes \frac{|1_+\ran|2_-\ran-|1_-\ran|2_+\ran}{\sqrt{2}}\nn\\
 +\frac{\lan\vr_1|e,\vR_1\ran \lan\vr_2|g,\vR_1\ran -\lan\vr_1|g,\vR_1\ran \lan\vr_2|e,\vR_1\ran}{\sqrt{2}}\otimes \frac{|1_+\ran|2_-\ran+|1_-\ran|2_+\ran}{\sqrt{2}}\bigg)\nn
 \eea 
each part being antisymmetric with respect to electron exchange. This triplet part can \textit{a priori} exist because the two electronic wave functions on the $\vR_1$ site now are different.

\noindent \textbf{(c)} We wish to mention that the above $|\Phi_g\ran$  and $|\Phi_{\sigma;\vR_\ell}\ran$  states do not exactly correspond to  Knox's formulation in two ways.

\noindent (\textit{i}) One (minor) difference is that we only consider semiconductor excited states with a total spin $S_z$ equal to zero, while Knox also considers the possibility for the excited electron  to have a spin $\sigma'$ different from the spin $\sigma$ of the removed ground-state electron. The resulting semiconductor excited states can then have a total spin $S_z$ different from zero. We do not see the interest of considering such states because the two physically relevant interactions that can lead to semiconductor excited states, namely the electron-photon interaction and the Coulomb interaction, conserve the spin; so, $|\Phi_g\ran$ can hardly be coupled to $\sigma'\neq\sigma$ excited states. Allowing them brings pairs in a spin-triplet state $(S=1,S_z=\pm1)$, while by only considering $\sigma'=\sigma$, the pairs have a total spin $S_z=0$. Yet, from them, we can still have spin-singlet and spin-triplet states $(S=(0,1),S_z=0)$ and disentangle their difference in order to properly answer the important question which is why interlevel Coulomb processes only exist between spin-singlet states.

\noindent (\textit{ii}) The major difference with Knox's book lies in the one-electron states that are used to describe the system. Knox considers that in the presence of spin, each atom or molecule ``contributes to a core and \textit{two} valence electrons''. The core includes the nucleus and a charge cloud representing the remaining electrons (\cite{Knox1963}, p.9). As a result, the core of a neutral atom or molecule has a charge $2|e|$. Using such a $2|e|$ core to define one-electron states does not correspond to the physical situation, as explained in the main part of the present manuscript, because the excited electron also feels the repulsion of the opposite-spin electron that remains in the ground level. So, the effective opposite charge felt by the excited electron to stay on the $\vR_\ell$ site, is more $|e|$ than $2|e|$. For this reason, we find it  reasonable to keep the same one-electron states to describe the system with or without spin, the excited-state energy having the same $\va_e$ value in both cases. As a strong support of this idea, we may remember that this exactly is what it is done in the case of Bloch states when the spin is introduced.

Actually, replacing the $2|e|$ core by an $|e|$ effective charge in the definition of the one-electron basis, does not change the structure of the calculations, except for the term associated with electron repulsion on the same lattice site. The existence of this on-site repulsion, which physically comes from the two-body electron-electron Coulomb interaction, was briefly mentioned by Knox: he then acknowledged the intrinsic problem associated with writing a many-body state as a product of one-body states. Such a problem does not exist when using the second quantization formalism because this formulation is valid whatever the chosen basis to define the electron creation operators: a physically bad basis would eventually lead to correct results, provided that we do not neglect terms improperly.

\subsubsection{Coulomb scatterings within the excited subspace}

The formation of Frenkel exciton waves relies on the capability of  the electronic excitation to transfer to a different lattice site. This cannot be done by the $H_0$ one-body part of the Hamiltonian, due to the tight-binding limit (\ref{ssa_8}) that is required to have the lowest excited subspace with excitations on the same lattice site only, as previously discussed. Indeed, $\lan\Phi_{ \sigma',\vR_{\ell'}}|H_0|\Phi_{\sigma;\vR_\ell}\ran$ contains integrals like 
\be\label{ssa_19}
\int d^3 r\lan e,\vR_{\ell'}|\vr\ran H_0 \lan \vr|e,\vR_\ell\ran
\ee
which reduce to zero in the absence of wave function overlap when $\ell'\not=\ell$. 

The interesting part of the Hamiltonian is concentrated into the two-body electron-electron interaction which reads in first quantization as
\be\label{ssa_20}
V_{e-e}=\frac{1}{2}\sum_i \sum_{j\not=i}\frac{e^2}{|\vr_i-\vr_j|}
\ee

 \begin{figure}[t]
\centering
\includegraphics[trim=0cm 12cm 6.5cm 0.5cm,clip,width=4in]{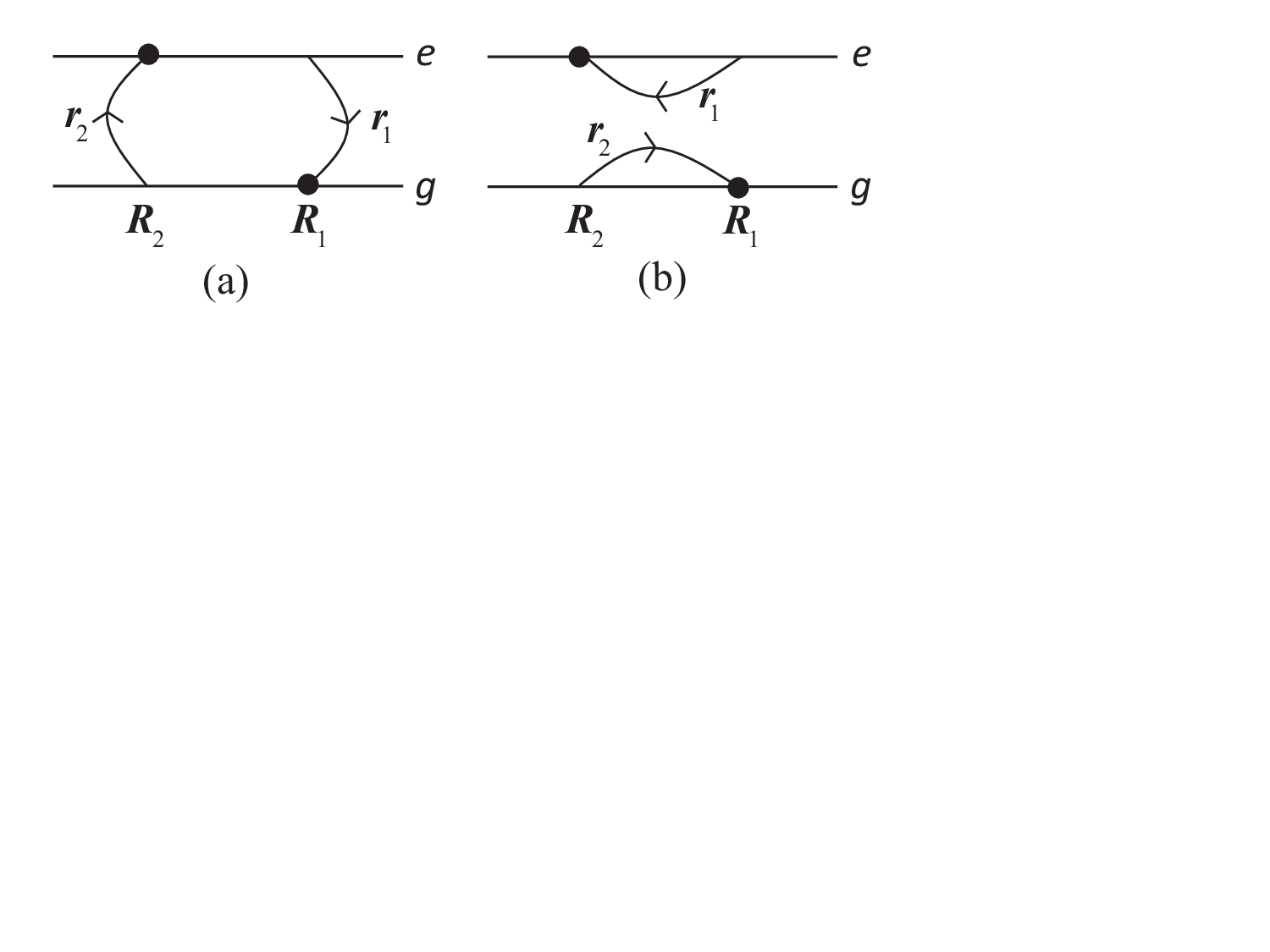}
\vspace{-0.7cm}
\caption{Coulomb processes between $\lan\Phi_{\vR_{2}}|$ and $|\Phi_{\vR_1}\ran$ in the case of just two lattice sites, $\vR_1$ and $\vR_2$. (a) \textit{Interlevel} processes on the same lattice site. (b) \textit{Intralevel} processes between the two lattice sites. These intralevel processes must be neglected in the tight-binding limit, in consistency with only considering excited states on the same lattice site.}
\label{fig16}
\end{figure}

\noindent  $\bullet$ To better catch that the tight-binding limit reduces the Coulomb interaction to \textit{interlevel} processes, let us first forget the electron spin. The possible coupling between excitations on different lattice sites, relies on the nonzero value of $\lan\Phi_{\vR_{\ell'}}|V_{e-e}|\Phi_{\vR_\ell}\ran$ for $\ell'\not=\ell$. The simplest way to catch the form of this scattering again  is to consider $N_s=2$ that is, $\vR_\ell=\vR_1$ and  $\vR_{\ell'}=\vR_2$. According to Eq.~(\ref{ssa_7}), we  find
\bea
\lan\Phi_{\vR_{2}}|V_{e-e}|\Phi_{\vR_1}\ran=\hspace{8.5cm}\label{ssa_21}\\
\frac{1}{2!}\iint d^3r_1d^3r_2 \begin{vmatrix}
  \lan\vr_1|g,\vR_1\ran & \lan\vr_1|e,\vR_2\ran \\
   \lan\vr_2|g,\vR_1\ran & \lan\vr_2|e,\vR_2\ran 
  \end{vmatrix}^\ast \frac{e^2}{|\vr_1-\vr_2|}\begin{vmatrix}
  \lan\vr_1|e,\vR_1\ran & \lan\vr_1|g,\vR_2\ran \\
   \lan\vr_2|e,\vR_1\ran & \lan\vr_2|g,\vR_2\ran 
  \end{vmatrix}\nn
\eea
which is equal to 
\bea
\frac{1}{2!}\iint d^3r_1d^3r_2 \frac{e^2}{|\vr_1-\vr_2|} \Big(\lan g,\vR_1|\vr_1\ran \lan e,\vR_2|\vr_2\ran  -\lan e,\vR_2|\vr_1\ran\lan g,\vR_1|\vr_2\ran \Big)\nn\\
\Big( \lan\vr_1|e,\vR_1\ran\lan\vr_2|g,\vR_2\ran - \lan\vr_1|g,\vR_2\ran \lan\vr_2|e,\vR_1\ran \Big)\label{ssa_22}
\eea
The above quantity contains two conceptually different terms. The term 
\be\label{ssa_23}
\iint d^3r_1d^3r_2 \frac{e^2}{|\vr_1-\vr_2|} \Big(\lan g,\vR_1|\vr_1\ran  \lan\vr_1|e,\vR_1\ran\Big) \Big( \lan e,\vR_2|\vr_2\ran \lan\vr_2|g,\vR_2\ran\Big)
\ee
corresponds to \textit{interlevel} Coulomb processes on the same lattice site\cite{Forster,Jones}: the $\vr_1$ electron goes from the $e$ level to the $g$ level on the same $\vR_1$ site (see Fig.~\ref{fig16}(a)). The other term
\be\label{ssa_24}
-\iint d^3r_1d^3r_2 \frac{e^2}{|\vr_1-\vr_2|} \Big(\lan e,\vR_2|\vr_1\ran  \lan\vr_1|e,\vR_1\ran\Big) \Big( \lan g,\vR_1|\vr_2\ran \lan\vr_2|g,\vR_2\ran\Big)
\ee
corresponds to \textit{intralevel} Coulomb processes between two lattice sites\cite{Dexter1953}: the $\vr_1$ electron stays in the same $e$ level, but moves from the $\vR_1$ site to the $\vR_2$ site (see Fig.~\ref{fig16}(b)). This second term has to be neglected in the tight-binding limit because it contains wave function overlaps between different sites. Knox mentions that this intralevel term, which appears in his Eq.~(3.7), decreases exponentially with the lattice site distance (see comment below his Eq.~(3.9)), due to the poor wave function overlap. We think that this term should be simply dropped from the $V_{e-e}$ matrix elements, in order to be consistent with the very first line on Frenkel excitons, \textit{i.e.}, when it is accepted that the excited subspace only contains excitations on the same lattice site.

By noting  that there are two terms similar to Eq.~(\ref{ssa_23}) that come from exchanging $(\vr_1,\vr_2)$ in Eq.~(\ref{ssa_22}), we ultimately find for $N_s=2$,
\bea
\lan\Phi_{\vR_{2}}|V_{e-e}|\Phi_{\vR_1}\ran=\nn\hspace{8cm}\\
\iint d^3r_1d^3r_2 \frac{e^2}{|\vr_1-\vr_2|} \Big(\lan g,\vR_1|\vr_1\ran  \lan\vr_1|e,\vR_1\ran\Big) \Big( \lan e,\vR_2|\vr_2\ran \lan\vr_2|g,\vR_2\ran\Big)\label{ssa_25}
\eea
This evidences that the Coulomb scatterings do reduce to interlevel processes on the same lattice site.

The above calculation performed for $N_s=2$ can be extended to larger $N_s$, by noting that the $\lan\vr_1|e,\vR_1\ran$ wave function will appear not only  with $\lan g,\vR_1|\vr_1\ran $ or $\lan g,\vR_2|\vr_1\ran$ as for $N_s=2$, but also  with $\lan g,\vR_3|\vr_1\ran$ and so on... Due to the absence of wave function overlap between different lattice sites, the unique term that contributes to the $V_{e-e}$ matrix element, still is $\lan g,\vR_1|\vr_1\ran  \lan\vr_1|e,\vR_1\ran$ which corresponds to the $\vr_1$ electron going from the $g$ to the $e$ level of the $\vR_1$ site.

 \begin{figure}[t]
\centering
\includegraphics[trim=0cm 12cm 7cm 1cm,clip,width=4in]{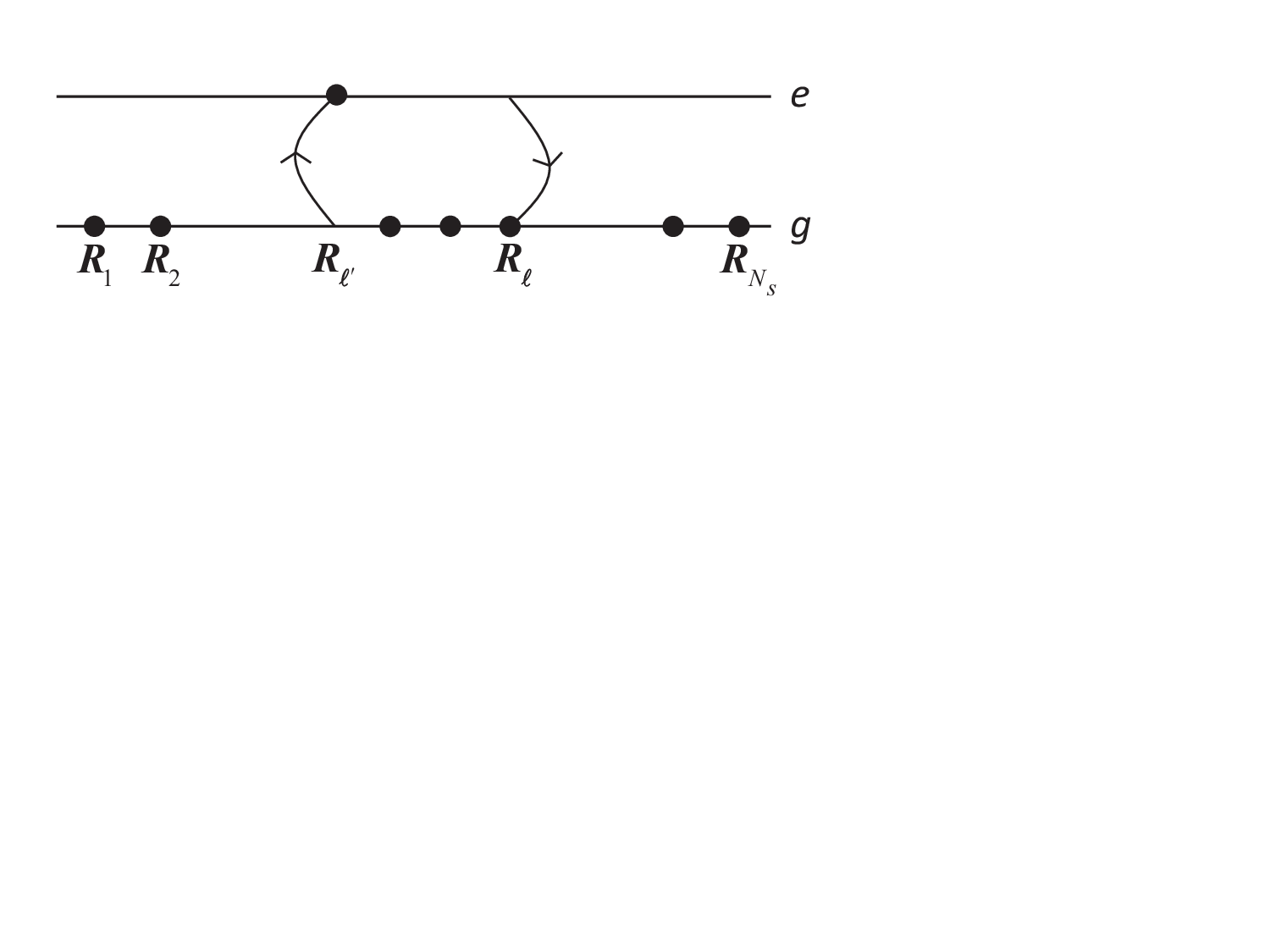}
\vspace{-0.7cm}
\caption{Interlevel Coulomb processes between $\lan\Phi_{\vR_{\ell'}}|$ and $|\Phi_{\vR_\ell}\ran$.}
\label{fig17}
\end{figure}

A careful  counting of the possible $\vr_i$ permutations ultimately leads, in the case of $N_s$ lattice sites, to 
\bea
\lan\Phi_{\vR_{\ell'}}|V_{e-e}|\Phi_{\vR_\ell}\ran=\nn\hspace{8cm}\\
\iint d^3r_1d^3r_2 \frac{e^2}{|\vr_1-\vr_2|} \Big(\lan g,\vR_\ell|\vr_1\ran  \lan\vr_1|e,\vR_\ell\ran\Big) \Big( \lan e,\vR_{\ell'}|\vr_2\ran \lan\vr_2|g,\vR_{\ell'}\ran\Big)\label{ssa_26}
\eea
which corresponds to the interlevel processes shown in Fig.~\ref{fig17}.

It should be noted that the above matrix element only depends on $\vR_{\ell'}-\vR_{\ell}$. This follows from the fact that the electronic wave functions are such that $\lan \vr_1 |\nu,\vR_\ell\ran = \lan \vr_1-\vR_\ell|\nu\ran$. So, for $\vr'_1=\vr_1-\vR_\ell$ and $\vr'_2=\vr_2-\vR_{\ell'}$, the above equation also reads as
\be
\iint d^3r'_1d^3r'_2 \frac{e^2}{|\vr'_1-\vr'_2+\vR_\ell -\vR_{\ell'} |} \Big(\lan g|\vr'_1\ran  \lan\vr'_1|e\ran\Big) \Big( \lan e|\vr'_2\ran \lan\vr'_2|g\ran\Big)\label{ssa_27}
\ee

\noindent $\bullet$ Although calculations including the spin degree of freedom are fundamentally performed in the same way as in the absence of spin, they definitely are more tricky because in the very end, we have to show that the interlevel Coulomb processes that control the $V_{e-e}$ matrix element, only exist for spin-singlet pairs. Since the difference between spin-singlet and spin-triplet is just a sign, we  have to be extremely careful when calculating quantities involving Slater determinants. Although the two electrons on the excited site can be written through their spin-singlet and spin-triplet combinations, as previously shown, we find it more transparent to handle the two electrons independently, instead of as spin-singlet and spin-triplet electron pairs, as done in Knox's book. 

Here again, let us consider $N_s=2$. The excited subspace is then made of  four  excited states,  $|\Phi_{+;\vR_1}\ran$,  $|\Phi_{-;\vR_1}\ran$,  $|\Phi_{+;\vR_2}\ran$, and  $|\Phi_{-;\vR_2}\ran$. According to Eq.~(\ref{ssa_16}), a possible Coulomb scattering between different lattice sites, reads
\bea
 \lan \Phi_{+;\vR_2}|V_{e-e}|\Phi_{+;\vR_1}\ran =\frac{1}{4!}\iiiint d^3r_1 d^3r_2d^3r_3 d^3r_4\hspace{3cm}\nn\\
\begin{vmatrix}
  \lan\vr_1|+,g,\vR_1\ran & \lan\vr_1|-,g,\vR_1\ran & \lan\vr_1|+,e,\vR_2\ran & \lan\vr_1|-,g,\vR_2\ran \\
  \vdots &\vdots &\vdots &\vdots \end{vmatrix}^\ast \nn\\
 V_{e-e}\begin{vmatrix}\lan\vr_1|+,e,\vR_1\ran & \lan\vr_1|-,g,\vR_1\ran & \lan\vr_1|+,g,\vR_2\ran & \lan\vr_1|-,g,\vR_2\ran \\
  \vdots &\vdots &\vdots &\vdots \end{vmatrix}\label{ssa_28}
\eea
Since the $|-,g,\vR_1\ran$ columns and the $|-,g,\vR_2\ran$ column of the above Slater determinants are the same, the nonzero contributions to the above integral reduce to terms made from the other columns, namely
\be\label{ssa_29}
\iint d^3r_1d^3r_2 \begin{vmatrix}
  \lan\vr_1|+,g,\vR_1\ran & \lan\vr_1|+,e,\vR_2\ran \\
   \lan\vr_2|+,g,\vR_1\ran & \lan\vr_2|+,e,\vR_2\ran 
  \end{vmatrix}^\ast \frac{e^2}{|\vr_1-\vr_2|}\begin{vmatrix}
  \lan\vr_1|+,e,\vR_1\ran & \lan\vr_1|+,g,\vR_2\ran \\
   \lan\vr_2|+,e,\vR_1\ran & \lan\vr_2|+,g,\vR_2\ran 
  \end{vmatrix}
\ee
which also reads, due to the $(\vr_1\longleftrightarrow \vr_2)$ permutation,
\begin{align}
2 \iint d^3r_1d^3r_2 \frac{e^2}{|\vr_1-\vr_2|}\bigg[ \Big(\lan +,g,\vR_1|\vr_1\ran \lan\vr_1|+,e,\vR_1\ran \Big) \Big(\lan +,e,\vR_2|\vr_2\ran \lan\vr_2|+,g,\vR_2\ran \Big)\nn\\
- \Big(\lan +,e,\vR_2|\vr_1\ran \lan\vr_1|+,e,\vR_1\ran \Big)\Big(\lan +,g,\vR_1|\vr_2\ran \lan\vr_2|+,g,\vR_2\ran \Big)  \bigg] \label{ssa_30}
\end{align}
the second term being equal to zero in the tight-binding limit for it contains wave function overlap between different sites, $\vR_1$ and $\vR_2$.

A careful counting of the $\vr_i$ permutations in Eq.~(\ref{ssa_28}) ultimately gives
\bea\label{ssa_31}
 \lan \Phi_{+;\vR_2}|V_{e-e}|\Phi_{+;\vR_1}\ran=\hspace{7.5cm}\\
\iint d^3r_1d^3r_2 \frac{e^2}{|\vr_1-\vr_2|}\Big(\lan +,g,\vR_1|\vr_1\ran \lan\vr_1|+,e,\vR_1\ran \Big) \Big(\lan +,e,\vR_2|\vr_2\ran \lan\vr_2|+,g,\vR_2\ran \Big)\nn
\eea

In the same way, the nonzero contributions to 
\bea
 \lan \Phi_{+;\vR_2}|V_{e-e}|\Phi_{-;\vR_1}\ran =\frac{1}{4!}\iiiint d^3r_1 d^3r_2d^3r_3 d^3r_4\hspace{3cm}\nn\\
\times\begin{vmatrix}
  \lan\vr_1|+,g,\vR_1\ran & \lan\vr_1|-,g,\vR_1\ran & \lan\vr_1|+,e,\vR_2\ran & \lan\vr_1|-,g,\vR_2\ran \\
  \vdots &\vdots &\vdots &\vdots \end{vmatrix}^\ast \nn\\
 \times  V_{e-e}\begin{vmatrix}\lan\vr_1|+,g,\vR_1\ran & \lan\vr_1|-,e,\vR_1\ran & \lan\vr_1|+,g,\vR_2\ran & \lan\vr_1|-,g,\vR_2\ran \\
  \vdots &\vdots &\vdots &\vdots \end{vmatrix}\label{ssa_32}
\eea
  in which the $|+,g,\vR_1\ran$ column and the  $|-,g,\vR_2\ran$ column now are the same, are made from the other two columns, namely
\be\label{ssa_33}
\iint d^3r_1d^3r_2 \begin{vmatrix}
  \lan\vr_1|-,g,\vR_1\ran & \lan\vr_1|+,e,\vR_2\ran \\
   \lan\vr_2|-,g,\vR_1\ran & \lan\vr_2|+,e,\vR_2\ran 
  \end{vmatrix}^\ast \frac{e^2}{|\vr_1-\vr_2|}\begin{vmatrix}
  \lan\vr_1|-,e,\vR_1\ran & \lan\vr_1|+,g,\vR_2\ran \\
   \lan\vr_2|-,e,\vR_1\ran & \lan\vr_2|+,g,\vR_2\ran 
  \end{vmatrix}
\ee
The above quantity, equal to
\bea
2\iint d^3r_1d^3r_2 \frac{e^2}{|\vr_1-\vr_2|}\bigg[ \Big(\lan -,g,\vR_1|\vr_1\ran \lan\vr_1|-,e,\vR_1\ran \Big) \Big(\lan +,e,\vR_2|\vr_2\ran \lan\vr_2|+,g,\vR_2\ran \Big)\nn\\
- \Big(\lan +,e,\vR_2|\vr_1\ran \lan\vr_1|-,e,\vR_1\ran \Big)\Big(\lan -,g,\vR_1|\vr_2\ran \lan\vr_2|+,g,\vR_2\ran \Big)  \bigg] \label{ssa_34}
\eea
also has a second term equal to zero, but for a deeper reason: its spin part $\lan +|-\ran$ is equal to zero, independently from the tight-binding limit, as in the case of Eq.~(\ref{ssa_30}). So, we recover the previous result (\ref{ssa_31}), which extends to any $|\Phi_{\sigma;\vR}\ran$ state. 

\begin{figure}[t]
\centering
\includegraphics[trim=0cm 12cm 6.5cm 0cm,clip,width=4in]{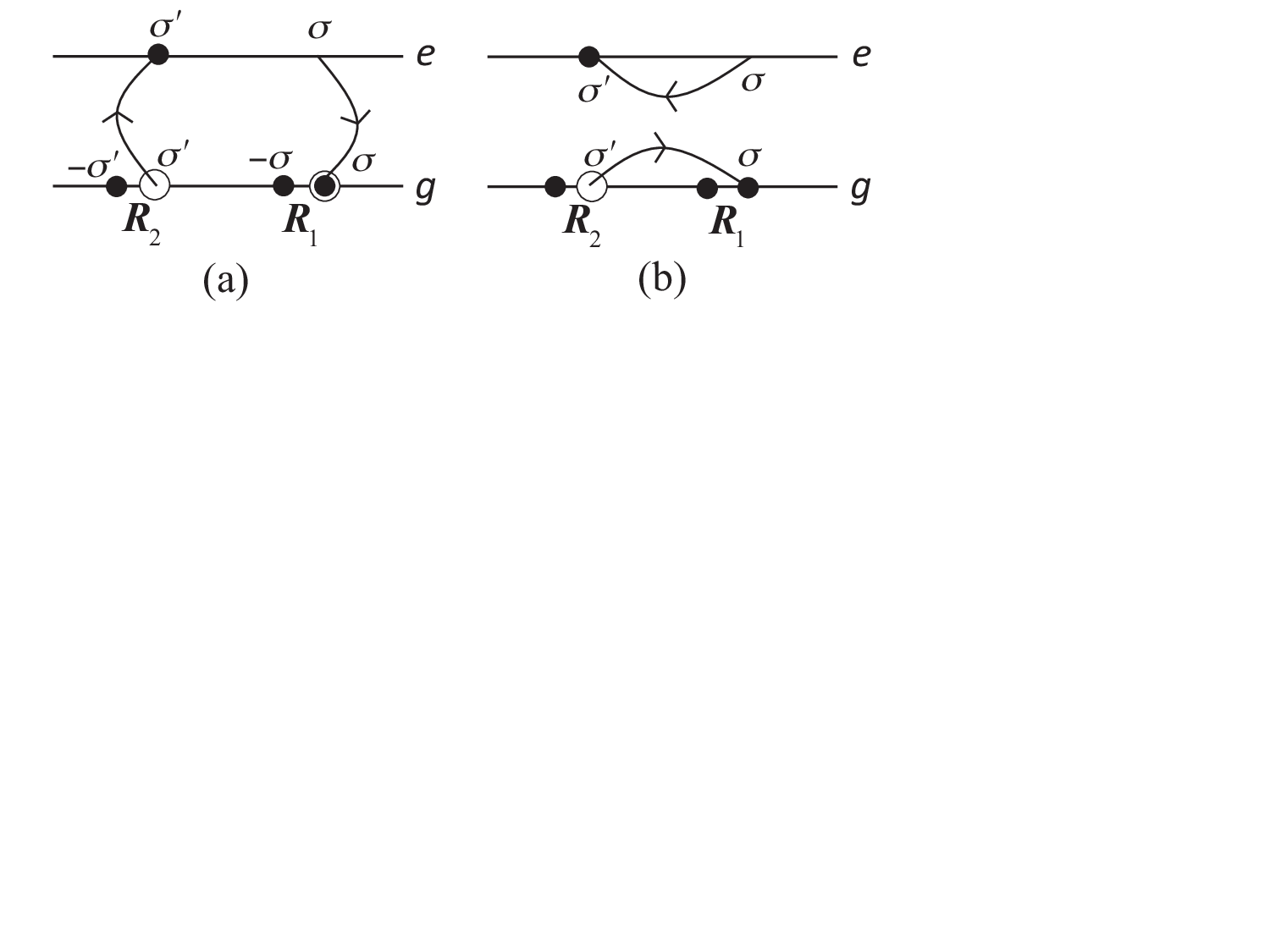}
\vspace{-0.7cm}
\caption{The \textit{interlevel} Coulomb processes between lattice sites $(\vR_2,\vR_1)$ shown in (a) exist whatever the electron spins $(\sigma',\sigma)$, while the  \textit{intralevel} Coulomb processes shown in (b), would require the two spins to be equal, $\sigma'=\sigma$, since electrons keep their spin under a Coulomb scattering. These intralevel processes have to be neglected, as a consequence of the tight-binding limit.}
\label{fig18}                                                                  
\end{figure}

We thus end with
\be\label{ssa_35}
\lan \Phi_{+;\vR_2}|V_{e-e}|\Phi_{+;\vR_1}\ran=\lan \Phi_{\sigma';\vR_2}|V_{e-e}|\Phi_{\sigma;\vR_1}\ran
\ee
whatever $(\sigma',\sigma)$. The $V_{e-e}$ matrix elements that differ from zero for $\vR_1\not=\vR_2$  in the tight-binding limit, physically come from interlevel Coulomb processes on the same lattice site, as given in Eq.~(\ref{ssa_31}). The physical reason for these matrix elements to be independent of $(\sigma',\sigma)$ is that they correspond to on-site processes (see Fig.~\ref{fig18}(a)). Note that the second term of Eq.~(\ref{ssa_34}) associated with the intralevel Coulomb interaction, that corresponds to process between sites, would impose a link between the spins of the $(\vR_1,\vR_2)$ excited electrons, in order to differ from zero (see Fig.~\ref{fig18}(b)). 

The last step is to show that such interlevel Coulomb processes only exist for lattice sites hosting \textit{electron} pairs in a spin-singlet state. This feature follows from the fact that the two electrons on the $\vR_1$ site of  $|\Phi_{+;\vR_1}\ran$  are in the electronic states $|+,e,\vR_1\ran$ and $|-,g,\vR_1\ran$, these states being occupied by electrons with opposite spins. Their associated wave function must read, due to its antisymmetry, 
\bea\label{ssa_36}
|\Phi_{+;\vR_1}\ran\!\!\!&\Longrightarrow&\!\!\! \lan \vr_1|+,e,\vR_1\ran \lan \vr_2|-,g,\vR_1\ran-\lan \vr_1|-,g,\vR_1\ran \lan \vr_2|+,e,\vR_1\ran\\
\!\!\!&& \!\!\! = \lan \vr_1|e,\vR_1\ran \lan \vr_2|g,\vR_1\ran\otimes |1_+,2_-\ran-\lan \vr_1|g,\vR_1\ran \lan \vr_2|e,\vR_1\ran\otimes |1_-,2_+\ran\nn
\eea
In the same way, the wave function associated with the two electrons  on the $\vR_1$ site of  $|\Phi_{-;\vR_1}\ran$ reads
\bea\label{ssa_37}
|\Phi_{-;\vR_1}\ran\!\!\!&\Longrightarrow&\!\!\! \lan \vr_1|+,g,\vR_1\ran \lan \vr_2|-,e,\vR_1\ran-\lan \vr_1|-,e,\vR_1\ran \lan \vr_2|+,g,\vR_1\ran\\
\!\!\!&& \!\!\!=  \lan \vr_1|g,\vR_1\ran \lan \vr_2|e,\vR_1\ran\otimes |1_+,2_-\ran-\lan \vr_1|e,\vR_1\ran \lan \vr_2|g,\vR_1\ran\otimes |1_-,2_+\ran\nn
\eea
As a result, the wave functions obtained from their sum and difference  correspond to a spin-singlet and a spin-triplet \textit{electron} pair on the $\vR_1$ site: indeed, for the sum, we find
\bea
|\Phi_{+;\vR_1}\ran+|\Phi_{-;\vR_1}\ran\Longrightarrow  \Big( \lan \vr_1|e,\vR_1\ran \lan \vr_2|g,\vR_1\ran+\lan \vr_1|g,\vR_1\ran \lan \vr_2|e,\vR_1\ran\Big)\nn\\
 \Big(|1_+,2_-\ran- |1_-,2_+\ran\Big)\label{ssa_38}
\eea
while for the difference, we get
\bea
|\Phi_{+;\vR_1}\ran-|\Phi_{-;\vR_1}\ran\Longrightarrow  \Big( \lan \vr_1|e,\vR_1\ran \lan \vr_2|g,\vR_1\ran-\lan \vr_1|g,\vR_1\ran \lan \vr_2|e,\vR_1\ran\Big)\nn\\
 \Big(|1_+,2_-\ran+|1_-,2_+\ran\Big)\label{ssa_39}
\eea
We can check that, as expected, these two combinations are antisymmetric with respect to the $(\vr_1\longleftrightarrow \vr_2)$ exchange.

This leads us to introduce the spin-singlet $(S=0,S_z=0)$ and spin-triplet $(S=1,S_z=0)$ combinations of excited states with electronic excitation on the  $\vR_1$ site,
\be\label{ssa_40}
|\Phi_{S;\vR_1}\ran=\frac{|\Phi_{+;\vR_1}\ran+(-1)^S|\Phi_{-;\vR_1}\ran}{\sqrt{2}}
\ee
for $S=(0,1)$, these combinations being such that
\be\label{ssa_41}
\lan \Phi_{S_2;\vR_2}|\Phi_{S_1;\vR_1}\ran=\delta_{S_2,S_1}\delta_{\vR_2,\vR_1}
\ee
in the tight-binding limit. It is then possible to show, with the help of Eqs.~(\ref{ssa_35},\ref{ssa_38},\ref{ssa_39}), that
\bea
\lan \Phi_{S_2;\vR_2}|V_{e-e}|\Phi_{S_1;\vR_1}\ran&=&\frac{\Big(1+(-1)^{S_1}\Big)\Big(1+(-1)^{S_2}\Big)}{4}\lan \Phi_{+;\vR_2}|V_{e-e}|\Phi_{+;\vR_1}\ran\nn\\
&=&\delta_{S_2,S_1}\delta_{S_1,0}\lan \Phi_{+;\vR_2}|V_{e-e}|\Phi_{+;\vR_1}\ran\label{ssa_42}
\eea
with $\lan \Phi_{+;\vR_2}|V_{e-e}|\Phi_{+;\vR_1}\ran$ given in Eq.~(\ref{ssa_31}). The above result proves that the interlevel Coulomb processes that control the electron-electron scattering only exist between lattice sites hosting spin-singlet electron pairs, an important result that is not at all obvious, and \textit{a priori}  far more tricky to derive when using Slater determinants than operators as in the second quantization formalism.

%\noindent \textbf{(c) Link with Knox's book}

\noindent $\bullet$ Let us now relate the above calculations to the results found in Knox's book\cite{Knox1963}. The intrinsic difficulty when using the first quantization formalism, is to speak in terms of holes correctly, because the Slater determinants are written in terms of \textit{electron} states. Of course, one can always say that the excitation of an electron leaves a hole in the ground state and that the spin and wave vector of this hole are opposite to the ones of the electron that has been removed. However, there are some tricky sign changes when turning from electron to hole. These sign changes can be handled in a secure way within the second quantization formalism, but their handling is far more problematic within the first quantization description of the problem. Let us discuss two particular aspects of this fundamental problem.

\noindent (\textit{1}) A major confusing point when using Slater determinants concerns spin-singlet and spin-triplet pairs: is Knox speaking in terms of \textit{electron} pairs, as we have done above to get Eq.~(\ref{ssa_40}) because the natural language when using Slater determinants is in terms of electrons in the ground or excited level? Or is Knox speaking in terms of \textit{electron-hole} pairs? If the latter is true, the question then is: how the change from electron-electron pair to electron-hole pair has been performed? The spin-singlet versus spin-triplet question is of importance because spin-singlet states are the only ones that suffer the interlevel Coulomb processes responsible for the Frenkel exciton formation. The $\delta_M=(1,0)$ prefactor that appears in various excited-state energies (Knox\cite{Knox1963}, Eqs.~(2.18, 2.29, 2.35, 3.7)), correctly tells that interlevel Coulomb processes exist for spin-singlet states only. The trouble is that these spin-singlet and spin-triplet states, associated with the correct $\delta_M$ prefactor, were not precisely defined in the book. This is a pity because there is a sign difference between the spin of the removed electron and the spin of the corresponding hole!

To clarify this important point, it is necessary to go back to the variables that are used when writing Slater determinants: these must be \textit{electron} variables, just as the ones we have used to define the spin-singlet combination in Eq.~(\ref{ssa_38}) and the spin-triplet combination in Eq.~(\ref{ssa_39}). As better seen from Eq.~(\ref{ssa_36}), the notation $|1_+,2_-\ran$ refers to $\lan \vr_1|+,e,\vR_1\ran\lan \vr_2|-,g,\vR_1\ran$, that is, two electrons $(\vr_1,\vr_2)$, with  spins $+$ and $-$, that respectively occupy the excited and ground levels of the $\vR_1$ lattice site. These spins refer to \textit{electrons}; so, Eqs.~(\ref{ssa_38}, \ref{ssa_39}) refer to spin-singlet and spin-triplet combinations of the two \textit{electrons} that occupy the $\vR_1$ site. The tricky point is that, when turning from these two electrons to one electron and one hole, we  \textit{de facto} drop one of the two electrons, namely, the one that remains in the ground state. The proper way to reformulate one \textit{electron-electron} pair into one \textit{electron-hole} pair is to say that if the $\vR_1$ lattice site is occupied by a $(+,e)$ electron along with a $(-,g)$ electron, this means that the $(+,g)$ electron has been removed from the $|\Phi_g\ran$ ground state. Removing the $(+,g)$ electron from the $\vR_1$ site corresponds to creating a down-spin hole ($-$) on this site. So, the $\vR_1$ site has a $(-)$ hole and a $(+)$ electron. Consequently, the formulation in terms of one electron-hole pair on the $\vR_1$ lattice site, instead of the two electrons on this site, follows from
\be\label{ssa_42'}
|1_+,2_-\ran=|e_+,h_-\ran
\ee 
 
\begin{figure}[t]
\centering
\includegraphics[trim=0cm 6cm 0cm 7cm,clip,width=5in]{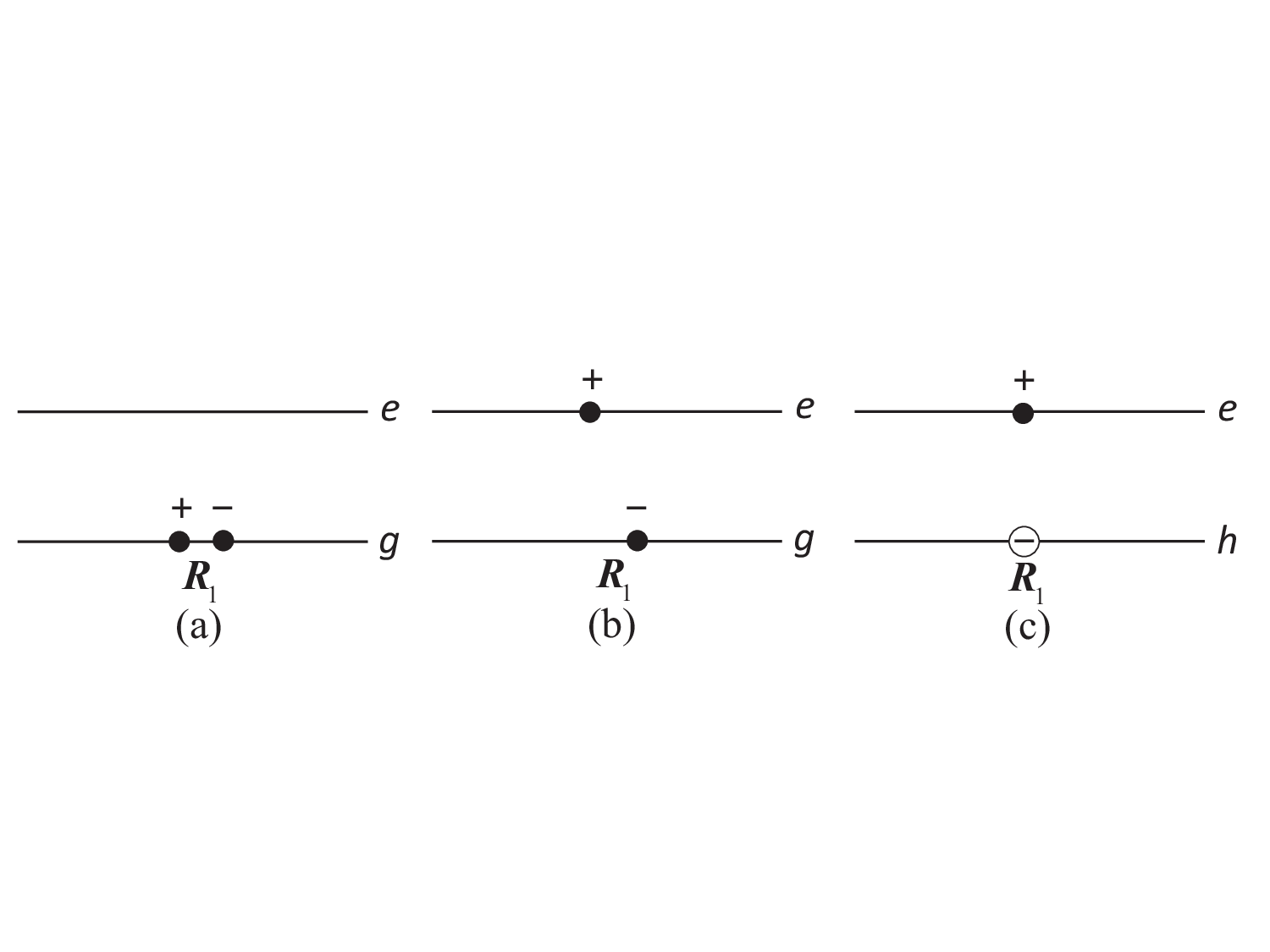}
\vspace{-0.7cm}
\caption{Occupation of the $\vR_1$ lattice site: (a) in the case of the $|\Phi_g\ran$ ground state, two up-spin and down-spin electrons are in the ground level $g$ of all lattice sites; (b) in the case of the excited state $|\Phi_{+;\vR_1}\ran$, one up-spin electron is in the excited level and one down-spin electron is in the ground level $g$; (c) the same $|\Phi_{+;\vR_1}\ran$ excited state, when described in terms of electron and hole, has an up-spin electron and a down-spin hole. }
\label{fig19}                                                                  
\end{figure}

The above equation readily proves that the spin-singlet and spin-triplet states that we have defined in terms of electron-electron pair on the $\vR_1$ lattice site, are identical to the ones defined in terms of electron-hole pair on this site (see Fig.~\ref{fig19}). While this identity might be obvious to Knox, we find that it warrants a precise derivation.

\noindent (\textit{2}) Another confusion associated with the difficulty to speak in terms of holes when using Slater determinants, shows up in the brute-force way that the hole appears in the definition of the excited subspace (see Knox\cite{Knox1963}, Eq.~(2.17)). In this part of Knox's book, the one-particle states are taken as Bloch states. The excited state, considered in this equation, is said to correspond to ``a valence electron, with spin $-\sigma$ and wave vector $\vk_h$, that is excited to a conduction state with wave vector $\vk_e$ '', the spin possibly changing to $\sigma'$. The  Slater determinant of the corresponding excited state is said to contain the following two-electron wave function 
\be\label{ssa_43}
\psi_{v,\vk_h,\sigma}\psi_{c,\vk_e,\sigma'}
\ee
Beside the fact that the physically relevant excited subspace is made of states having a  total spin equal to zero, so that $\sigma'$ should be equal to $-\sigma$, the trouble with the above equation is the physical meaning of its wave vectors; they do not correspond to the ones associated with the aforementioned excited state: indeed, starting from the valence state $(v,\vk_v)$ occupied by up-spin and down-spin electrons in the $|\Phi_g\ran$ ground state, the corresponding Slater determinant should contain the following wave function product
\be\label{ssa_44}
\psi_{v,\vk_v,\sigma}\psi_{v,\vk_v,-\sigma}
\ee
A possible excited state is obtained by replacing one of the two valence Bloch states $(v,\vk_v)$ by a conduction state $(c,\vk_c)$. The Slater determinant of the corresponding excited state then contains the wave function product 
\be\label{ssa_45}
\psi_{v,\vk_v,\sigma}\psi_{c,\vk_c,-\sigma}
\ee
Since the hole that appears in the valence band when the $(v,\vk_v,-\sigma)$ electron is removed, has a wave vector $\vk_h=-\vk_v$, the above wave function product should read in terms of electron and hole wave vectors as
 \be\label{ssa_46}
\psi_{v,-\vk_h,\sigma}\psi_{c,\vk_e,-\sigma}
\ee
for $\vk_e=\vk_c$: this differs from Eq.~(\ref{ssa_43}) through a sign difference in the hole wave vector.

The above points evidence the intrinsic difficulty to properly speak in terms of electron-hole pairs when using the first quantization formalism because, in this formalism, the many-body wave functions appear through Slater determinants that are written in terms of \textit{electron} wave functions.

 \subsubsection{Frenkel excitons}
 
 The preceding pages were dealing with the derivation of the Coulomb interaction in the excited subspace by using the first quantization formalism, that is, states written through wave functions in terms of Slater determinants, instead of through operators as done in the main part of the present manuscript. It is clear that handling these Slater determinants is definitely quite heavy. Moreover, turning to the physically relevant formulation in terms of system excitations, that is, electron-hole pairs, is  delicate when signs matter.

In the preceding pages, we have shown that the matrix elements
\be       
\lan \Phi_{S_2;\vR_2}| V_{e-e}|\Phi_{S_1;\vR_1}\ran 
\ee
of the electron-electron Coulomb interaction  $V_{e-e}$ between the excited states, in which the two electrons of the lattice sites $\vR_2$ or $\vR_1$  are in the spin combination $S_2$ or $S_1$, differ from zero for singlet states only, $(S_2=S_1=0)$; the nonzero contribution to this matrix element comes from interlevel Coulomb processes on the same lattice site. This is just what has been found in the main part of the manuscript. In this main part, we also have taken into account the possibility for the spatial wave functions to be degenerate. This is easy to include when using Slater determinants: in the case of degenerate excited level, one just has to replace the excited state $|e,\vR_\ell\ran$ by $|\mu,e,\vR_\ell\ran$. The  $V_{e-e}$ matrix, calculated through Slater determinants, ends with the same expression as the one derived from using the operator formalism of the second quantization. Its diagonalization then follows in exactly the same way.

 To conclude this section, we can say that the unique but major problem with Knox's approach to Frenkel excitons, is to derive the $V_{e-e}$ matrix for many-body states written as Slater determinants in terms of electrons when electron-hole pairs in spin-singlet states matter.

\section{Frenkel exciton many-body effects \label{sec8}}

\subsection{Composite boson nature}

Frenkel excitons are linear combinations of two fermions: one electron in an electronic excited state and one electron absence in the electronic ground state of the same lattice site, or one hole in the semiconductor ground state. We have shown (see Eqs.~(\ref{42}, \ref{56})) that, in the absence of spin and spatial degeneracy, the linear combination of on-site excitations that corresponds to a Frenkel exciton with wave vector $\vK_n$, reads as
\be\label{mb_1}
|\Phi_{\vK_n}\ran=\frac{1}{\sqrt{N_s}}\sum_{\ell=1}^{N_s}e^{i\vK_n\cdot\vR_\ell} \hat{a}^\dag_{e,\ell}\hat{a}_{g,\ell}|\Phi_g\ran= \frac{1}{\sqrt{N_s}}\sum_{\ell=1}^{N_s}e^{i\vK_n\cdot\vR_\ell} \hat{a}^\dag_{\ell}\hat{b}^\dag_{\ell}|0\ran
\ee 
This leads us to assign to the Frenkel exciton, a creation operator that appears in terms of electron and hole as
\be\label{mb_2}
\hat{B}^\dag_{\vK_n}=\frac{1}{\sqrt{N_s}}\sum_{\ell=1}^{N_s}e^{i\vK_n\cdot\vR_\ell} \hat{a}^\dag_{\ell}\hat{b}^\dag_{\ell}
\ee

Being made of fermion pairs, the Frenkel excitons have a bosonic nature. This is seen from the fact that their creation operators commute
\be\label{mb_3}
\left[\hat{B}^\dag_{\vK_{n'}},\hat{B}^\dag_{\vK_n}\right]_-=0
\ee
which readily follows from the commutation relations between different electronic states $\Big[\hat{a}_{g,\ell'}, \hat{a}^\dag_{e,\ell}\Big]_+=0$ for $\hat{a}_{g,\ell'}=\hat{b}^\dag_{\ell'}$.

As for any fermion pair, their composite nature shows up through the fact that the commutator of their destruction and creation operators differs from the one of elementary bosons, as seen by calculating 
\be\label{mb_4}
\left[\hat{B}_{\vK_{n'}},\hat{B}^\dag_{\vK_n}\right]_-=\delta_{n',n}-\hat{D}_{\vK_{n'},\vK_n}
\ee
The difference, that we called ``deviation-from-boson operator'', is precisely given by\cite{Monicbook,Agrabook}
\be\label{mb_5}
\hat{D}_{\vK_{n'},\vK_n}=\frac{1}{\sqrt{N_s}}\sum_{\ell=1}^{N_s}e^{i(\vK_n- \vK_{n'})\cdot\vR_\ell}\big( \hat{a}^\dag_{\ell}\hat{a}_{\ell} +\hat{b}^\dag_{\ell}\hat{b}_{\ell}  \big)
\ee
While this operator gives zero when acting on the electron-hole-pair vacuum $|0\ran$, as consistent with the orthogonality of Frenkel excitons that are semiconductor eigenstates,
\be\label{mb_6}
\lan 0| \hat{B}_{\vK_{n'}}\hat{B}^\dag_{\vK_n}  |0\ran=\delta_{n',n}
\ee
the $\hat{D}_{\vK_{n'},\vK_n}$ operator renders the norm of the $N$-Frenkel exciton state smaller and smaller when $N$ increases\cite{Nnorm}. Indeed, the iteration
\be
\label{mb_7}
\left[\hat{D}_{\vK_{n'_1},\vK_{n_1}},\hat{B}^\dag_{\vK_n}\right]_-=\frac{2}{N_s} \hat{B}^\dag_{\vK_n+\vK_{n_1}-\vK_{n'_1}}
\ee 
that follows from Eqs.~(\ref{mb_2},\ref{mb_5}), leads to
\be
\label{mb_8}
\left[\hat{D}_{\vK_{n'_1},\vK_{n_1}},\big(\hat{B}^\dag_{\vK_n}\big)^N\right]_-=\frac{2N}{N_s}\big(\hat{B}^\dag_{\vK_n}\big)^{N-1} \hat{B}^\dag_{\vK_n+\vK_{n_1}-\vK_{n'_1}}
\ee
from which we get
\be
\label{mb_9}
\lan 0| \big(\hat{B}_{\vK_n}\big)^N\big(\hat{B}^\dag_{\vK_n}\big)^N  |0\ran= N! F_N
\ee
where $F_N$ decreases with $N$ as
\be\label{mb_10}
\frac{F_{N+1}}{F_N}=1-\frac{N}{N_s}
\ee

This norm decrease physically comes from the Pauli exclusion principle: because of this state blocking, the ground state for semiconductors hosting Frenkel excitons hosts one electron only in each electronic ground level of the $N_s$ lattice sites (in the absence of spin and spatial degeneracies). As a result, the maximum number of electrons that can be excited is $N_s$; so is the maximum number of Frenkel excitons a semiconductor sample of $N_s$ sites can handle, in agreement with Eq.~(\ref{mb_9}): indeed, due to Eq.~(\ref{mb_10}), the norm of the $N$-Frenkel exciton state is equal to zero for $N>N_s$, which means that such state does not exist.

\subsection{Dimensionless many-body parameter}

The effect of the Pauli exclusion principle that appears in Eq.~(\ref{mb_9}) corresponds to what we have called a ``moth-eaten effect'': as more and more ``little'' electron-hole pairs are used to construct a $N$-Frenkel exciton state, less pair states will be available for  the formation of an additional exciton. For $N=N_s$,  no pair state is left to add a new Frenkel exciton to the $\big(\hat{B}^\dag_{\vK_n}\big)^N  |0\ran$ state: the ``moths'' have eaten all available electron-hole pair states.

\begin{figure}[t]
\centering
\includegraphics[trim=0cm 12.8cm 16cm 1cm,clip,width=2.3in]{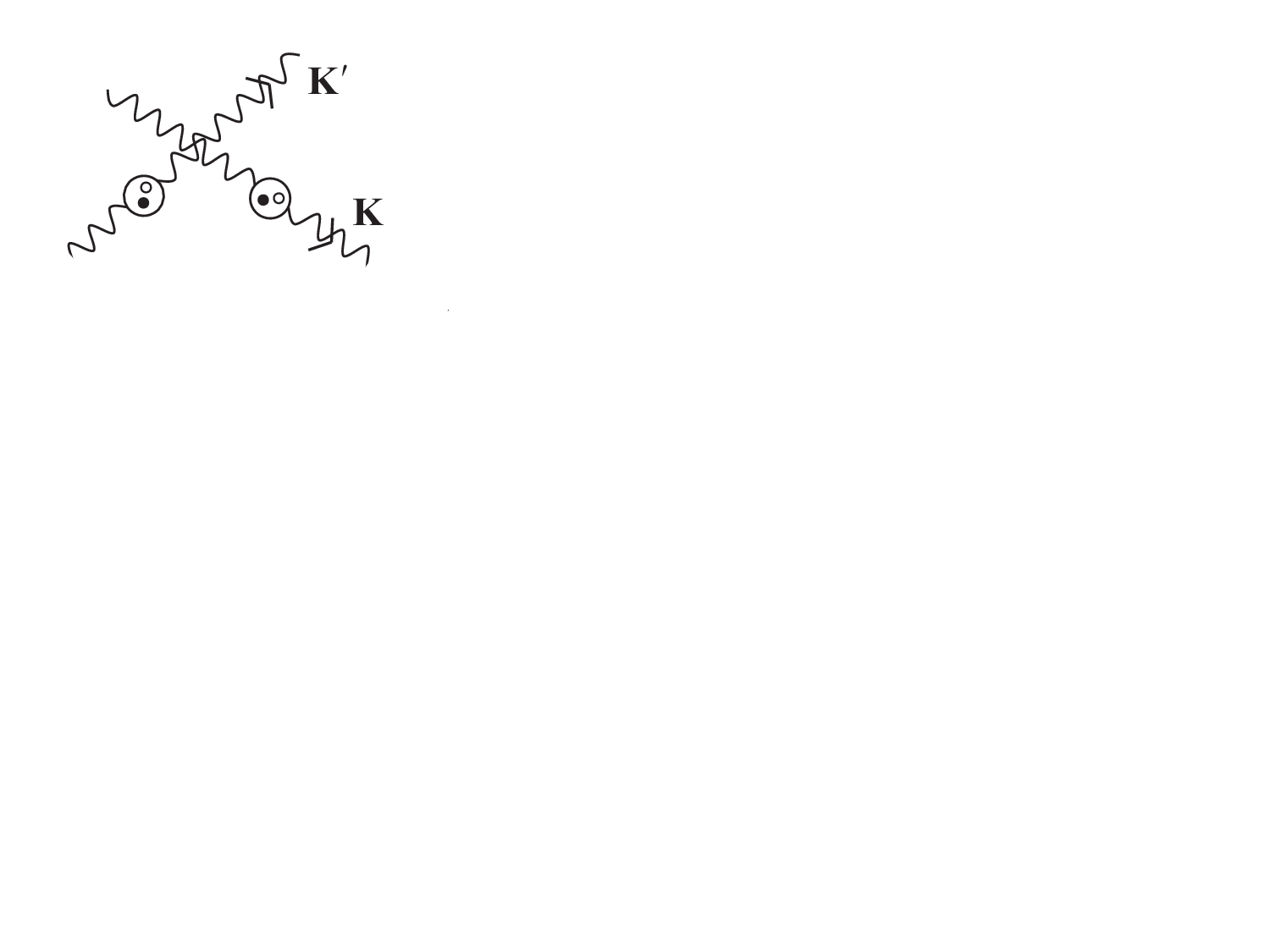}
\vspace{-0.7cm}
\caption{The dimensionless parameter (Eq.~(\ref{mb_12})) that controls Wannier exciton many-body effects can be physically understood as coming from Coulomb interaction when the two exciton waves $(\vK,\vK')$ overlap, due to their finite spatial extension, or Bohr radius. }
\label{fig21}                                                                  
\end{figure}

In view of Eq.~(\ref{mb_10}), the dimensionless parameter that controls many-body effects driven by the Pauli exclusion principle, corresponds in the case of Frenkel excitons to
\be\label{mb_11}
\eta=\frac{N}{N_s}
\ee
This parameter  appears  quite different from the one that was first written for Wannier excitons, namely
\be\label{mb_12}
\eta=N\left(\frac{a_{_X}}{L}\right)^3
\ee 
where $a_{_X}$ is the Bohr radius associated with the relative-motion extension of the electron-hole pair in a Wannier exciton. This parameter was physically understood as coming from the Coulomb interaction that occurs when two Wannier excitons, with wave vectors $(\vK,\vK')$ and spatial extension $a_{_X}$, overlap (see Fig.~\ref{fig21}). Indeed, the Wannier exciton wave extends over the sample volume $L^3$, while the volume over which the electron and the hole of an exciton can interact is the exciton volume $a^3_{_X}$. As Frenkel excitons have a relative-motion extension that reduces to zero, since they are made of on-site excitations, the above expression could lead us to na\"{i}vely conclude that the many-body parameter for Frenkel excitons should reduce to zero.

Since this is hard to accept, we are forced to reconsider the physical understanding of the many-body parameter for Wannier excitons (\ref{mb_12}) in view of its expression for Frenkel exciton given in Eq.~(\ref{mb_11}). To relate these two expressions, we note that $N_s$ in Eq.~(\ref{mb_11}) is the maximum number of Frenkel excitons that a semiconductor sample having $N_s$ lattice sites can accommodate, as previously explained in view of the state norm given in Eq.~(\ref{mb_9}). With this idea in mind, we then note that $(L/a_{_X})^3$ also is the maximum number of Wannier excitons a semiconductor sample having a $L^3$ volume can accommodate. Indeed, for an exciton number such that $Na^3_{_X}\sim L^3$, the electron-hole pairs of the $N$ Wannier excitons would fill the whole sample, and dissociate into an electron-hole plasma, known as Mott dissociation. So, $(L/a_{_X})^3$ also is of the order of the maximum number of Wannier excitons that a $L^3$ semiconductor sample can contain.

This leads us to conclude that both the many-body effects for Frenkel excitons and Wannier excitons are  controlled by the Pauli exclusion principle through a dimensionless parameter that reads\cite{Monicbook} 
\be\label{mb_13}
\eta=\frac{N}{N_{\rm max}}
\ee
where $N_{\rm max}$ is the maximum exciton number the sample can contain, that is, the number of lattice sites, $N_s$, in the case of Frenkel excitons, and $\sim(L/a_{_X})^3$ for a sample volume $L^3$ in the case of Wannier excitons.

Interestingly, this understanding also extends to Cooper pairs that  are composite bosons made of up-spin and down-spin electrons inside an energy layer $\Omega$ extending over two-phonon energy around the Fermi level, over which the BCS attractive potential  acts: indeed, we find that the energy of $N$ Cooper pairs decreases with the pair number as $\eta$ given in Eq.~(\ref{mb_13}), with $N_{\rm max}=\rho \Omega$, where $\rho$ is the electron density of states in the $\Omega$ layer; so, $\rho \Omega$ is nothing but the maximum number of electron pairs that can be involved in the BCS coupling\cite{PogosovCP}.

\subsection{Frenkel exciton scatterings}

The interaction scatterings fundamentally dictate how two particles change under an interaction. As pointed out in the case of Wannier excitons, two composite bosons made of charged fermions feel each other not only through the Coulomb interaction, but also through the Pauli exclusion principle between their fermionic components, the associated scatterings being somewhat unusual because they are dimensionless.

The Frenkel exciton many-body physics\cite{monic_Fmb} has to be simpler than the one for Wannier excitons. The fundamental reason is that Frenkel excitons are made of electronic excitations on lattice sites; so, they depend on a single index, their wave vector $\vK$. By contrast, Wannier excitons are made of a delocalized electron $\vk_e$ and a delocalized hole $\vk_h$; so, they depend on two indices, a wave vector and a relative-motion index. Nevertheless, the structure of the composite boson many-body formalism for Frenkel excitons closely follows the one that we have constructed for Wannier excitons\cite{Monicbook,Physrep,monicPRL2010}.

\begin{figure}[t]
\centering
\includegraphics[trim=0cm 12cm 11cm 2cm,clip,width=3.5in]{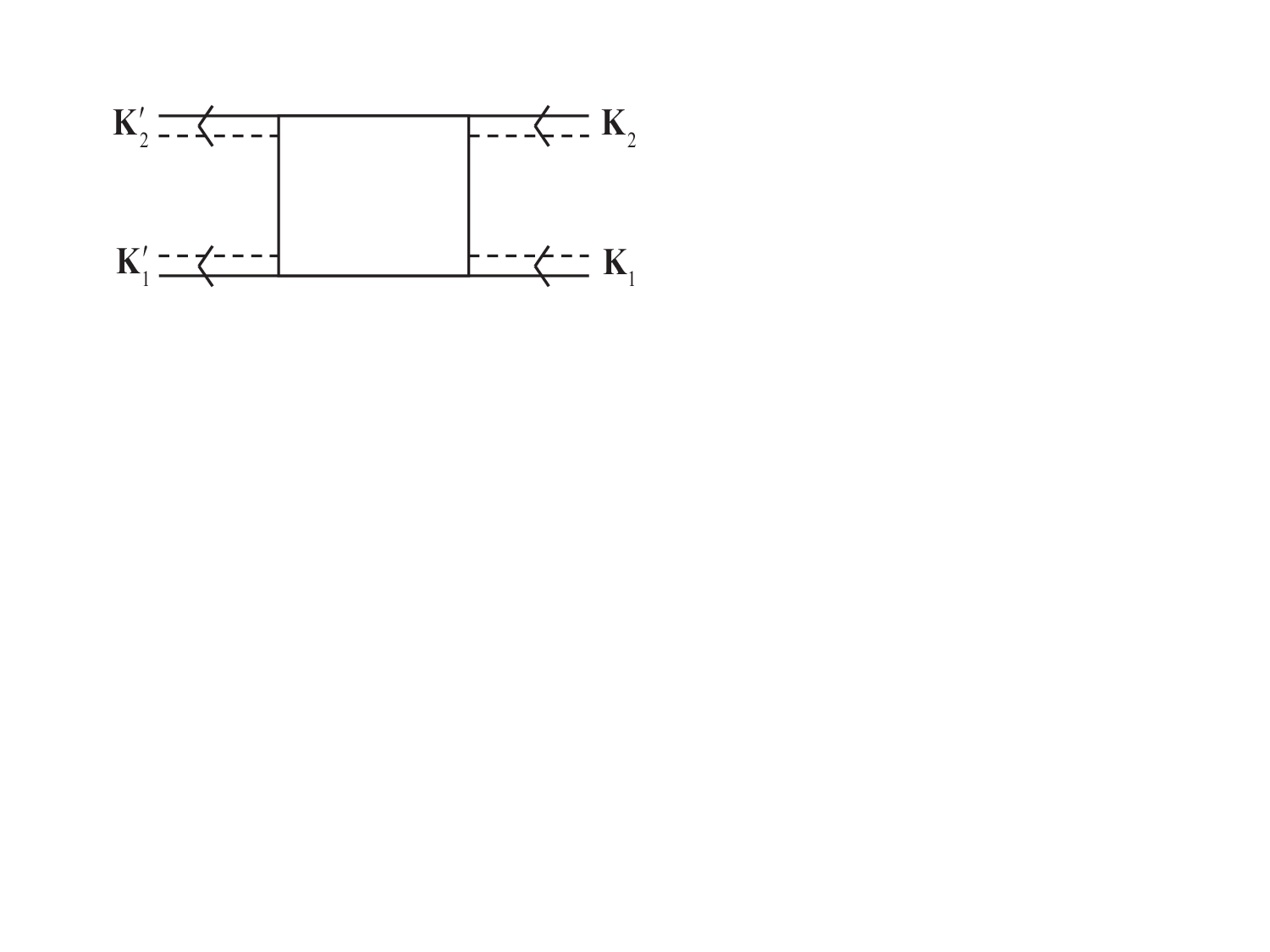}
\vspace{-0.7cm}
\caption{The scattering between Frenkel excitons $(\vK_1,\vK_2)$ that end as $(\vK'_1,\vK'_2)$. }
\label{fig22}                                                                  
\end{figure}

The problem is to determine the physics that transforms two Frenkel excitons $(\vK_1,\vK_2)$ into two Frenkel excitons $(\vK'_1,\vK'_2)$ (see Fig.~\ref{fig22}). This change can be induced either by a fermion exchange between two Frenkel excitons\cite{ShiauPRA} or by  Coulomb interactions between their charged components\cite{Toshich,Chernyak2}.

\subsubsection{Pauli scattering between two Frenkel excitons}

The Pauli exclusion principle leads to scatterings between Frenkel excitons that follow from the deviation-from-boson operator in the commutator of Eq.~(\ref{mb_4}). The fermion exchange scatterings $\lambda\left(\begin{smallmatrix}
\vK'_2& \vK_2\\ \vK'_1& \vK_1\end{smallmatrix}\right)$ induced by this operator, appear through a second commutation relation
\be\label{mb_14}
\left[\hat{D}_{\vK'_1,\vK_1},\hat{B}^\dag_{\vK_2}\right]_-=\sum_{\vK'_2}\Big(\lambda\left(\begin{smallmatrix}
\vK'_2& \vK_2\\ \vK'_1& \vK_1\end{smallmatrix}\right)+\lambda\left(\begin{smallmatrix}
\vK'_1& \vK_2\\ \vK'_2& \vK_1\end{smallmatrix}\right)\Big)\hat{B}^\dag_{\vK'_2}
\ee
In the case of single-index Frenkel excitons, this scattering reduces, with the help of Eqs.~(\ref{mb_2},\ref{mb_5}), to
\be\label{mb_15}
\lambda\left(\begin{smallmatrix}
\vK'_2& \vK_2\\ \vK'_1& \vK_1\end{smallmatrix}\right)=\frac{1}{N_s}\,\delta_{\vK'_1+\vK'_2,\vK_1+\vK_2}
\ee
It just corresponds to wave vector conservation which is ubiquitous in any fermion exchange between composite bosons.

\subsubsection{Coulomb scattering between two Frenkel excitons}

Since the electrons and holes of two Frenkel excitons are charged particles, these excitons also feel each other through the Coulomb interaction. To derive the associated scattering, we follow the same procedure as the one for Wannier excitons: the commutation of the Frenkel exciton creation operator with the semiconductor Hamiltonian written in terms of electrons and holes, contains a na\"{i}ve term that corresponds to the exciton energy, plus an operator $\hat{V}_\vK^\dag$ that we called ``Coulomb creation operator'' 
\be\label{mb_16}
\left[\hat{H}_{eh},\hat{B}^\dag_{\vK}\right]_-=E_\vK \hat{B}^\dag_{\vK}+\hat{V}_\vK^\dag
\ee
This operator gives zero when acting on the electron-hole pair vacuum $|0\ran$, as seen from the above equation projected onto $|0\ran$, namely
\be\label{mb_17}
\left[\hat{H}_{eh},\hat{B}^\dag_{\vK}\right]_-|0\ran=\hat{H}_{eh}\hat{B}^\dag_{\vK}|0\ran=E_\vK \hat{B}^\dag_{\vK}|0\ran+\hat{V}_\vK^\dag|0\ran
\ee
By contrast, it leads to energy-like scatterings when acting on Frenkel excitons. The corresponding scatterings formally follow from a second commutation relation
\be\label{mb_18}
\left[\hat{V}_{\vK_2}^\dag,\hat{B}^\dag_{\vK_1}\right]_-=\sum_{\vK'_1,\vK'_2}\xi\left(\begin{smallmatrix}
\vK'_2& \vK_2\\ \vK'_1& \vK_1\end{smallmatrix}\right)\hat{B}^\dag_{\vK'_1}\hat{B}^\dag_{\vK'_2}
\ee

Since the Coulomb physics between Frenkel excitons results from electron-hole pair exchange between lattice sites, it may not be surprising to find that the precise calculation of the above commutators, leads to a Coulomb scattering that simply reads in terms of the Pauli scattering for fermion exchange defined in Eq.~(\ref{mb_15}), namely
\be\label{mb_19}
\xi\left(\begin{smallmatrix}
\vK'_2& \vK_2\\ \vK'_1& \vK_1\end{smallmatrix}\right)=-\left(\mathcal{V}_{\vK'_1}+\mathcal{V}_{\vK'_2}\right) \lambda\left(\begin{smallmatrix}
\vK'_2& \vK_2\\ \vK'_1& \vK_1\end{smallmatrix}\right)
\ee
where $\mathcal{V}_{\vK}$ is the $\vK$-dependent part of the Frenkel exciton energy, as necessary for this scattering to be an energy-like quantity.

\subsubsection{Hamiltonian mean value}

By using the above commutators, we can calculate the Hamiltonian mean value in the $N$-Frenkel exciton ground state
\be\label{mb_20}
\lan \hat{H}_{eh}\ran_N=\frac{\lan 0| \big(\hat{B}_{\bf{K}}\big)^N \hat{H}_{eh} \big(\hat{B}^\dag_{\bf{K}}\big)^N  |0\ran}{\lan 0| \big(\hat{B}_{\bf{K}}\big)^N  \big(\hat{B}^\dag_{\bf{K}}\big)^N  |0\ran}
\ee
To do it, we first push the Hamiltonian and the $\hat{V}_\vK^\dag$ operators to the right with the help of the two commutators (\ref{mb_17},\ref{mb_18}), in order to end with $ \hat{H}_{eh}|0\ran=0$ and $\hat{V}_\vK^\dag|0\ran=0$. We are left with scalar products of Frenkel exciton states. To calculate them, we push the operator $\hat{B}_\vK$ and $\hat{D}_{\vK',\vK}$ to the right with the help of the two commutators (\ref{mb_4},\ref{mb_14}), in order to end with $\hat{B}_\vK|0\ran=0$ and  $\hat{D}_{\vK',\vK}|0\ran=0$.

For state in which all the Frenkel excitons have a zero wave vector, we find
\be\label{mb_21}
\lan \hat{H}_{eh}\ran_N\simeq N E_{\bf{0}}+\frac{N(N-1)}{2}\,\xi \left(\begin{smallmatrix}
\bf{0}& \bf{0}\\ \bf{0}& \bf{0}\end{smallmatrix}\right)
\ee

More details on these many-body calculations can be found in the book ``Excitons and Cooper pairs''\cite{Monicbook}, and also in Refs.~\cite{monic_Fmb,Pogosov_mb}.

\section{Conclusion}

Besides their applied interest in the physics of inorganic semiconductors and in the physics of ``Open Systems'',  Frenkel excitons have an utmost fundamental interest: like Wannier excitons, they correspond to electronic excitation extending over the whole sample. However, being made of on-site electronic excitations, Frenkel excitons have no relative-motion extension; so, they depend on one parameter only, their wave vector $\textbf{K}_n$. This renders their many-body physics far simpler, but mostly this gives rise to a totally different Coulomb physics. 

In the case of Wannier exciton made of one delocalized electron in the conduction band and one delocalized electron-absence in the valence band, the Coulomb interaction acts in two different ways: its intraband processes, responsible for binding a conduction electron to a valence hole, control the exciton relative-motion extension, while the interband Coulomb processes that only exist between optically bright electron-hole pairs, split their degenerate state into one longitudinal and two transverse modes. 

The situation is totally different in the case of Frenkel excitons because these excitons are made of on-site electronic excitations. The tight-binding approximation, valid for semiconductors hosting Frenkel excitons, renders negligible the intralevel processes between sites, that are the analog of intraband processes. So, we are left with  interlevel processes in which one electron-hole pair recombines on one site while another pair is created on another site: this fundamentally corresponds to an electron-hole \textit{pair} exchange between two sites --- not an ``electron-hole exchange'', which would be meaningless for different quantum particles. These interlevel processes do both: they delocalize the electronic excitations on the $\vR_\ell$ lattice sites as $\textbf{K}_n$ excitonic waves extending over the whole sample; they also split the degenerate exciton level into one longitudinal and two transverse modes, just as for Wannier excitons. Note that, since the interlevel Coulomb processes only act between spin-singlet electron-hole pairs, just like the electron-photon interaction, the Frenkel excitons are by construction optically bright --- in the tight-binding limit. Yet, the exciton longitudinal mode is not coupled to light because the photon  field is transverse.

The presentation of Frenkel excitons we give in the present manuscript, makes use of the tight-binding approximation that we take in the limit of zero wave function overlap between different lattice sites. This allows us to introduce clean fermionic operators for the one-electron basis in terms of which we write the Frenkel exciton problem in second quantization. The second quantization formalism not only avoids using the cumbersome Slater determinants for many electron states when written in the first quantization, but it also allows properly introducing the concept of hole that goes along with many tricky minus signs, which are difficult to follow when using Slater determinants: indeed, these determinants have been introduced for electron states, to take care of the sign change that appears when exchanging two electrons, not one electron and one hole.

In a last part, we give an overview of how to handle many-body effects between Frenkel excitons, through a formalism appropriate to composite bosons having no spatial extension, as required because these excitons are made of on-site excitations, in contrast to Wannier excitons made of delocalized conduction electron and valence-electron absence but ending as bound electron-hole pairs extending over a Bohr radius. We show that the dimensionless parameter for many-body effects between  composite bosons like Frenkel excitons, Wannier excitons and also Cooper pairs, is entirely controlled by the Pauli exclusion principle between their fermionic components, whatever the mechanism that binds their fermion pairs.

We wish to say that the present work does not address two important parts of the exciton physics. 

 \noindent (\textit{i}) We do not consider the interaction between Frenkel excitons and photons. Like Wannier excitons, Frenkel excitons can form polaritons, that is, linear combinations of one photon and one exciton.  This formation occurs in the so-called ``strong coupling'' regime in which the exciton couples to a photon over a characteristic time much smaller than the time over which its wave vector would change by collisions: in this regime, the exciton keeps re-emitting a photon having the same wave vector; so, a coupled photon-exciton state can develop. 
 
\noindent (\textit{ii}) The second important part is the spin-orbit interaction\cite{Cooper}. This interaction splits the $(2\times3)$ spin and spatial degeneracies of the energy levels. The relevant electron states one then has to consider are these spin-orbit eigenstates. The procedure to  derive Frenkel excitons will, in the same way, follow with writing the matrix for the one electron-hole pair Hamiltonian in the degenerate subspace and diagonalizing this matrix. However, due to the intrinsic mixture of spin and spatial degeneracies induced by the spin-orbit interaction, the physics associated with the singlet-triplet splitting and the one associated with the longitudinal-transverse splitting will be less transparent to pick out.

\appendix
%\numberwithin{equation}{section}

\section{Derivation of the commutation relations (\ref{8},\ref{9}) \label{app:M1}}

 We here derive the anticommutation relations between electron  operators $\hat{a}^\dag_{\nu,\ell}$ defined in Eq.~(\ref{7}), namely
 \be
|\nu,\vR_\ell\ran=\hat{a}^\dag_{\nu,\ell}|v\ran
\ee
where $|v\ran$ is the vacuum state. 

\noindent $\bullet$ To obtain the anticommutation relation $\left[\hat{a}_{\nu',\ell'},\hat{a}^\dag_{\nu,\ell}\right]_+$, we start with the $\hat{a}_{\nu',\ell'}\hat{a}^\dag_{\nu,\ell}|v\ran$ state that we calculate in two ways. 

\noindent (\textit{1}) Since this state contains no electron, it reduces to vacuum. So, 
\bea
\hat{a}_{\nu',\ell'}\hat{a}^\dag_{\nu,\ell}|v\ran &=&|v\ran\lan v| \hat{a}_{\nu',\ell'}\hat{a}^\dag_{\nu,\ell}|v\ran\nn\\
&=& |v\ran \lan \nu',\vR_{\ell'}|\nu,\vR_\ell\ran=\delta_{\nu',\nu}\delta_{\ell',\ell}|v\ran
\eea

\noindent (\textit{2}) In the other way, we rewrite the above state as
\be
\hat{a}_{\nu',\ell'}\hat{a}^\dag_{\nu,\ell}|v\ran = \left(\left[\hat{a}_{\nu',\ell'},\hat{a}^\dag_{\nu,\ell}\right]_+ - \hat{a}^\dag_{\nu,\ell}\hat{a}_{\nu',\ell'}\right)|v\ran
\ee
 the second term being equal to zero, since $\hat{a}_{\nu',\ell'}|v\ran=0$.

From these two calculations, we readily get
\be
\left[\hat{a}_{\nu',\ell'},\hat{a}^\dag_{\nu,\ell}\right]_+=\delta_{\nu',\nu}\delta_{\ell',\ell}
\ee

\noindent $\bullet$ To obtain $\Big[\hat{a}^\dag_{\nu',\ell'},\hat{a}^\dag_{\nu,\ell}\Big]_+$, we first note that  $|\nu,\vR_\ell\ran$  can be written on the electron basis $|\nu,\vR_{\ell_0}\ran$ of the $\vR_{\ell_0}$ sites as
\bea
\hat{a}^\dag_{\nu,\ell}|v\ran&=&|\nu,\vR_\ell\ran=\left(\sum_{\nu_1}|\nu_1,\vR_{\ell_0}\ran\lan\nu_1,\vR_{\ell_0}| \right)|\nu,\vR_\ell\ran\nn\\
&=&\left(\sum_{\nu_1}\hat{a}^\dag_{\nu_1,\ell_0}\lan\nu_1,\vR_{\ell_0}|\nu,\vR_\ell\ran \right)|v\ran
\eea
from which we get
\bea
\hat{a}^\dag_{\nu',\ell'}\hat{a}^\dag_{\nu,\ell}&=&\sum_{\nu_1\nu_2}\hat{a}^\dag_{\nu_2,\ell_0}\hat{a}^\dag_{\nu_1,\ell_0}\lan\nu_2,\vR_{\ell_0}|\nu',\vR_{\ell'}\ran \lan\nu_1,\vR_{\ell_0}|\nu,\vR_\ell\ran\nn\\
&=& -\hat{a}^\dag_{\nu,\ell}\hat{a}^\dag_{\nu',\ell'}
\eea 
 since the electron operators $\hat{a}^\dag_{\nu,\ell_0}$ anticommute; so, $\hat{a}^\dag_{\nu_2,\ell_0}\hat{a}^\dag_{\nu_1,\ell_0}=-\hat{a}^\dag_{\nu_1,\ell_0}\hat{a}^\dag_{\nu_2,\ell_0}$. As a result, we end with 
 \be
 \left[\hat{a}^\dag_{\nu',\ell'},\hat{a}^\dag_{\nu,\ell}\right]_+=0
 \ee

 \section{$\hat{H}$ Hamiltonian in the $|\Phi_g\ran$ subspace\label{app:M2}}    
  
  The parts of $\hat{H}$ that act on the system ground state $|\Phi_g\ran$, reduce to $\hat{H}_{0,g}+\hat{V}_{gg}+V_{i-i}$ because  $|\Phi_g\ran$ is made of ground-level electrons only.
  
 \noindent $\bullet$ Using $\hat{H}_{0,g}$ in Eq.~(\ref{40_2}), we get, since $\hat{a}^\dag_{g,\ell}\hat{a}_{g,\ell}|\Phi_g\ran=|\Phi_g\ran$ as all $N_s$ lattice sites are occupied by a ground-level electron,
 \bea
 \hat{H}_{0,g}|\Phi_g\ran&=& (\va_g+t_{g,g})\sum_{\ell=1}^{N_s} \hat{a}^\dag_{g,\ell}\hat{a}_{g,\ell}|\Phi_g\ran\nn\\
 &=& {N_s}(\va_g+t_{g,g})|\Phi_g\ran\label{40_8}
 \eea

  \noindent $\bullet$ In the same way, since  $\hat{a}^\dag_{g,\ell_1}\hat{a}^\dag_{g,\ell_2}\hat{a}_{g,\ell_2}\hat{a}_{g,\ell_1}|\Phi_g\ran=(1-\delta_{\ell_1,\ell_2})|\Phi_g\ran$, the $\hat{V}_{gg}$ Coulomb interaction (\ref{40_5}) acting on  $|\Phi_g\ran$ gives, due to lattice periodicity,
  \bea
  \hat{V}_{gg}|\Phi_g\ran = \left(\frac{1}{2}\sum_{\ell_1=1}^{N_s} \sum_{\ell_2\not=\ell_1}^{N_s} \mathcal{V}_{\vR_{\ell_1}-\vR_{\ell_2}}\left(\begin{smallmatrix}
g& g\\ g& g\end{smallmatrix}\right)\right)|\Phi_g\ran= \frac{{N_s}}{2}\sum_{\vR\not=\bf0} \mathcal{V}_{\vR}\left(\begin{smallmatrix}
g& g\\ g& g\end{smallmatrix}\right)|\Phi_g\ran
  \eea

 \noindent $\bullet$ The above results readily give, for the one-body part of $\hat{H}$, 
 \be\label{28}
 \lan \Phi_g|\hat{H}_{0,g} |\Phi_g\ran=N_s(\va_g+t_{g,g})
 \ee
while for the two-body Coulomb part, we get
\be
\lan \Phi_g|\hat{V}_{gg} |\Phi_g\ran=\frac{{N_s}}{2}\sum_{\vR\not=\bf0}\mathcal{V}_{\vR}\left(\begin{smallmatrix}
g& g\\ g& g\end{smallmatrix}\right)\label{31}
\ee

\noindent $\bullet$ By adding the constant term that corresponds to the ion-ion interaction given in Eq.~(\ref{32}), we end with the ${N_s}$-electron ground-state energy, defined in Eq.~(\ref{26}), as
\be\label{33}
E'_g={N_s}(\va_g+v_{gg})
\ee
where the energy contribution from the Coulomb interactions is given, according to Eqs.~(\ref{28},\ref{31}) and Eqs.~(\ref{16},\ref{32}), by
\bea
v_{gg}&=&t_{g,g}+\frac{1}{2}\sum_{\vR\not=\bf0}\mathcal{V}_{\vR}\left(\begin{smallmatrix}
g& g\\ g& g\end{smallmatrix}\right)+\frac{1}{2}\sum_{\vR\not=\bf0}\frac{e^2}{|\vR|}\nn\\
&=&\sum_{\vR\not=\bf0}\bigg(\frac{1}{2}\iint_{L^3}d^3rd^3r' |\lan\vr|g\ran|^2|\lan\vr'|g\ran|^2\frac{e^2}{|\vR+\vr-\vr'|}\label{34}\\
&&-\int_{L^3}d^3r |\lan\vr|g\ran|^2\frac{e^2}{|\vR-\vr'|}+\frac{1}{2}\frac{e^2}{|\vR|}  \bigg)\nn
\eea
By noting that
\be\label{35}
1=\int_{L^3}d^3r |\lan\vr|g\ran|^2
\ee
for normalized ground-level wave function, we can rewrite $v_{gg}$ in a compact form as
\be
v_{gg}=\frac{1}{2}\iint_{L^3}d^3rd^3r'\, |\lan\vr|g\ran|^2|\lan\vr'|g\ran|^2\sum_{\vR\not=\bf0}\left(\frac{e^2}{|\vR|}+\frac{e^2}{|\vR+\vr-\vr'|}-\frac{2e^2}{|\vR-\vr'|} \right)\label{36}
\ee

The above equation evidences that it is necessary to keep the electron interaction with all the other ions through $t_{g,g}$, as well as the interaction $V_{i-i}$ between all the ions, to possibly cancel the overextensive contribution coming from $\hat{V}_{e-e}$ alone. Indeed, for $|\vr|$ kept small by the $\lan\vr|g\ran$ wave function, the leading term of the  $\hat{V}_{e-e}$ scattering scales as
\be\label{37}
\sum_{\vR\not=\bf0}\frac{1}{|\vR+\vr-\vr'|}\simeq \sum_{\vR\not=\bf0}\frac{1}{|\vR|}\simeq \int_{L^3}d^3R\frac{1}{R}\propto L^2
\ee
The contribution of the electron-electron interaction to the   ground-state energy would then be volume infinite compared to $\va_{g}$. This once more proves that  considering a neutral system, with the same amount of negative electron charges and positive ion charges, is crucial to avoid spurious singularities.

\section{$\hat{H}$ Hamiltonian in the $|\Phi_{\vR_\ell}\ran$ subspace\label{app:M3}}

We now consider $\hat{H}$ acting on $|\Phi_{\vR_\ell}\ran$ given in Eq.~(\ref{38}), with the electron on the $\vR_\ell$ ion being in the excited level, while all the other lattice sites are occupied by a ground-level electron.

\noindent $\bullet$ Using the one-body part $\hat{H}_{0,e}$ of the Hamiltonian given in Eq.~(\ref{40_3}), we readily find
\bea
\hat{H}_{0,e}|\Phi_{\vR_\ell}\ran&=&(\va_e+t_{e,e})\sum_{\ell_1=1}^{N_s} \hat{a}^\dag_{e,\ell_1}\hat{a}_{e,\ell_1}|\Phi_{\vR_\ell}\ran\nn\\
&=&(\va_e+t_{e,e}) |\Phi_{\vR_\ell}\ran\label{37_1}
\eea
while from the $\hat{H}_{0,g}$ part given in Eq.~(\ref{40_2}), we get
\bea
\hat{H}_{0,g}|\Phi_{\vR_\ell}\ran&=&(\va_g+t_{g,g})\sum_{\ell_1=1}^{N_s} \hat{a}^\dag_{g,\ell_1}\hat{a}_{g,\ell_1}|\Phi_{\vR_\ell}\ran\nn\\
&=&({N_s}-1)(\va_g+t_{g,g}) |\Phi_{\vR_\ell}\ran\label{37_2}
\eea
since the $|\Phi_{\vR_\ell}\ran$ state contains  $({N_s}-1)$ ground-level electrons.

\noindent $\bullet$ From the Coulomb interaction $\hat{V}^{(intra)}_{eg}$ given in Eq.~(\ref{40_6}), in which each electron stays in its  level, we get
  \bea
\hat{V}^{(intra)}_{eg}|\Phi_{\vR_\ell}\ran&=&\sum_{\ell_1=1}^{N_s} \mathcal{V}_{\vR_{\ell_1}-\vR_{\ell}}\left(\begin{smallmatrix}
e& e\\ g& g\end{smallmatrix}\right)\hat{a}^\dag_{g,\ell_1}\hat{a}_{g,\ell_1}|\Phi_{\vR_\ell}\ran\nn\\
&=& \sum_{\ell_1\not=\ell}^{N_s} \mathcal{V}_{\vR_{\ell_1}-\vR_{\ell}}\left(\begin{smallmatrix}
e& e\\ g& g\end{smallmatrix}\right)|\Phi_{\vR_\ell}\ran
%\nn\\
=|\Phi_{\vR_\ell}\ran\sum_{\vR\not=\bf0}\mathcal{V}_{\vR}\left(\begin{smallmatrix}
e& e\\ g& g\end{smallmatrix}\right)\label{37_3}
 \eea
 Indeed, the $\vR_\ell$ ion is occupied by an excited-level electron, so that $\hat{a}_{g,\ell_1}|\Phi_{\vR_\ell}\ran=0$ for $\ell_1=\ell$.

\noindent $\bullet$ Turning to the Coulomb interaction $\hat{V}^{(inter)}_{eg}$ given in Eq.~(\ref{40_7}), in which each electron changes its level, we first note that
\be\label{37_4}
\hat{a}^\dag_{e,\ell_1}\hat{a}^\dag_{g,\ell_2}\hat{a}_{e,\ell_2}\hat{a}_{g,\ell_1}|\Phi_{\vR_\ell}\ran=\delta_{\ell_2,\ell}(1-\delta_{\ell_1,\ell})|\Phi_{\vR_{\ell_1}}\ran 
\ee
 since $\hat{a}_{g,\ell}|\Phi_{\vR_{\ell}}\ran=0$. This leads to
  \be\label{37_5}
 \hat{V}^{(inter)}_{eg}|\Phi_{\vR_\ell}\ran=\sum_{\ell_1\not=\ell}  \mathcal{V}_{\vR_{\ell_1}-\vR_{\ell}}\left(\begin{smallmatrix}
g& e\\ e& g\end{smallmatrix}\right)|\Phi_{\vR_{\ell_1}}\ran
 \ee
 
\noindent $\bullet$ The $\hat{V}_{gg}$ interaction, given in Eq.~(\ref{40_5}), also acts on $|\Phi_{\vR_\ell}\ran$ because this state contains $({N_s}-1)$ ground-level electrons. Its contribution reads
\bea\label{37_6}
\hat{V}_{gg}|\Phi_{\vR_\ell}\ran&=&\bigg[ \frac{1}{2}\sum_{\ell_1\not=\ell}\,\sum_{\ell_2\not=(\ell_1,\ell)}\!\!\mathcal{V}_{\vR_{\ell_1}-\vR_{\ell_2}}\left(\begin{smallmatrix}
g& g\\ g& g\end{smallmatrix}\right)\bigg]\!|\Phi_{\vR_\ell}\ran\nn\\
&=&\left(\frac{{N_s}}{2}-1\right)\sum_{\vR\not=\bf0}\mathcal{V}_{\vR}\left(\begin{smallmatrix}
g& g\\ g& g\end{smallmatrix}\right)|\Phi_{\vR_\ell}\ran
\eea

\noindent $\bullet$ The above results give the Coulomb parts of the $\hat{H}$ matrix elements in the $|\Phi_{\vR_\ell}\ran$ subspace as
\bea
\lan \Phi_{\vR_{\ell'}}| \hat{V}_{gg} |\Phi_{\vR_\ell}\ran=\delta_{\ell',\ell}\left(\frac{{N_s}}{2}-1\right)\sum_{\vR\not=\bf0}\mathcal{V}_{\vR}\left(\begin{smallmatrix}
g& g\\ g& g\end{smallmatrix}\right)\\
\lan \Phi_{\vR_{\ell'}}| \hat{V}^{(intra)}_{eg} |\Phi_{\vR_\ell}\ran=\delta_{\ell',\ell}\sum_{\vR\not=\bf0}\mathcal{V}_{\vR}\left(\begin{smallmatrix}
e& e\\ g& g\end{smallmatrix}\right)\label{37_7}\hspace{1cm}\\
\lan \Phi_{\vR_{\ell'}}| \hat{V}^{(inter)}_{eg} |\Phi_{\vR_\ell}\ran=(1-\delta_{\ell',\ell})\,\mathcal{V}_{\vR_{\ell'}-\vR_\ell}\left(\begin{smallmatrix}
g& e\\ e& g\end{smallmatrix}\right)\label{37_8}
\eea

\noindent $\bullet$ Using them, we obtain the diagonal part of the $\hat{H}$ Hamiltonian in the $|\Phi_{\vR_\ell}\ran$ subspace as
\be\label{37_9}
\lan \Phi_{\vR_\ell}| \hat{H} |\Phi_{\vR_\ell}\ran= E'_e=E'_g+\va_e-\va_g+v_{eg}
\ee
whatever $\ell$, the Coulomb part 
\be\label{54}
v_{eg}=t_{e,e}-t_{g,g}+\sum_{\vR\not=\bf0}\Big(\mathcal{V}_{\vR}\!\left(\begin{smallmatrix}
e& e\\ g& g\end{smallmatrix}\right)-\mathcal{V}_{\vR}\!\left(\begin{smallmatrix}
g& g\\ g& g\end{smallmatrix}\right)\Big)
\ee
being precisely given by
\be\label{app:54_1}
v_{eg}=\iint_{L^3}d^3rd^3r' \Big(|\lan\vr'|e\ran|^2-|\lan\vr'|g\ran|^2\Big)|\lan\vr|g\ran|^2
\sum_{\vR\not=\bf0}\left(\frac{e^2}{|\vR+\vr-\vr'|}-\frac{e^2}{|\vR-\vr'|}\right)
\ee 
The above equation again shows that the electron interaction with all the other ions, that produces the $e^2/|\vR-\vr'|$ term, goes to cancel the dominant $1/|\vR|$ contribution of the first term that comes from electron-electron interaction, so that  no overextensive contribution remains in $v_{eg}$.

\noindent $\bullet$ The nondiagonal parts of the $\hat{H}$ Hamiltonian in the $|\Phi_{\vR_\ell}\ran$ subspace only come  from Coulomb processes. They reduce to
\be\label{37_10}
\lan \Phi_{\vR_{\ell'}}| \hat{H} |\Phi_{\vR_\ell}\ran=\lan \Phi_{\vR_{\ell'}}| \hat{V}^{(inter)}_{eg} |\Phi_{\vR_\ell}\ran= \mathcal{V}_{\vR_{\ell'}-\vR_\ell}\!\left(\begin{smallmatrix}
g& e\\ e& g\end{smallmatrix}\right)\equiv \mathcal{V}^{(e,g)}_{\ell',\ell}
\ee

\noindent $\bullet$ As a result, the ($N_s\times N_s$) matrix for the $\hat{H}$ Hamiltonian in this subspace  appears as
\be
\left(\begin{matrix}
E'_e& \cdots & \mathcal{V}^{(e,g)}_{\ell',\ell} \\ \vdots& \ddots & \vdots\\
\mathcal{V}^{(e,g)}_{\ell',\ell} &\cdots & E'_e \end{matrix}\right)
\ee
%with $\mathcal{V}^{(ge)}_{n',n}= \mathcal{V}_{\vR_{n'}-\vR_n}\!\left(\begin{smallmatrix}
%g& e\\ e& g\end{smallmatrix}\right)$.

 \section{$\hat{H}$ Hamiltonian in the ground and excited subspaces with spin\label{app:D}}
 
 \noindent $\bullet$ The  system ground-state energy in the presence of Coulomb interactions between electrons, between electrons and the other ions, and between ions, reads
 \be\label{108}
 \lan \Phi_g| \hat{H}|\Phi_g\ran=2 N_s (\va_g+v_{gg})
 \ee
 with the Coulomb part given by
 \be\label{109}
 v_{gg}=t_{g,g}+\frac{1}{2}\mathcal{V}_{\vR=\bf0}\left(\begin{smallmatrix}
 g& g\\ g& g\end{smallmatrix}\right)+ \sum_{\vR\not=\bf0}\mathcal{V}_\vR\left(\begin{smallmatrix}
 g& g\\ g& g\end{smallmatrix}\right)+\frac{V_{i-i}}{2N_s}
 \ee
 We can check that the overextensive terms of this Coulomb part cancel out exactly.

     \begin{figure}[t]
\centering
\includegraphics[trim=1cm 4.5cm 10cm 7.5cm,clip,width=3in]{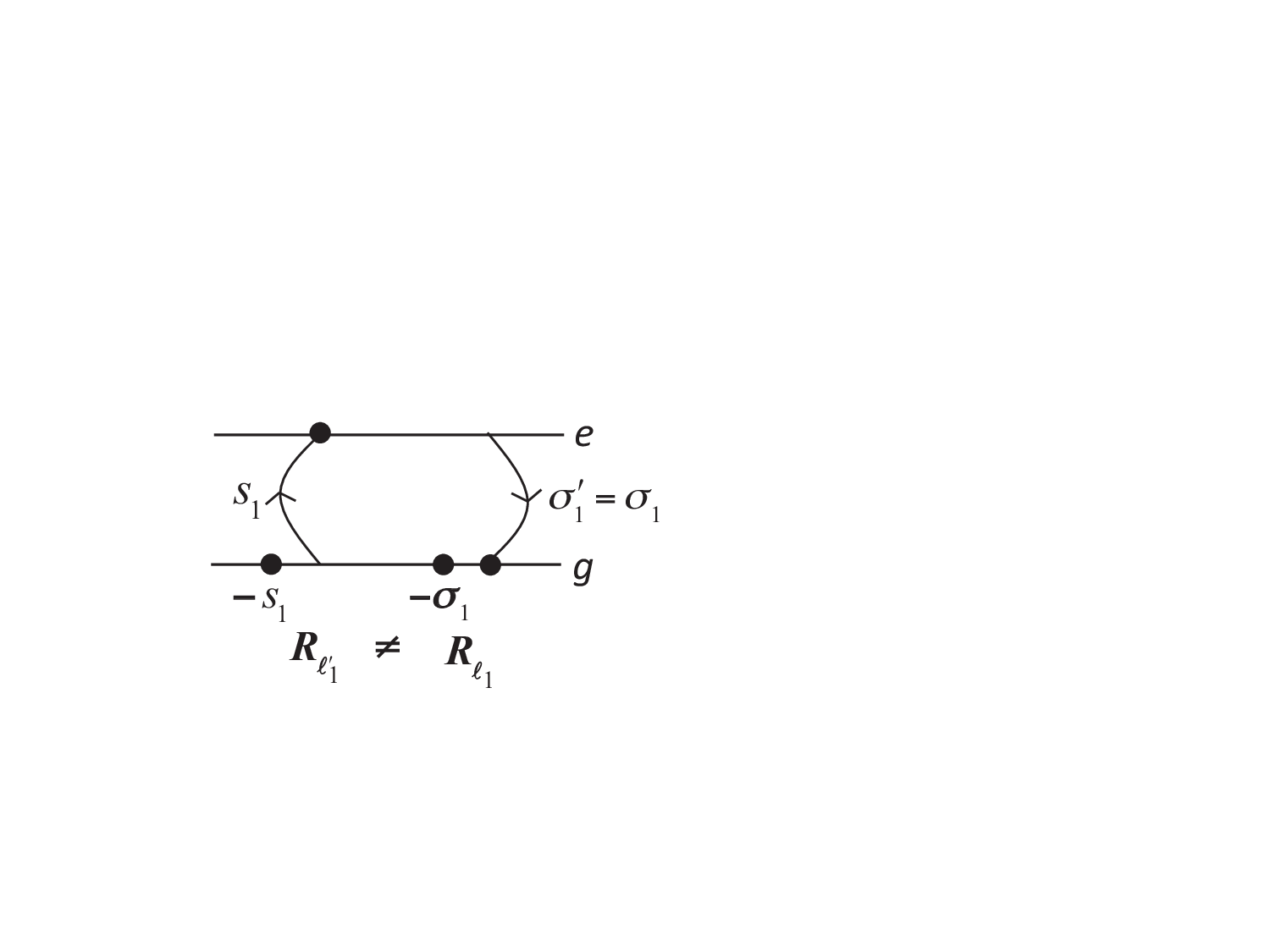}
\vspace{-0.7cm}
\caption{The $\hat{V}^{(inter)}_{eg}$ interaction only acts between $|\Phi_{\sigma'_1,\sigma_1;\vR_{\ell_1}}\ran$ states with $\sigma'_1=\sigma_1$, because Coulomb interaction conserves the spin (see Eq.~(\ref{110})).   }
\label{fig11}
\end{figure}

  \noindent $\bullet$ For the sake of generality, we here consider that the excited electron may not have the same spin as the absent ground-level electron. So, the excited subspace is made of states $|\Phi_{\sigma',\sigma;\vR_{\ell}}\ran=\hat{a}^\dag_{\sigma',e,\ell}\hat{a}_{\sigma,g,\ell} |\Phi_g\ran$. The terms of $\hat{H}$ in this subspace, $\lan \Phi_{\sigma'_2,\sigma_2;\vR_{\ell_2}}| \hat{H}|\Phi_{\sigma'_1,\sigma_1;\vR_{\ell_1}}\ran$, that are responsible for  Frenkel exciton formation, come from the interlevel Coulomb interaction. By using Eq.~(\ref{107}), we find that the state $\hat{V}^{(inter)}_{eg}|\Phi_{\sigma'_1,\sigma_1;\vR_{\ell_1}}\ran$ contains two types of terms: one term, shown in Fig.~\ref{fig11}, 
 \be\label{110}
\delta_{\sigma'_1,\sigma_1}\sum_{s_1} \sum_{\ell'_1\not=\ell_1}\mathcal{V}_{\vR_{\ell'_1}-\vR_{\ell_1}}\left(\begin{smallmatrix}
 g& e\\ e& g\end{smallmatrix}\right)|\Phi_{s_1,s_1;\vR_{\ell'_1}}\ran
 \ee
 comes from processes on different lattice sites $\ell'_1\not=\ell'_2$, but for $\sigma_1=\sigma'_1$ only.
    \begin{figure}[t]
\centering
\includegraphics[trim=4.5cm 4.5cm 3.5cm 7cm,clip,width=3.5in]{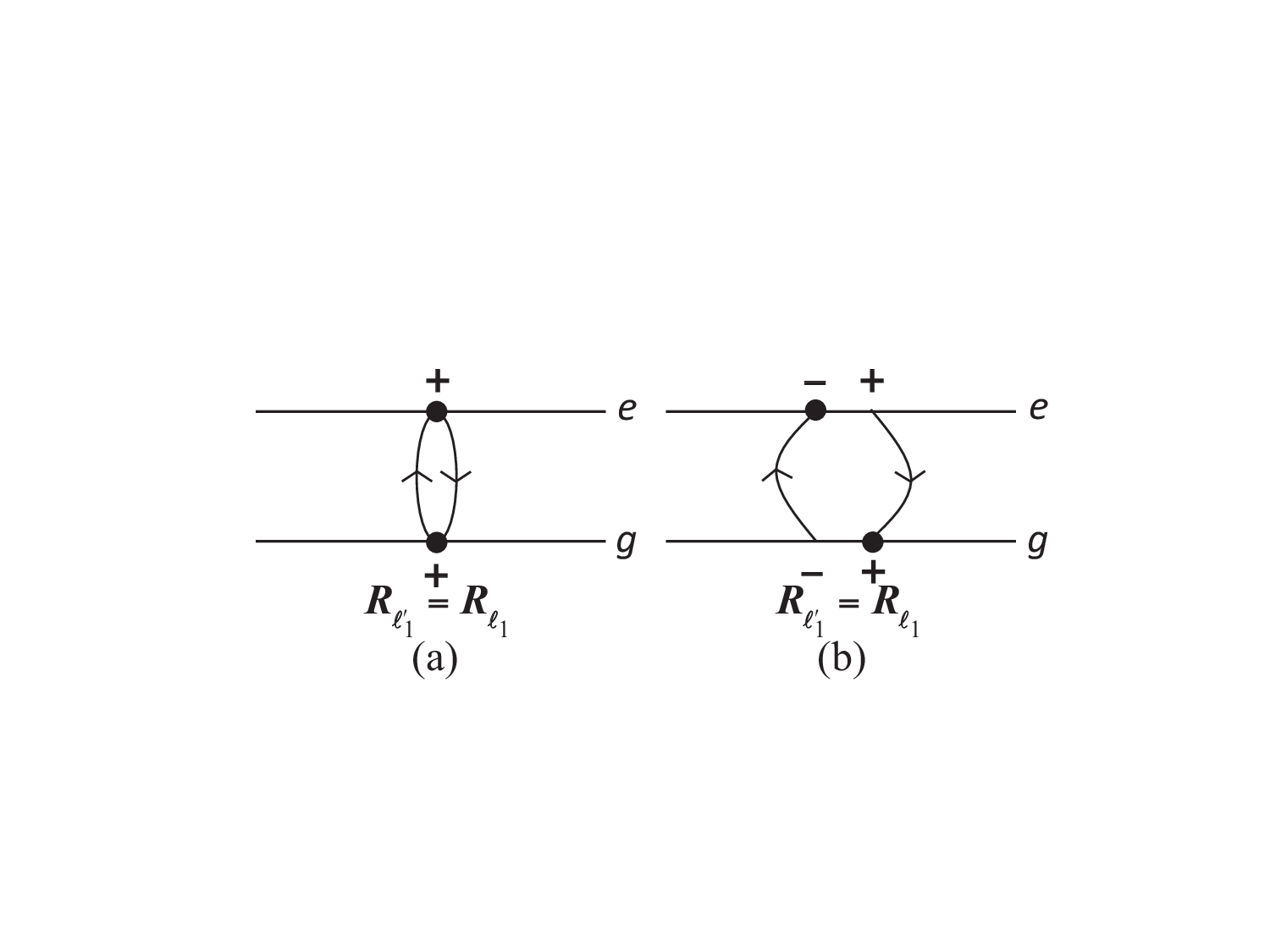}
\vspace{-0.7cm}
\caption{The $\hat{V}^{(inter)}_{eg}$ interaction inside a lattice site, as given in Eq.~(\ref{111}). }
\label{fig12}                                                                  
\end{figure}
 We can also have $\ell'_1=\ell'_2$, which corresponds to Coulomb processes on the same site. These on-site processes can involve  electrons having the same spin (see Fig.~\ref{fig12}(a)) or different spins (see Fig.~\ref{fig12}(b)). The resulting states read
\be\label{111}
\mathcal{V}_{\vR=\bf0}\left(\begin{smallmatrix}
 g& e\\ e& g\end{smallmatrix}\right)|\Phi_{-\sigma_1,-\sigma'_1;\vR_{\ell_1}}\ran
 \,\,\, \,\,\, \,\,\,
\textrm{for}\,\,  \,\,\, \sigma'_1=\pm \sigma_1
\ee

 The $\hat{H}$ Hamiltonian in the ($(2\times 2)\times N_s$) degenerate subspace made of $|\Phi_{\sigma',\sigma;\vR_\ell}\ran$ states, has the same $4\times4$ submatrix $h(\ell_1=\ell_2)$ on the main diagonal, and a set of ($4\times4$) submatrices $h(\ell_1\not=\ell_2)$ out of the diagonal. The $h(\ell_1=\ell_2)$ submatrix, which does not depend on $\ell_1$, appears in the $(++,+-,-+,--)$ basis as
 \be\label{112}
 h(\ell_1=\ell_2)=\left(\begin{matrix}
 E'_e& 0 & 0 & \mathcal{V}^{(e,g)}\\ 0& E'_e+\mathcal{V}^{(e,g)} & 0 &0\\
0& 0& E'_e+\mathcal{V}^{(e,g)} & 0 \\
\mathcal{V}^{(e,g)}&0 &0 &  E'_e\end{matrix}\right)
 \ee
 with $\mathcal{V}^{(e,g)}=\mathcal{V}_{\vR=\bf0}\left(\begin{smallmatrix}
 g& e\\ e& g\end{smallmatrix}\right)$ and 
 \bea
 E'_e &=& \lan \Phi_g| \hat{H}|\Phi_g\ran +(\va_e-\va_g)+(t_{e,e}-t_{g,g})
 +\Big(\mathcal{V}_{\vR=\bf0}\left(\begin{smallmatrix}
 e& e\\ g& g\end{smallmatrix}\right)-\mathcal{V}_{\vR=\bf0}\left(\begin{smallmatrix}
 g& g\\ g& g\end{smallmatrix}\right)\Big)\nn\\
 &&+2\sum_{\vR\not=\bf0}\Big(\mathcal{V}_{\vR}\left(\begin{smallmatrix}
 e& e\\ g& g\end{smallmatrix}\right)-\mathcal{V}_{\vR}\left(\begin{smallmatrix}
 g& g\\ g& g\end{smallmatrix}\right)\Big)\label{113}
 \eea
 
The ($4\times4$) off-diagonal submatrices appear in this basis as
\be\label{114}
h(\ell_1 \not= \ell_2)=\mathcal{V}_{\vR_{\ell_1}-\vR_{\ell_2}}\left(\begin{smallmatrix}
 g& e\\ e& g\end{smallmatrix}\right)\left(\begin{matrix}
1& 0 & 0 & 1\\ 0& 0 & 0 &0\\
0& 0& 0 & 0 \\
1&0 &0 &  1\end{matrix}\right)
\ee

\section{Derivation of Eq.~(\ref{98}) from Eq.~(\ref{97})\label{app:B}}

We consider an \textit{arbitrary} set of orthonormal vectors $({\bf x},{\bf y},{\bf z})$ and we take $\boldsymbol{\mu}$ and  $\boldsymbol{\mu}'$ as any of these three vectors. For $R_\mu=\vR\cdot \boldsymbol{\mu}$, the sum 
\bea
\label{app:1}
 S_\vK(\mu',\mu)&=&\sum_{\vR\not=\bf0}\frac{e^{i\vK\cdot \vR}}{R^3}\left(\delta_{\mu',\mu}-3 \frac{R_{\mu'}R_\mu}{R^2}\right)\nn\\
 &=&\sum_{\vR\not=\bf0}\frac{e^{i\vK\cdot \vR}}{R^3}\left(\boldsymbol{\mu}'\cdot \boldsymbol{\mu}-3 \frac{(\vR\cdot \boldsymbol{\mu}')\, (\vR\cdot \boldsymbol{\mu})}{R^2}\right)
 \eea
is singular  in the $\vK\rightarrow \bf0$ limit: its $\vK=\bf0$ value differs from the $\vK\rightarrow \bf0$ limit, and this limit depends on the $\vK$ direction.

\subsection{Calculation of $S_\vK(\mu',\mu)$ for $\vK=\bf0$}

We first show that $S_\vK(\mu',\mu)$ cancels for $\vK=\bf0$. To do it, we introduce the cubic crystal axes $({\bf X}, {\bf Y}, {\bf Z})$ and we expand the three vectors $(\vR,\boldsymbol{\mu}',\boldsymbol{\mu})$ on these axes 
\bea
\vR&=&R_X {\bf X}+R_Y {\bf Y}+R_Z {\bf Z}\label{app:2}\\
\boldsymbol{\mu}&=& \mu_X {\bf X}+\mu_Y {\bf Y}+\mu_Z {\bf Z}\label{app:3}
\eea
with a similar expansion for $\boldsymbol{\mu}'$. As $\vR\cdot \boldsymbol{\mu}=R_X \mu_X+ R_Y\mu_Y+R_Z\mu_Z$, the product $(\vR\cdot \boldsymbol{\mu}')( \vR\cdot \boldsymbol{\mu})$ in Eq.~(\ref{app:1}) reads 
\bea
(\vR\cdot \boldsymbol{\mu}') (\vR\cdot \boldsymbol{\mu})&=&\Big(R_X^2 \mu'_X\mu_X + R_Y^2 \mu'_Y\mu_Y +R_Z^2 \mu'_Z\mu_Z\Big)\nn\\
&&+ \Big(R_XR_Y (\mu_X\mu'_Y+\mu_Y\mu'_X) + \cdots\Big)\label{app:4}
\eea
As a result, $S_\vK(\mu',\mu)$ taken for $\vK=\bf0$ contains the following terms, which, for $\vR$ written in terms of 
$(R_X,R_Y,R_Z)$, reduce through symmetry, either to
\bea
\sum_{\vR\not=\bf0}\frac{R_X^2}{R^5}\!\!\!\!&=&\!\!\!\!\sum_{\vR\not=\bf0}\frac{R_Y^2}{R^5}=\sum_{\vR\not=\bf0}\frac{R_Z^2}{R^5}
=\frac{1}{3}\sum_{\vR\not=\bf0}\frac{R_X^2{+}R_Y^2{+}R_Z^2}{R^5}=\frac{1}{3}\sum_{\vR\not=\bf0}\frac{1}{R^3}\label{app:5}
\eea
or to
\bea
\sum_{\vR\not=\bf0}\frac{R_XR_Y}{R^5}\!\!\!\!&=&\!\!\!0\label{app:6}
\eea

When used into Eq.~(\ref{app:1}), this leads to
\bea
 S_{\vK=\bf0}(\mu',\mu)&=&\sum_{\vR\not=\bf0}\frac{1}{R_3}\left(\boldsymbol{\mu}'\cdot \boldsymbol{\mu}{-}\mu'_X\mu_X {-} \mu'_Y\mu_Y {-} \mu'_Z\mu_Z\right)\label{app:7}\\
 &=&0\nn
\eea

\subsection{Calculation  of $S_\vK(\mu',\mu)$  in the $\vK\rightarrow\bf0$ limit}

Since $S_{\vK=\bf0}(\mu',\mu)=0$, we can rewrite $S_\vK(\mu',\mu)$ as 
\be\label{app:8}
 S_\vK(\mu',\mu)=\sum_{\vR\not=\bf0}\frac{e^{i\vK\cdot \vR}-1}{R^3}\left(\delta_{\mu',\mu}-3 \frac{R_{\mu'}R_\mu}{R^2}\right)
 \ee
which readily shows that the $S_\vK(\mu',\mu)$ limit for $\vK\rightarrow\bf0$ is controlled by its large-$\vR$ terms. As a result, we can replace the $S_\vK(\mu',\mu)$ discrete sum over the vectors $\vR$ of a cubic lattice with cell size $a_c$, by an integral according to
\be\label{app:9}
\sum_\vR f(\vR)\simeq \frac{1}{a_c^3}\int d^3R \,\, f(\vR)
\ee

A convenient way to calculate $S_{\vK}(\mu',\mu)$ in the small $\vK$ limit is to introduce the orthonormal vectors $({\bf x}_\vK, {\bf y}_\vK,{\bf z}_\vK)$ with ${\bf z}_\vK=\vK/K$ and to expand the three vectors $(\vR,\boldsymbol{\mu}',\boldsymbol{\mu})$ on these vectors as
\bea
\vR&=& R(\sin\theta \cos\varphi\, {\bf x}_\vK+ \sin\theta \sin\varphi\, {\bf y}_\vK+ \cos\theta\, {\bf z}_\vK)\label{app:10}\\
\boldsymbol{\mu}&=& \alpha \,{\bf x}_\vK  +\beta\,{\bf y}_\vK +\gamma\, {\bf z}_\vK  \label{app:11}
\eea
with a similar expansion for $\boldsymbol{\mu}'$. This gives
\bea
\label{app:12}
\lim_{\vK \rightarrow \bf0}S_{\vK}(\mu',\mu)\simeq \frac{1}{a_c^3}\int_0^\infty R^2 dR \int_0^\pi \sin\theta d\theta \int_0^{2\pi}d\varphi \frac{e^{iKR \cos\theta}}{R^3} 
\Big(\alpha'\alpha+\beta'\beta+\gamma'\gamma 
\nn\hspace{0.5cm} \\
 -3(\alpha'\sin\theta\cos\varphi+\beta'\sin\theta \sin\varphi+\gamma'\cos\theta)
 %\,\,\,\,\,\,\,\,\,\,\,\,\,\,\,\,\,\,\,\,\,\,\,\,\,\,\,\,\,\,\,\,\,\,\,\,
 (\alpha\sin\theta\cos\varphi+\beta\sin\theta \sin\varphi+\gamma\cos\theta)\Big)\nn
\eea
The integration over $\varphi$ reduces the above equation to
\bea\label{app:13}
\lim_{\vK \rightarrow \bf0}S_{\vK}(\mu',\mu)&\simeq& \frac{2\pi}{a_c^3}\int_0^\infty \frac{ dR}{R} \int_0^\pi \sin\theta d\theta\, e^{iKR \cos\theta}\\
&&\times\Big((\alpha'\alpha+\beta'\beta+\gamma'\gamma)-\frac{3}{2} (\alpha'\alpha+\beta'\beta)\sin^2\theta -3 \gamma'\gamma \cos^2\theta\Big)\nn
\eea
So, by setting $x=KR$, this also reads
\bea
\lim_{\vK \rightarrow \bf0}S_{\vK}(\mu',\mu)\!\!\!&\simeq&\!\!\! \frac{\pi}{a_c^3} (- \alpha'\alpha-\beta'\beta+ 2\gamma'\gamma)\int_0^\infty \frac{dx}{x} \int_0^\pi \sin\theta d\theta \, e^{ix\cos\theta}
 (1-3\cos^2\theta)\nn\\
\!\!\!&=& \!\!\!\frac{4\pi}{3a_c^3}(- \alpha'\alpha-\beta'\beta+ 2\gamma'\gamma)\label{app:14}
\eea
as obtained from
\be\label{app:15}
\int_0^\infty \frac{dx}{x}\int_{-1}^1 dt(1-3t^2)\, e^{ixt}=\frac{4}{3}
\ee

Since $\gamma=\vK\cdot\boldsymbol{\mu}/K=K_\mu/K$, while $(- \alpha'\alpha-\beta'\beta+ 2\gamma'\gamma)=3\gamma'\gamma-( \alpha'\alpha+\beta'\beta+ \gamma'\gamma)$, we ultimately find
 \be\label{app:16}
 \lim_{\vK\rightarrow \bf0}S_\vK(\mu',\mu)=-\frac{4\pi}{3a_c^3}\left(\delta_{\mu',\mu}-3 \frac{K_{\mu'}K_\mu}{K^2}\right)
 \ee
Note the sign change between the two terms of $S_\vK(\mu',\mu)$ in Eq. (\ref{app:1}) and the two terms of its $\vK\rightarrow\bf0$ limit. 

\section{Large $\vR$ expansion of $v_{\mu',\mu}$ \label{app:A}}

The large-$\vR$ expansion to the Coulomb part of $v_{\mu',\mu}$ defined in Eq.~(\ref{87}) as
 \bea
 \label{F.1}
 v_{\mu',\mu}&=& \iint_{L^3}d^3rd^3r' \,|\lan \vr|g\ran|^2
\Big(\lan \mu',e|\vr'\ran \lan \vr'| \mu,e\ran- \delta_{\mu',\mu} |\lan \vr'|g\ran|^2 \Big)\nn\\
&&\times
\sum_{\vR\not=\bf0}\left[\frac{e^2}{|\vR+\vr-\vr'|}-\frac{e^2}{|\vR-\vr'|}\right]
\eea
reads as
\bea
\frac{1}{|\vR+\vr-\vr'|}-\frac{1}{|\vR-\vr'|}\simeq 
-\frac{1}{2R}\left(\frac{2\vR}{R}\cdot\frac{\vr}{R}+\frac{-2\vr\cdot\vr'+\vr^2}{R^2}\right)\nn\hspace{2cm}\\
+\frac{3}{8R}\left[-2\left(\frac{2\vR}{R}\cdot\frac{\vr'}{R}\right)\left(\frac{2\vR}{R}\cdot\frac{\vr}{R}\right)+\left(\frac{2\vR}{R}\cdot\frac{\vr}{R}\right)^2 \right]\label{app:01}
\eea
When used into  Eq.~(\ref{F.1}), we find that, as $\int d^3 r |\lan \vr|g\ran|^2 \vr=0$, the nonzero terms of this large-\textbf{R} limit reduce to
\bea
 v_{\mu',\mu}&\simeq& \iint_{L^3}d^3rd^3r' |\lan \vr|g\ran|^2\Big(\lan \mu',e|\vr'\ran \lan \vr'| \mu,e\ran- \delta_{\mu',\mu} |\lan \vr'|g\ran|^2 \Big)\nn\\
 &&\sum_{\vR\not=\bf0}\frac{1}{2R^3}\left[-\vr^2{+}\frac{3}{R^2}\left(R_X^2{r}_X^2{+}R_Y^2{r}_Y^2{+}R_Z^2{r}_Z^2\right)\right] \label{app:02}
\eea
for $\vR$ written as $R_X {\bf X}+R_Y {\bf Y}+R_Z {\bf Z}$ in terms of the cubic crystal axes 
$({\bf X}, {\bf Y}, {\bf Z})$. Since for cubic symmetry,
\be
\sum_{\vR\not=\bf0}\frac{R_X^2}{R^5}=\sum_{\vR\not=\bf0}\frac{R_Y^2}{R^5}=\sum_{\vR\not=\bf0}\frac{R_Z^2}{R^5}=\frac{1}{3}\sum_{\vR\not=\bf0}\frac{1}{R^3}
\ee
we readily find that Eq.~(\ref{app:02}) reduces to zero. So, $ v_{\mu',\mu}$ scales as $\mathcal{O}(R^{-4})$ for $\vR$ large.

\section{Coulomb interaction $\hat{V}_{gg}$ in terms of holes\label{app:G}}

\noindent $\bullet$ Equation (\ref{146}) gives the Coulomb interaction $\hat{V}_{gg}$  between ground-level electrons, in terms of electron operators. To write it in terms of holes, we first note that
\bea
\hat{a}^\dag_{s_1,\mu'_1,g,\ell_1}\hat{a}^\dag_{s_2,\mu'_2,g,\ell_2} \hat{a}_{s_2,\mu_2,g,\ell_2}  \hat{a}_{s_1,\mu_1,g,\ell_1}\nn\hspace{3cm}\\
=\hat{a}^\dag_{s_1,\mu'_1,g,\ell_1}\Big( \delta_{\mu'_2,\mu_2}-\hat{a}_{s_2,\mu_2,\ell_2} \hat{a}^\dag_{s_2,\mu'_2,\ell_2} \Big)  \hat{a}_{s_1,\mu_1,g,\ell_1}
\eea
 which becomes in terms of hole operators  
\bea
\Big(\delta_{\mu'_1,\mu_1}\delta_{\mu'_2,\mu_2}-\delta_{s_1,s_2}\delta_{\mu'_1,\mu_2}\delta_{\mu'_2,\mu_1}\delta_{\ell_1,\ell_2}\Big)\nn\hspace{3.5cm}\\
+\bigg[\delta_{s_1,s_2}\delta_{\ell_1,\ell_2}\Big(\delta_{\mu'_1,\mu_2}\hat{b}^\dag_{-s_1,\mu_1,\ell_1}\hat{b}_{-s_2,\mu'_2,\ell_2} +(1\longleftrightarrow 2) \Big) \nn\\
-\Big(\delta_{\mu'_1,\mu_1} \hat{b}^\dag_{-s_2,\mu_2,\ell_2} \hat{b}_{-s_2,\mu'_2,\ell_2}+(1\longleftrightarrow 2)\Big)\bigg] \nn\\
+  \hat{b}^\dag_{-s_1,\mu_1,\ell_1}\hat{b}^\dag_{-s_2,\mu_2,\ell_2} \hat{b}_{-s_2,\mu'_2,\ell_2} \hat{b}_{-s_1,\mu'_1,\ell_1}
\eea
When used into $\hat{V}_{gg}$ given in Eq.~(\ref{146}), we get this operator as
\be\label{app:153}
\hat{V}_{gg}=
V_g-\sum \mathcal{V}_g(\mu',\mu)\,\hat{b}^\dag_{s,\mu,\ell}\hat{b}_{s,\mu',\ell}+\cdots
\ee
 The constant part $V_g$ is given by 
 \bea
V_g&=& \sum \big(\delta_{\mu'_1,\mu_1}\delta_{\mu'_2,\mu_2}-\delta_{s_1,s_2}\delta_{\mu'_1,\mu_2}\delta_{\mu'_2,\mu_1}\delta_{\ell_1,\ell_2}\big) \mathcal{V}_{\vR_{\ell_1}-\vR_{\ell_2}}\left(\begin{smallmatrix}
\mu'_2,g&\mu_2, g\\ \mu'_1,g&\mu_1, g\end{smallmatrix}\right)\nn\\
&=&N_s \sum_{\mu_1,\mu_2} \Big( \mathcal{V}_{\vR=\bf0}\left(\begin{smallmatrix}
\mu_2,g&\mu_2, g\\ \mu_1,g&\mu_1, g\end{smallmatrix}\right)   +       2\sum_{\vR\not=\bf0}\mathcal{V}_{\vR}\left(\begin{smallmatrix}
\mu_2,g&\mu_2, g\\ \mu_1,g&\mu_1, g\end{smallmatrix}\right)   \Big)\nn\\
&&
+N_s \sum_{\mu_1\not=\mu_2}\Big(\mathcal{V}_{\vR=\bf0}\left(\begin{smallmatrix}
\mu_2,g&\mu_2, g\\ \mu_1,g&\mu_1, g\end{smallmatrix}\right)-\mathcal{V}_{\vR=\bf0}\left(\begin{smallmatrix}
\mu_1,g&\mu_2, g\\ \mu_2,g&\mu_1, g\end{smallmatrix}\right) \Big)
%\label{154}\\
\eea
while the prefactors $\mathcal{V}_g(\mu',\mu)$ in the one-body part of $\hat{V}_{gg}$ are given by
\bea
\mathcal{V}_g(\mu',\mu)&=&
 \sum_{\mu_1}  \Big(  \mathcal{V}_{\vR=\bf0}\left(\begin{smallmatrix}
\mu',g&\mu, g\\ \mu_1,g&\mu_1, g\end{smallmatrix}\right)  +2\sum_{\vR\not=\bf0}\mathcal{V}_{\vR}\left(\begin{smallmatrix}
\mu',g&\mu, g\\ \mu_1,g&\mu_1, g\end{smallmatrix}\right) 
 \Big)\nn\\
&&+ \sum_{\mu_1\not=(\mu',\mu)}\Big(\mathcal{V}_{\vR=\bf0}\left(\begin{smallmatrix}
\mu',g&\mu, g\\ \mu_1,g&\mu_1, g\end{smallmatrix}\right)-\mathcal{V}_{\vR=\bf0}\left(\begin{smallmatrix}
\mu',g&\mu_1, g\\ \mu_1,g&\mu, g\end{smallmatrix}\right) \Big)\label{155}
\eea

\end{document}